# The link between unemployment and real economic growth in developed countries

Ivan O. Kitov, Institute of Geosphere Dynamics, RAS
Oleg O. Kitov

**Abstract**
Ten years ago we presented a modified version of Okun's law for the biggest developed economies and reported its excellent predictive power. In this study, we revisit the original models using the estimates of real GDP per capita and unemployment rate between 2010 and 2019. The initial results show that the change in unemployment rate can be accurately predicted by variations in the rate of real economic growth. There is a discrete version of the model which is represented by a piecewise linear dependence of the annual increment in unemployment rate on the annual rate of change in real GDP per capita. The lengths of the country-dependent time segments are defined by breaks in the GDP measurement units associated with definitional revisions to the nominal GDP and GDP deflator (dGDP). The difference between the CPI and dGDP indices since the beginning of measurements reveals the years of such breaks. Statistically, the link between the studied variables in the revised models is characterized by the coefficient of determination in the range from $R^2$=0.866 (Australia) to $R^2$=0.977 (France). The residual errors can be likely associated with the measurement errors, e.g. the estimates of real GDP per capita from various sources differ by tens of percent. The obtained results confirm the original finding on the absence of structural unemployment in the studied developed countries.

Key words: unemployment, GDP, modelling, Okun's law
JEL classification: J65

## Introduction

The COVID-19 pandemic is an extremely active exogenous non-economic force affecting the quasi-stationary evolution of developed economies. In the beginning of 2021, one of extensively discussed topics is related to the dramatic fall in the unemployment rate forced by lockdowns and closure of all except vital businesses. In the United States, the rate unemployment jumped to 12.9% in the second quarter of 2020 and dropped to 8.9% in the third quarter. In the fourth quarter, the Bureau of Labor Statistics (BLS) reported further fall in the rate of unemployment due to quick economic recovery. This pattern is quite different from the unemployment persistence observed during the 2008-2009 financial crises.

The observed large-amplitude changes in unemployment and real economic growth provide the best opportunity to assess the performance of multiple economic theories and to test the hypothesis of tangible structural changes in the labour market. Do we really observe major changes in the overall economic organization when significant parts of it become unnecessary as in the COVID-19 pandemic? This and many other theoretical assumptions and experimental facts have to be validated by successful economic models in order to pass consistency and quantitative accuracy tests. In this paper, the modified Okun's law is tested against the new data with the 2020 data playing the principal role for the USA.

We re-estimate and validate the model developed in (Kitov&Kitov, 2011). It was revised once in (Kitov, 2011) using the data on GDP per capita provided by the Conference Board (2011) and data on unemployment from the OECD (2011). New data sets are crucial for testing the statistical agreement between the change in unemployment rate and real GDP per capita. Therefore, the annual readings between 2010 and 2019, as well as quarterly estimates in 2020, are able to improve, and thus, validate the model. They can also destroy the previously achieved statistical significance of the underlying relationships and reject the model.

One of important features in the modified Okun's law is the introduction of discontinuities in the linear relationship in order to accommodate the artificial changes in the definitions of real GDP and unemployment. Real shifts in the relationship between unemployment and real economic growth are not excluded, however. Such a structural, i.e. inherently economic, break should be detached from the revisions to GDP and unemployment definitions. It is demonstrated that the definitional revisions introduced by economic agencies like the Bureau of Economic Analysis (BEA) and the BLS just change the units of measurements like one changes the speed measurement units from km/h to mph when crossing the Canada-US border, with the physical speed retained.

The modified Okun's law is estimated for the biggest developed countries: the United States, France, Germany, the United Kingdom, France, Australia, Canada, and Spain. The revised results have high statistical significance and suggest the absence of structural unemployment in the studied developed countries.

## 1. Model

According to the gap version of Okun's law, there exists a negative relation between the output gap, $(Y^p-Y)/Y^p$, where $Y^p$ is potential output at full employment and $Y$ is actual output, and the deviation of actual unemployment rate, $u$, from its natural rate, $u^n$. It is important to highlight that Okun's law implies constant potential output and natural rate of unemployment. The modified model is dynamic and the same rate of unemployment may be the result of various trajectories in the real GDP evolution. The initial conditions may also be defining when major non-economic events significantly change population pyramid and economic structure. For example, the economic transition from socialism to capitalism in East European and Former Soviet Union countries followed various paths with different levels of real GDP per capita and rates of unemployment in the beginning of capitalistic development. The reunification of Germany is another example. Therefore, our model proposes a dynamic approach to the interpretation of the link between unemployment and real economic growth.

The real GDP or output depends not only on economic performance but also on the change in population as an extensive component, which is not necessary dependent on other macroeconomic variables. For example, the USA population was growing by 1% and more per year between 1960 and 2010 due to immigration and high birth rate. Since 2010, this growth rate is systematically below 1% per year. The population growth in developed European countries was close to zero or even negative in the 20th century. In Japan, the total population dropped by 1.6 million between 2008 and 2018.

Econometrically, it is mandatory to use macroeconomic variables of the same origin and dimension. The unemployment rate does not depend on the total population. Thus, we use real GDP per capita, $G$, and rewrite Okun's law in the following form:

$$du = a + bdlnG \qquad (1)$$

where $du$ is the change in the rate of unemployment per unit time (say, 1 year); $dlnG=dG/G$ is the relative change rate in real GDP per capita, $a$ and $b$ are empirical coefficients. Okun's law implies $b<0$. The intuition behind Okun's law is very simple. Everybody may feel that the rate unemployment is likely to rise when real economic growth is very low or negative. An economy needs fewer employees to produce the same or smaller real GDP also because of the labor productivity growth.

When integrated between $t_0$ and $t$, equation (1) can be rewritten in the following form:

$$u_t = u_0 + bln[G_t/G_0] + a(t-t_0) + c \quad (2)$$

where $u_t$ is the rate of unemployment at time $t$. The intercept $c \equiv 0$, as is clear for $t=t_0$. This form of the modified Okun's law explicitly links the current rate of unemployment with the initial rate, $u_0$. In the absence of economic growth, i.e. $G_t/G_0 = 1$, the rate of unemployment has to increase if $a > 0$ and to decrease for $a < 0$. It is also important that for $a<0$ the rate of real GDP per capita growth has to be $ln[G_t/G_0] > -a(t-t_0)/b$, where $b<0$. This condition means that economic growth not necessarily results in a decreasing rate of unemployment. For a weak economic growth, unemployment may grow.

The continuous form (2) is easily transformed in a discrete one, and the integral is replaced by a cumulative sum of the annual estimates of $dlnG$ with the appropriate initial conditions. By definition, the cumulative sum of the observed annual $du$'s is a discrete time series of the unemployment rates, $u_t$. Statistically, the use of levels, i.e. $u$ and $G$, instead of their differentials is superior due to potential suppression of the uncorrelated measurement errors.

Kitov (2011) showed the necessity of discontinuities in (1). Therefore, we introduced floating breaks in (2), with the precise years to be determined by the best fit. These are not actual structural breaks related to the change in economic structure, but artificial or definitional breaks. We also propose an independent procedure to assess the tentative break years. Thus, relationship (2) can be split into N segments. The integral form of the modified Okun's law should be also split into N time segments:

$$\begin{aligned}
u_t &= u_0 + b_1 ln[G_t/G_0] + a_1(t-t_0), \quad t<t_1 \\
u_t &= u_1 + b_2 ln[G_t/G_1] + a_2(t-t_1), \quad t_2 \geq t \geq t_1 \\
&\ldots \\
u_t &= u_{N-1} + b_N ln[G_t/G_{tN-1}] + a_N(t-t_{N-1}), \quad t_N \geq t \geq t_{N-1}
\end{aligned} \quad (3)$$

The model is obtained by statistical estimation of the break years and linear regression coefficients, which provide the lowermost RMS model residual for the entire studied period. There is also a possibility to introduce dummy variables when necessary. In some time series steps and spikes are not excluded. We did not use this option so far in the model.

Formally, relationship (3) suggests that a negative rate of unemployment can be reached when the rate of real GDP per capita growth is high during a longer period of time. This is a model limitation if to consider the real economic growth and unemployment as independent parameters. The real GDP growth, however, is limited by the available workforce, i.e. it needs the unemployed to get job and more people not in the labor force to get job as well. Both processes are constrained by the availability of work-capable people in a given economy. When the limit is reached, further workforce increase is not possible as well as high rate of economic growth as related to the workforce. The only source of further economic growth is the change in labor productivity. Therefore, the negative rate of unemployment is not possible because the real economic growth is limited by human resources. There exits an effective measure to increase the long-term rate of real economic growth - migration. The USA has been using this remedy for decades. In the 21st century, many developed countries in Europe also understood the importance of population growth. Correspondingly, migration to the EU from the countries with low GDP per capita is one of the most effective economic drivers now. A similar process is observed in Russia - millions of migrants from the FSU countries play an important role in economic development. At the same time, there are closed economies with decreasing population. These countries demonstrate poorer economic performance.

## 2. Countries

### *2.1. USA*

In 2011, we started with the USA model (Kitov&Kitov, 2011). The LSQR method applied to the integral form of the modified Okun's law (3) resulted in the following relationship:

$$du_p = -0.406dlnG + 1.113, 1979>t\geq 1951$$
$$du_p = -0.465dlnG + 0.866, 2010\geq t\geq 1979 \quad (4)$$

An artificial break around 1979 was found. It divides the whole 60-year-long interval into two practically equal segments. The agreement between the predicted and measured rates of unemployment was excellent with a standard model error of 0.55%. The average rate of unemployment for the same period was 5.75%, with the mean absolute annual increment of 1.1%. This is an accurate model of unemployment with $R^2$=0.89. Hence, our model explained 89% of the variability in the rate of unemployment between 1951 and 2010.

We noticed (Kitov, 2011) that the break year (1979) corresponds to the point in time when the estimates of CPI and GDP deflator started to diverge. Figure 1 displays three price indices: CPI (BLS), PCE, and GDP deflator (BEA) in the USA between 1929 and 2019. The CPI curve starts to deviate from the PCE and dGDP ones around 1979 expressing the introduction of new components in the GDP definition. The goods and products included in the CPI are a part of the GDP, and thus, the GDP deflator suffers the artificial break relative to the CPI, which is considered as a benchmark.

The difference between the CPI and dGDP in Figure 2 illustrates the revision to the GDP definition. Between 1979 and 2010, this difference can be accurately described by a linear function of time. In other words, the dGDP is measured in slightly different units since 1979, which are just proportional to the units used in the previous period. Figure 3 presents two original and two corrected CPI and dGDP curves. When multiplied by a factor of 1.22, the dGDP curve is almost fully converted into the CPI curve, as supposed by the linear link between the CPI and dGDP. The period after 2010, demonstrates the start of a new deviation between the CPI and corrected dGDP curves. This deviation is also seen in Figure 2, where the segment after 2010 has a different slope compared to the previous segment. A new multiplication factor is likely needed after 2010 to match the CPI and dGDP.

Figure 4 extends our study of the CPI and dGDP in the USA by data from the OECD database (OECD, 2020). The slope in the *CPI* and *dGDP* linear relationship for the OECD between 1979 and 2010 is 1.26, i.e. slightly larger than that for the BLS and BEA data. After 2010, the best fit between the *CPI* and *dGDP* curves is obtained with a coefficient 1.26*0.8 =1.0, i.e. the CPI and dGDP price indices have been growing in sync since 2010, as it was between 1929 and 1979. The shelf observed in the difference curve in Figure 2 has the same sense - the *CPI* and *dGDP* are almost identical. It is also worth noting that the estimates of the same economic variables are different for the US agencies and the OECD. This might be related to the definitions developed and applied in different sources, but for an accurate quantitative analysis based on statistical principles this facts manifests lower data quality and reliability in all sources.

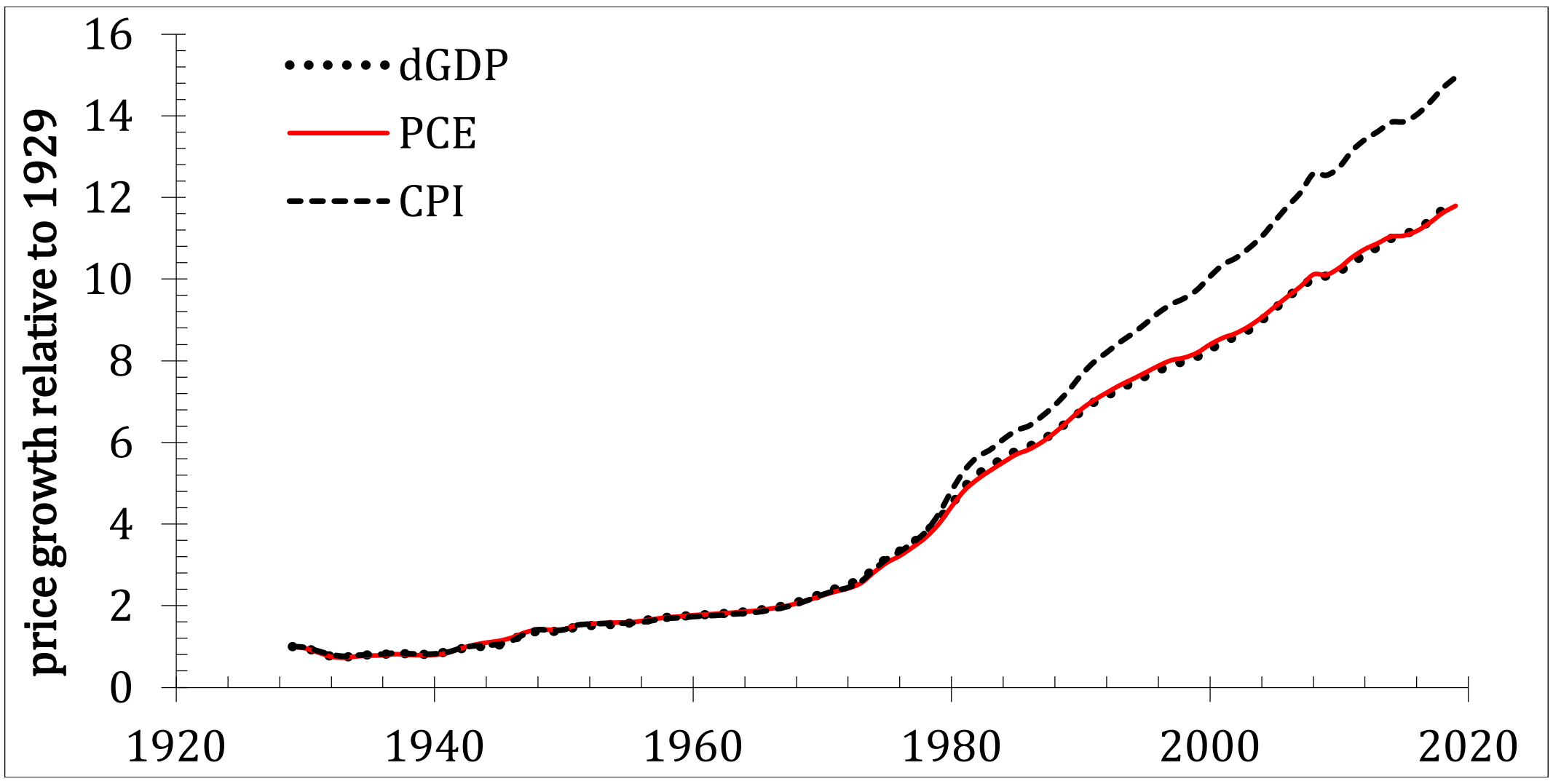

Figure 1. The CPI, dGDP, and PCE indices in the USA between 1929 and 2019. The CPI curve starts to diverge around 1979.

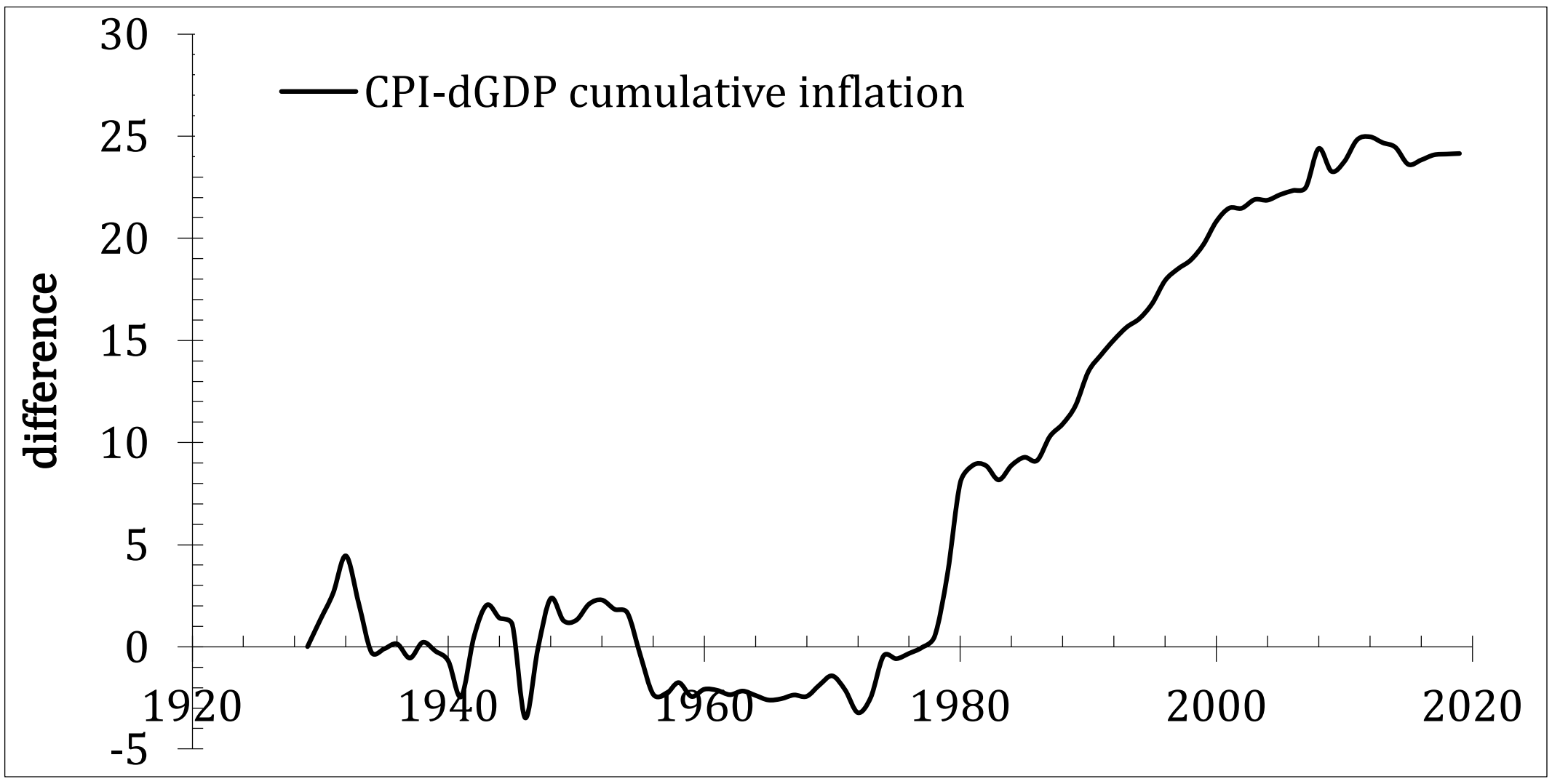

Figure 2. The difference between the CPI and dGDP illustrates the revision to the GDP definition.

One can also extend the dGDP definition used between 1979 and 2010 into the past using the same linear link: *dGDP*=0.8**CPI*. This corrected segment of the dGDP is shown in Figure 3. Obviously, since the *dGDP* is by 20% lower than the *CPI*, the total growth in real GDP per capita since 1929 is larger than that reported by the BEA. Therefore, the corrected dGDP can be used for a more accurate estimate of the real GDP per capita in the past. We present this extended analysis of the *dGDP* behavior in the USA in order to justify the breaks in the modified Okun's law and to assess the BLS, BEA, and OECD data quality as a crucial component of the statistical analysis.

Our initial model for the USA worked well and its performance can be further validated and improved by new data. Almost ten years passed since 2010, and now we have two opportunities to check the model: new readings for the previous years since 2010 and the extremely deep fall in the real GDP per capita accompanied by an unprecedented growth in the rate of unemployment in the USA, both induced by the COVID-19 pandemic. The latter is a dynamic effect of an external force.

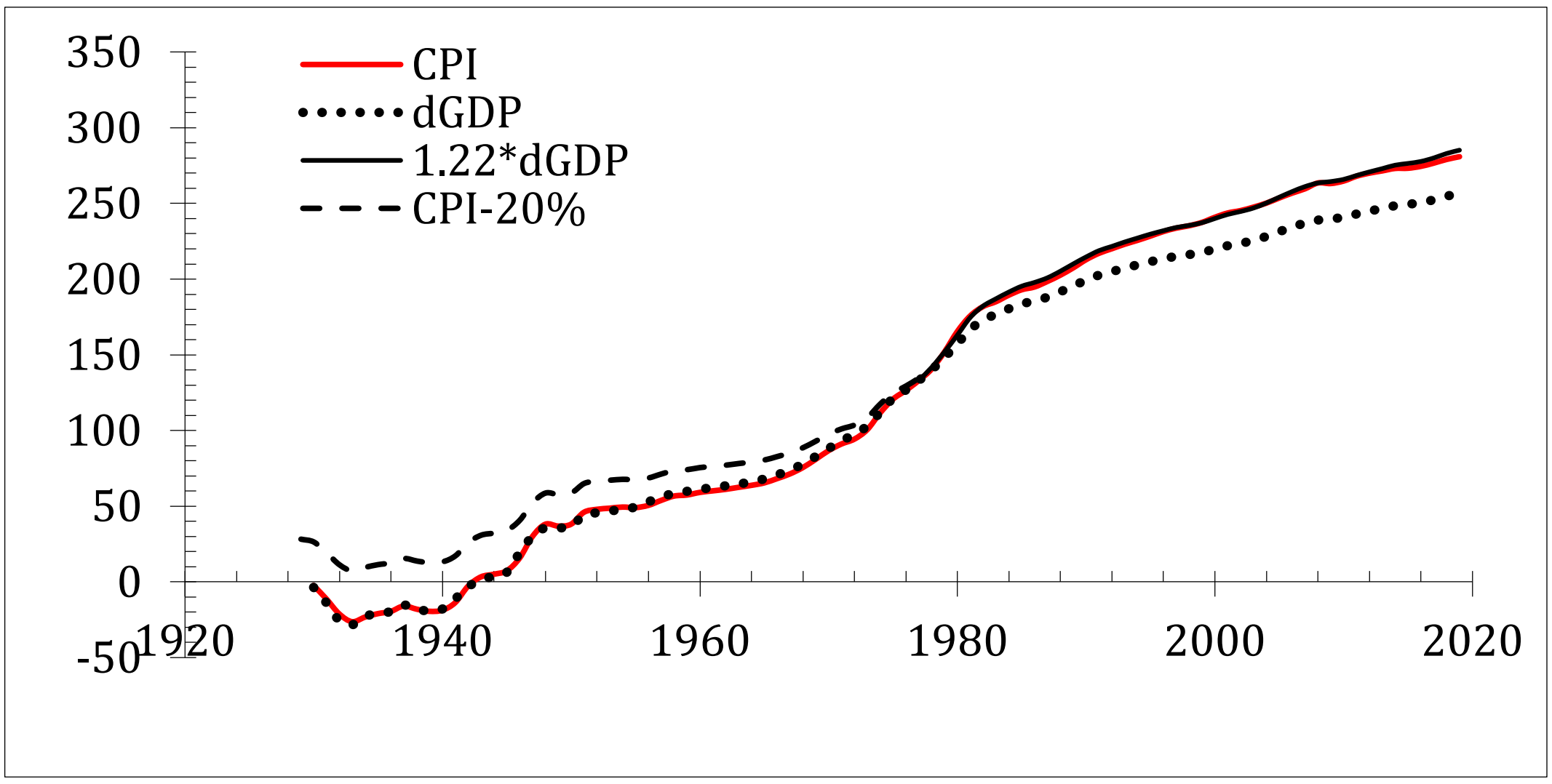

Figure 3. The *dGDP* curve corrected to the *CPI* as a benchmark after and before 1979. After 2010, the curves start to diverge again.

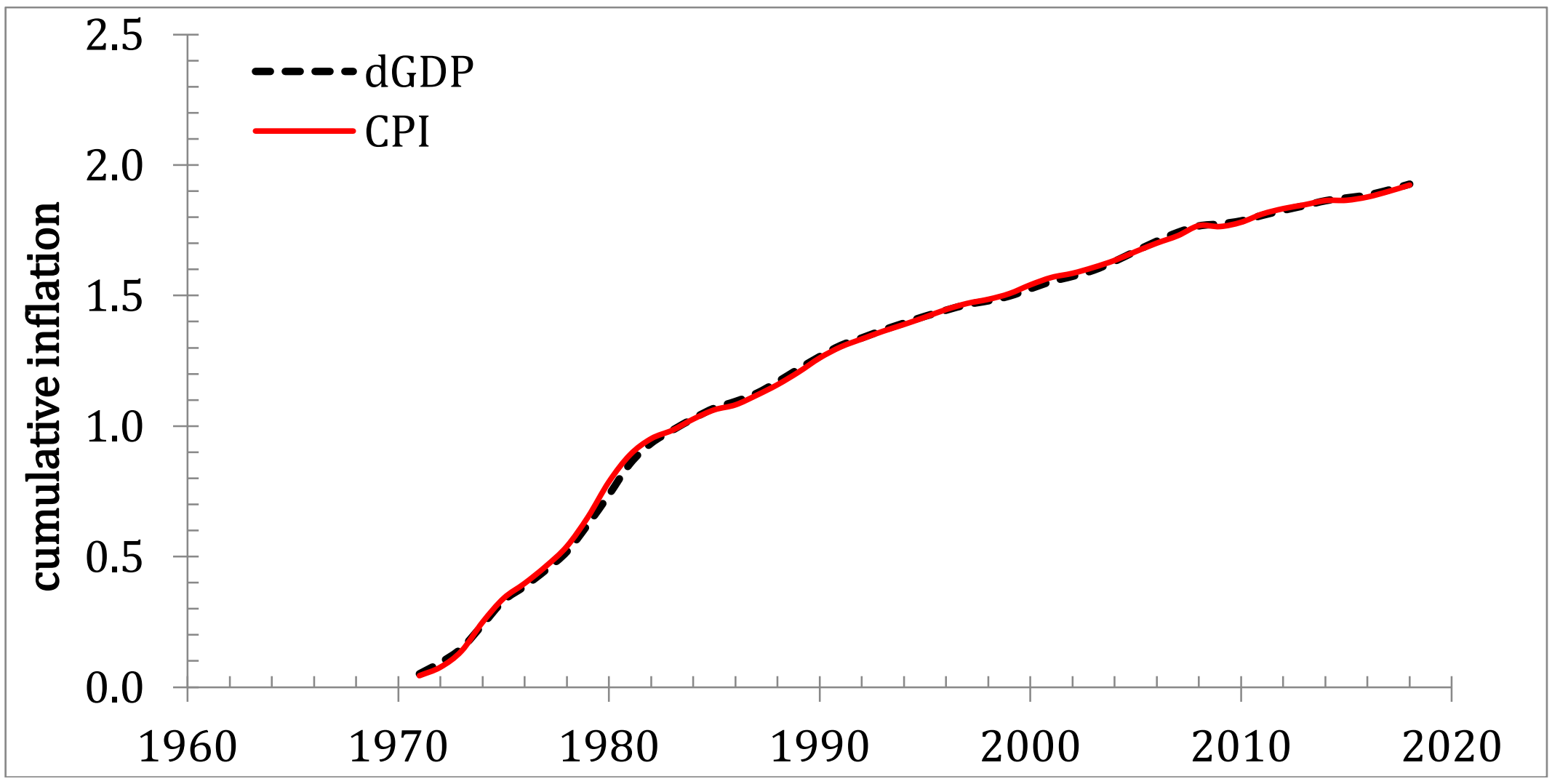

Figure 4. A new coefficient of 0.8 is used to fit the *CPI* and *dGDP* after 2010. The fit is good and the total factor for this period is 0.8*1.26=1.00. It seems the original GDP definition is used after 2010.

The main results of our meticulous inspection of the *CPI* and *dGDP* deviation are the presence of two breaks in data due to major revisions to the real GDP definition. One of these two breaks was used in our version of Okun's law for the USA as described by equation (4). The 2010 break may extend equation (4) to three different segments: 1951 to 1979, 1980 to 2010, and after 2010, with three different sets of coefficients. When the 1979-to-2010 set of coefficients is applied to the data after 2010 one obtains the curve shown in Figure 5, which does not match the measured rate of unemployment. Therefore, we apply a standard LSQR procedure to estimate a new set of coefficients for the period after 2010. This analysis gives the following model:

$$du_p = -0.406dlnG + 1.122, 1979>t\geq 1951$$
$$du_p = -0.465dlnG + 0.899, 2010\geq t\geq 1979$$
$$du_p = -0.260dlnG - 0.250, \quad t\geq 2010 \quad (5)$$

It is important to stress that the most recent time segment has a negative constant $a$=-0.25. This means that the $du_p$ in (5) is negative and the rate of unemployment decerases with time even in the absence of real economic growth. The fall in the unemployment rate from 9.63% in 2010 to 3.67% in 2019 was not accompanied by a stellar economic growth. The real GDP per capita was growing by 1.6% per year on average.

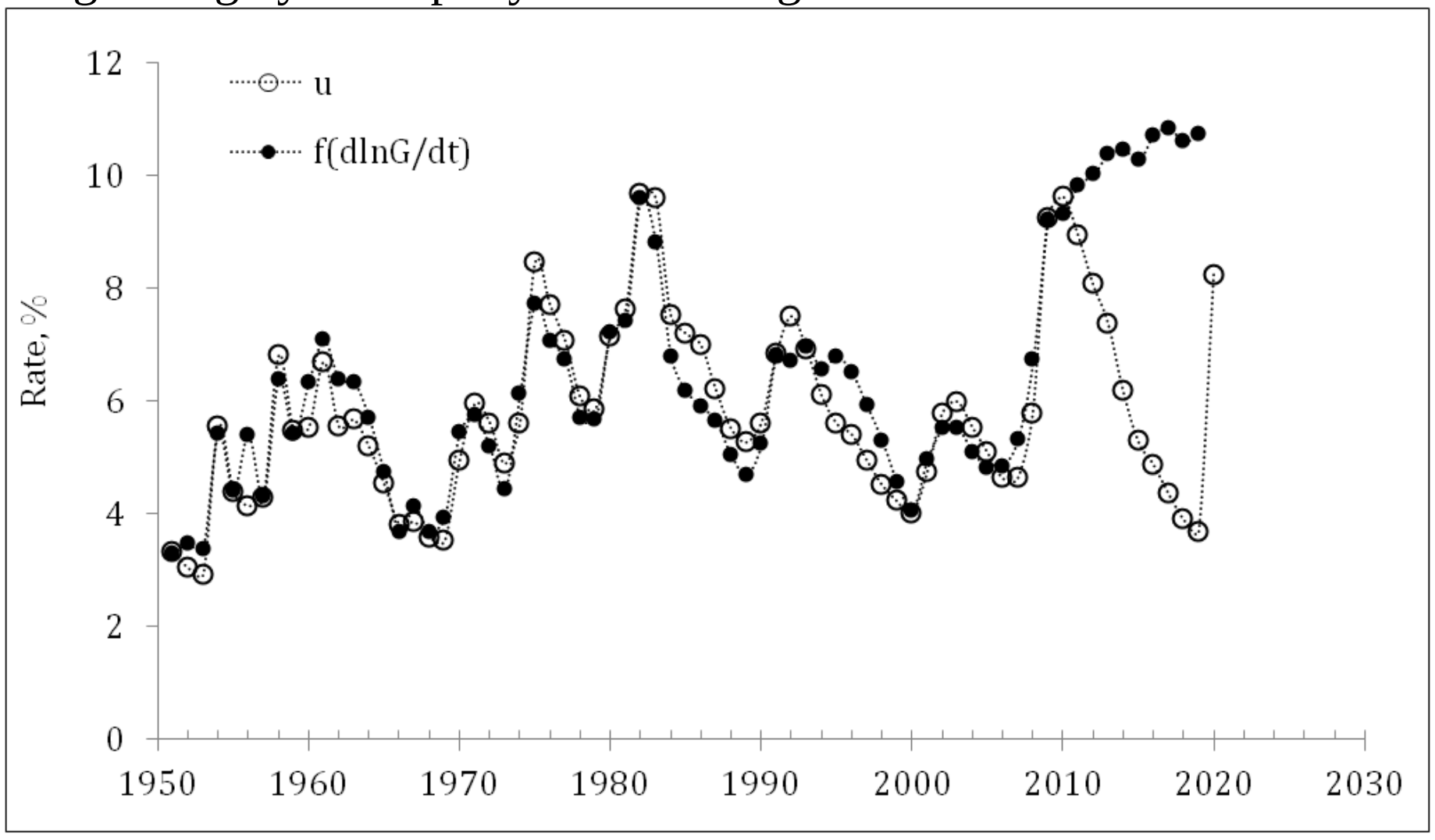

Figure 5. When the 1979-to-2010 set of coefficients are applied to the data after 2010 the predicted curve does not match the measured one.

Figure 6 illustrates the predictive power of model (5): the measured rate of unemployment USA between 1951 and 2019 is very close to the predicted rate, with the real GDP per capita published by the BEA (2020). The rate of unemployment is borrowed from the BLS (2020). Figure 7 presents the model residual errors as a function of time. with the standard deviation of 0.49% and the mean uneployment rate of 5.8% for the entire period. The avearge annual absolute change in the unemployment rate was 0.8%. Figure 8 depicts the linear regression of the measured and predcited time series with $R^2$=0.89. Hence, the new set of coeffcients also provides an excellent match between the measured and predicted values, i.e. the model is validated by the data between 2010 and 2019.

So far, data from the US sources were used. As Figure 4 demonstrates, various sources of economic data may give slightly different coeffcients in the regression analysis. Thus, it is instructive to use different sources. For the real GDP per capita, we use the Maddison Project Database (2020). Figure 9 illustartes the difference between the BEA and MPD estimates. Both time series are normalized to their respective values in 1970 in order to have the same reference year. In addition, the OECD and Total Economy Database estimates are presented for a broader view on the data quality. According to panel a) in Figure 9, the TED curve diverges from the other three since 1979. The BEA, OECD, and MPD curves are close. for example, the ratio of the OECD and BEA is close to 1.0, as panel b) shows. The MPD/BEA ratio indicates that the statistical estimates with these to time series may differ. The MPD/TED ratio reveals two important features: the deviation reaches ~15%, which is an extremely high value if to consider the accuracy of statistical estimates, and the difference since 1979 is a linear function of time - likely the legacy of the TED-realted cooperation between the Groningen Growth and Development Centre and the Conference Board.

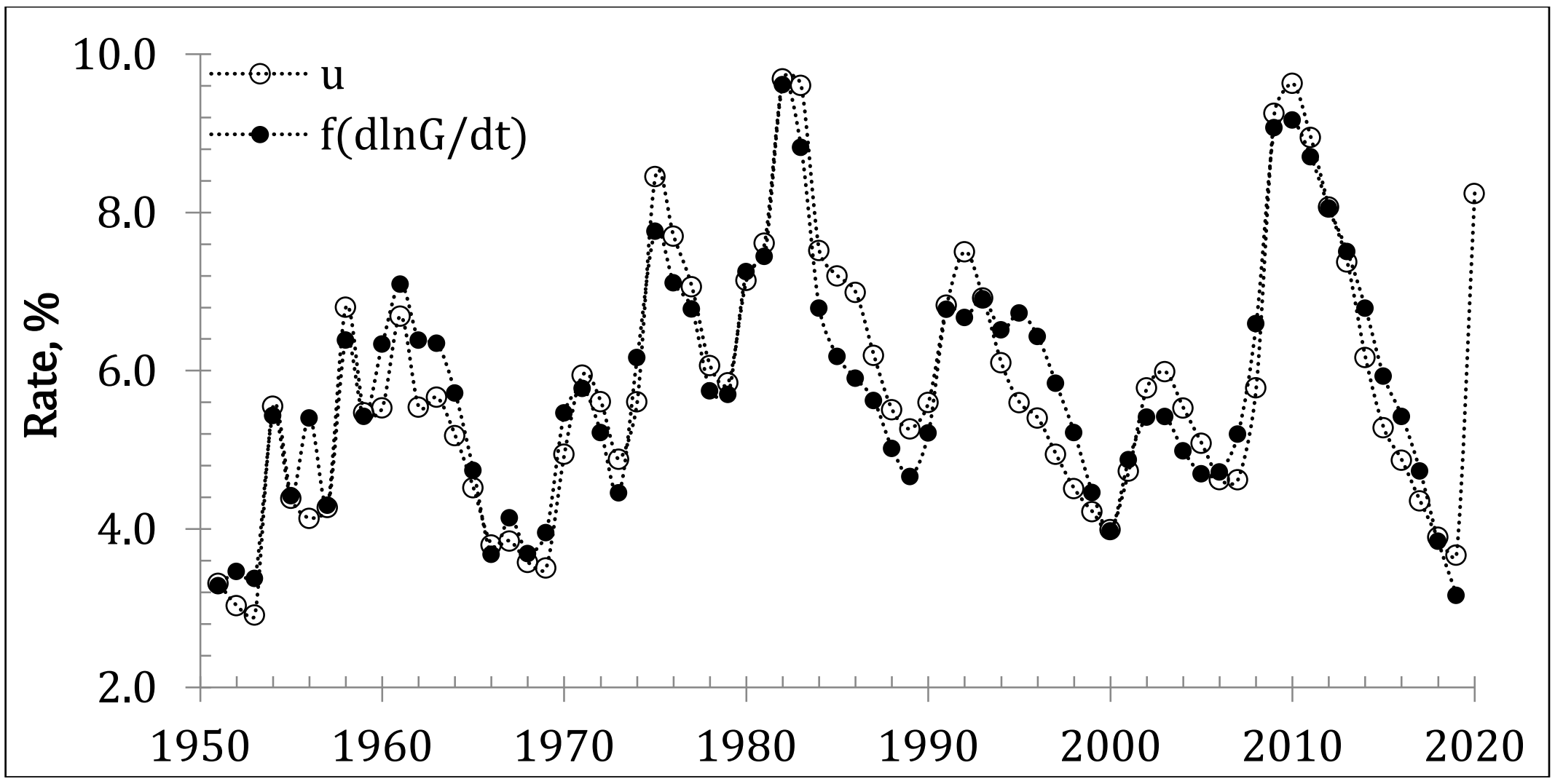

Figure 6. The measured rate of unemployment in the USA between 1951 and 2019 and the rate predicted by model (5) with the real GDP per capita published by the BEA.

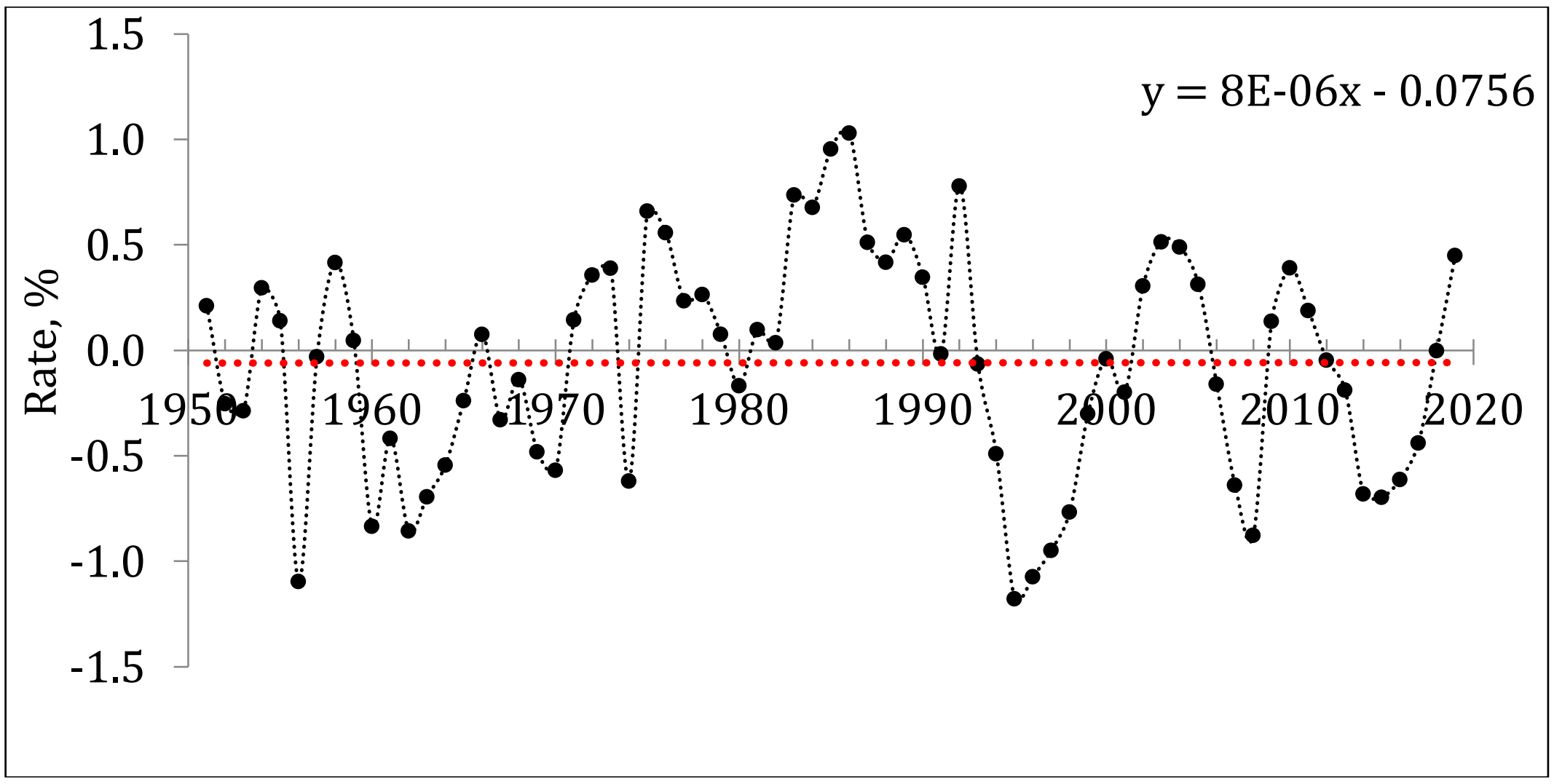

Figure 7. The model residual. Red dotted line is the linear regression line with a slope of $8*10^{-6}$.

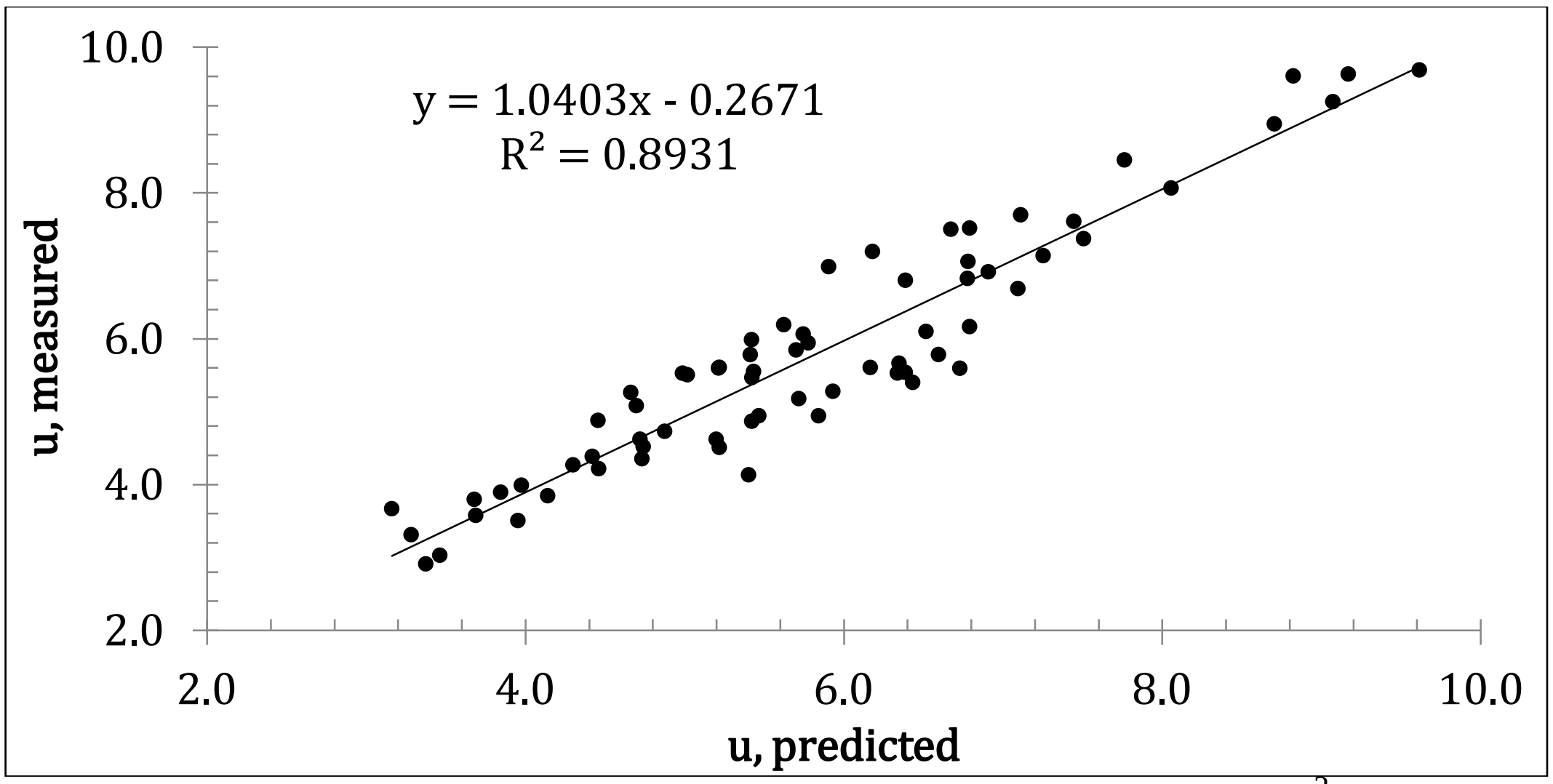

Figure 8. The linear regression of the measured and predcited time series, $R^2 = 0.89$.

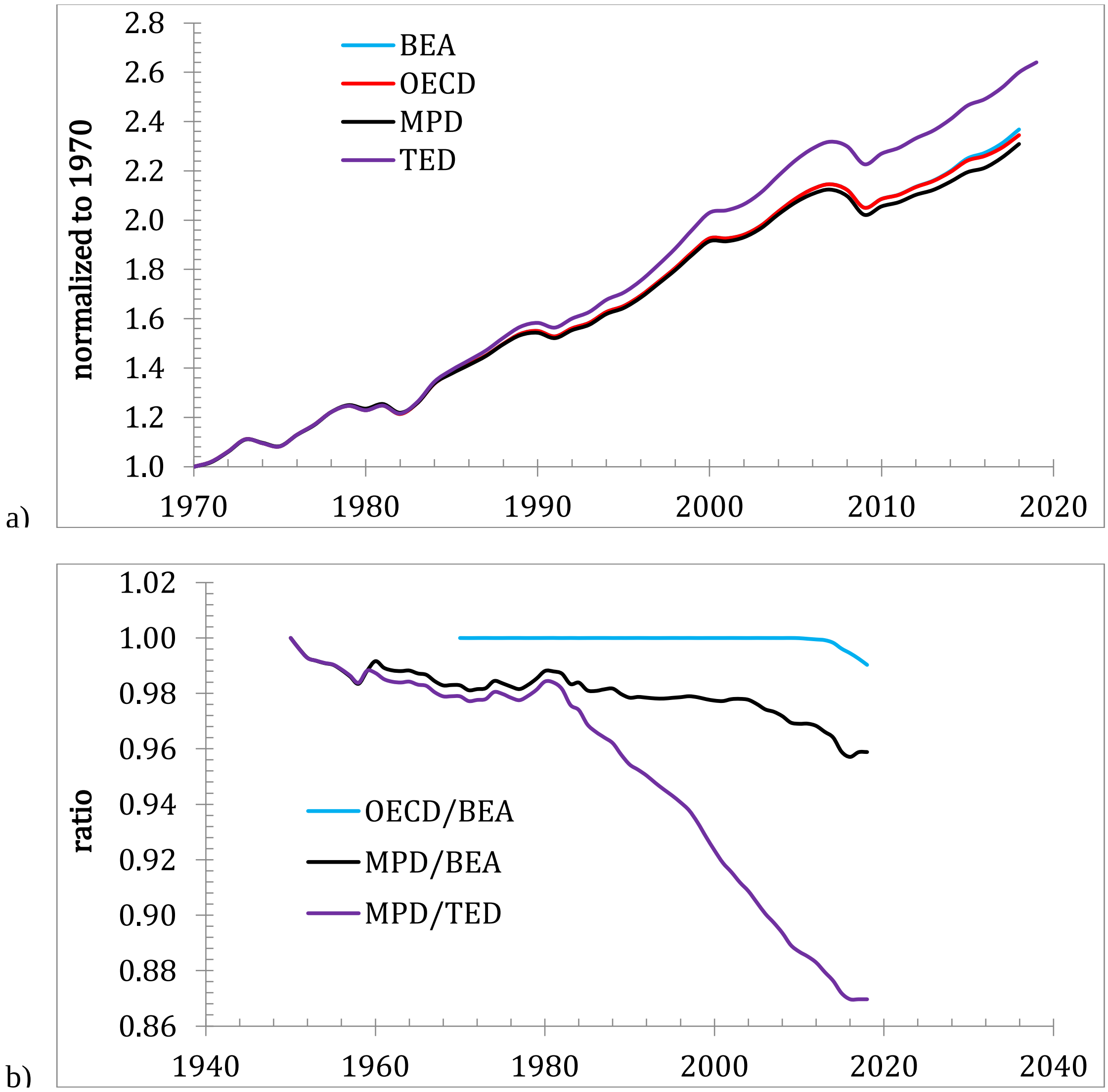

Figure 9. a) For times series (BEA, MPD, OECD, and TED) normalized to their respective levels in 1970. The TED curve is quite different from the other three. b) Pair-wise ratios of the curves in panel a).

The OECD readings of the unemloyemnt rate do not differ much (max difference of 0.02%) from the BLS data. Figure 9 shows that the MPD gives a slightly better fit with $R^2$=0.91. This is just a marginal improvement but it is important in terms of methodology of statistical estimates with not perfect data measurements.

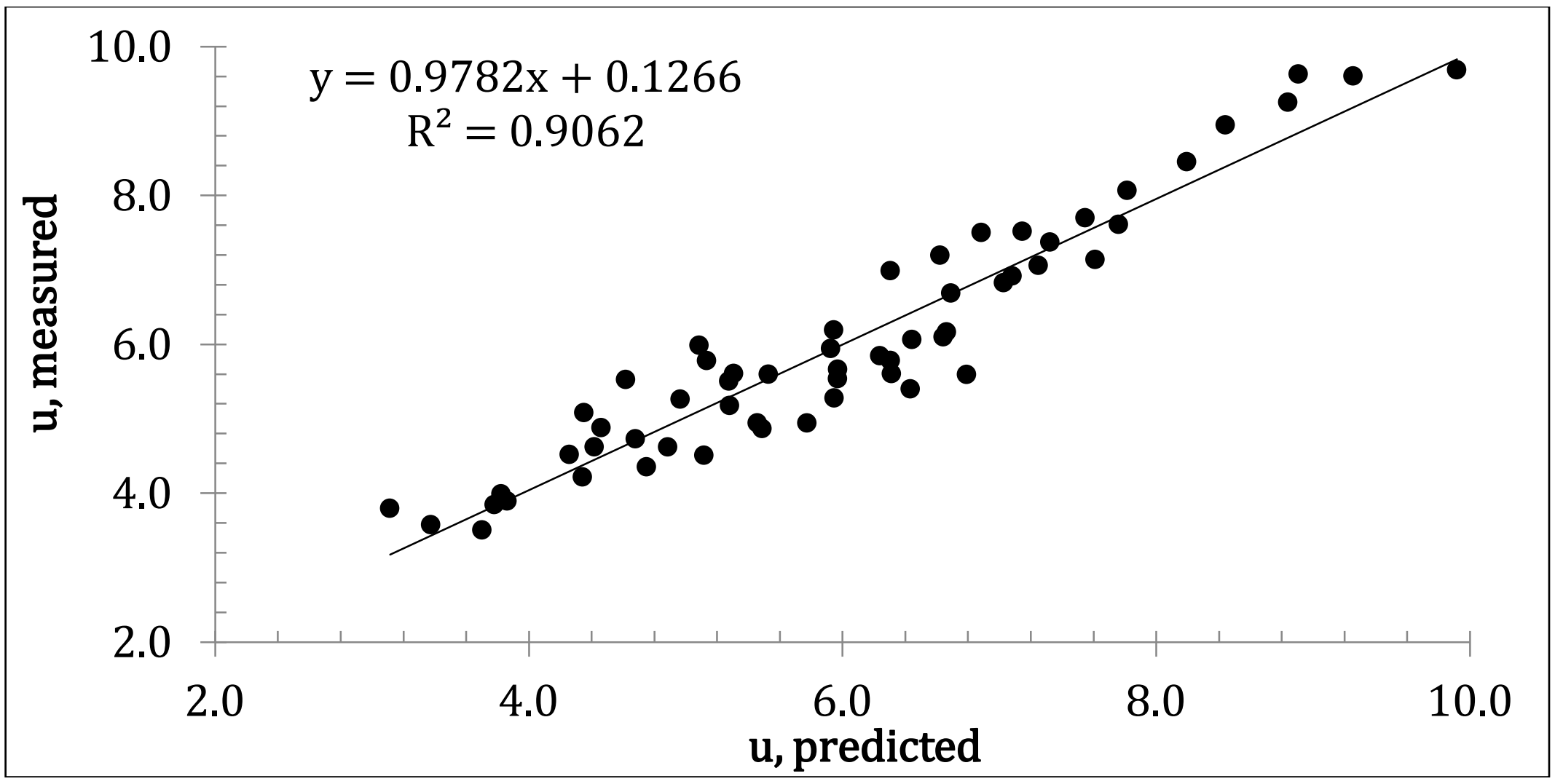

Figure 10. Same as in Figures 8 with the MPD estimates of the real GDPpc and the OECD data for unemployment.

The ultimate validation test would be the model prediction for 2020, when the rate of uneployment changes by 10% per quarter and the real GDP per capita falls by 35% in one quarter and then rebounds by 30%. Figure 11 presents the rate of unemployment predicted for the first three quarters of 2020. The spike in the second quarter is accurately predicted with model (5) estimated for the period between 2010 and 2019. It is an indicator that the model is still applicable. Interestingly, the third quarter demonstrates a large prediction error: 5.3% instead of the measured value of 8.8%. The predicted unemployment rate is obtained with the real GDPpc growth of 30% in the third quarter. The first GDP estimates for the third quarter might be highly overestimated. If the measured value of 8.8% is correct, the *GDPpc* growth has to be only 17% from the previous quarter. We will follow the BEA releases with updated GDP estimates as well as the BLS releases with new estimates of the unemployment rate.

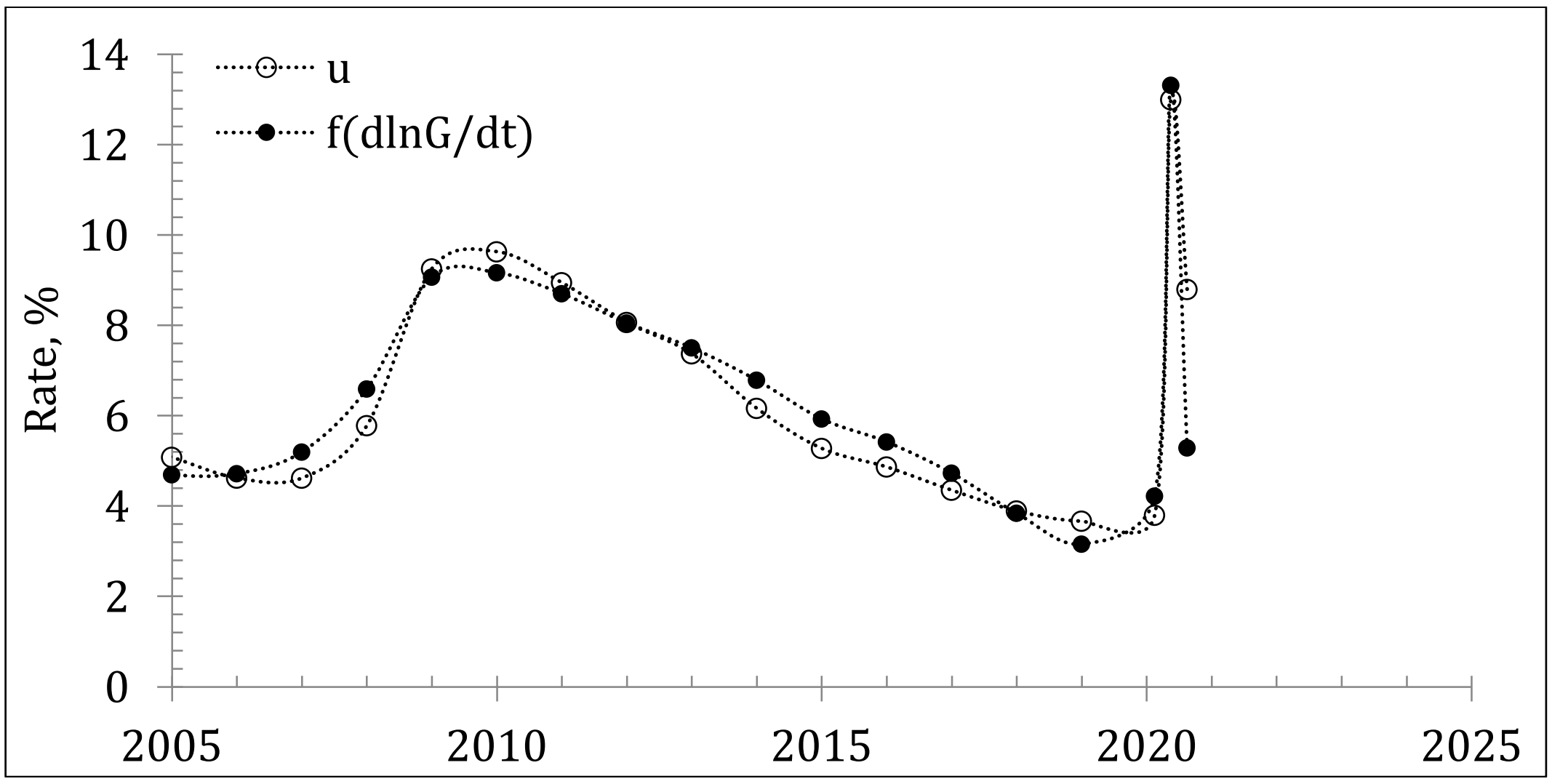

Figure 11. The rate of unemployment predicted for three quarters of 2020. The spike in the second quarter is accurately predicted with the model (5) estimated for the period after 2010.

We started with the USA for many reasons. The most important is the length and accuracy of economic data. For a quantitative model, data quality is one of defining requirements of successful statistical analysis. The results obtained with the unemployment and GDP readings since 2010 demonstrate that the original model predicts the measured rate of unemployment in the USA with high accuracy compared to the potential measurement errors as revealed by direct comparison of the real GDP per capita time series obtained from various sources. The difference in the estimates of economic variables published by the Conference Board (TED), BEA, BLS, OECD, and the Maddison Project Database can be a major methodological problem for our model in the countries with a shorter history of measurements and often definitional revisions.

### 2.2. *The United Kingdom*

As in hard sciences, standard modelling procedure is to check data quality. It was demonstrated in Section 2.1 the the GDP deflator (dGDP), which is a principal component of the real GDP estimation procedure, is prone to artificial breaks introduced by definitional revisions. Such breaks in the GDP time series look like the structural breaks related to inherent changes in economic performance.  When taken as real, the definitional breaks ruin the statistical performance of the mainstream economic models.

The UK has likely the most developed school in hard and soft sciences. Data quality is the essence of experimental sciences in support to theoretical consideration. Economics is not an exception and the quality and compatibility of measurements is generally retained in the UK. In panel a) of Figure 12, we present the evolution of the cumulative inflation (the sum of annual inflation rates) as defined by the *CPI* and *dGDP* between 1955 and 2018. Both variables are borrowed from the OECD database. In order to reduce the two time series to the same reference year and level, they are normalized to their respective values in 1955, and thus, both curves start from 1.0. One can see that these curves are close, but deviate from the very beginning.  In panel b), the rates of price inflation are shown as calculated using the CPI and dGDP indices. The only large deviation between the inflation curves was observed in 1994 and is likely artificial. This spike is pretending to be described with a dummy variable, which we did not use in the previous models.

Panel c) in Figure 12, depicts the difference between the cumulative and change rate curves in panels a) and b). The cumulative curves demonstrate approximately linear deviation since the late 1970s. Taking into account the quasi-linear deviation between the cumulative curves, we propose the following piece-wise model for the best fit between the *CPI* and *dGDP* curves:

$$CPI = 0.926 dGDP, \quad 1970 > t \geq 1960$$
$$CPI = 0.974 dGDP, \quad 2018 \geq t \geq 1971, \text{ except } 1994$$
$$CPI = 0.974 dGDP - 0.049, \quad t = 1994 \qquad (6)$$

Panel d) illustrates the match between the cumulative curves, which include a dummy variable of -0.049 in 1994. The spike is removed. The model residual is presented in panel e), where the standard deviation is 0.018 for the period between 1956 and 2018. The *CPI* and *dGDP* comparison indicates that there are potential breaks in the *dGDP* curve in the

early1970s, the late 1980s, and likely between 2009 and 2012. There is no sharp transition between the segments like that observed in 1979 in the USA.

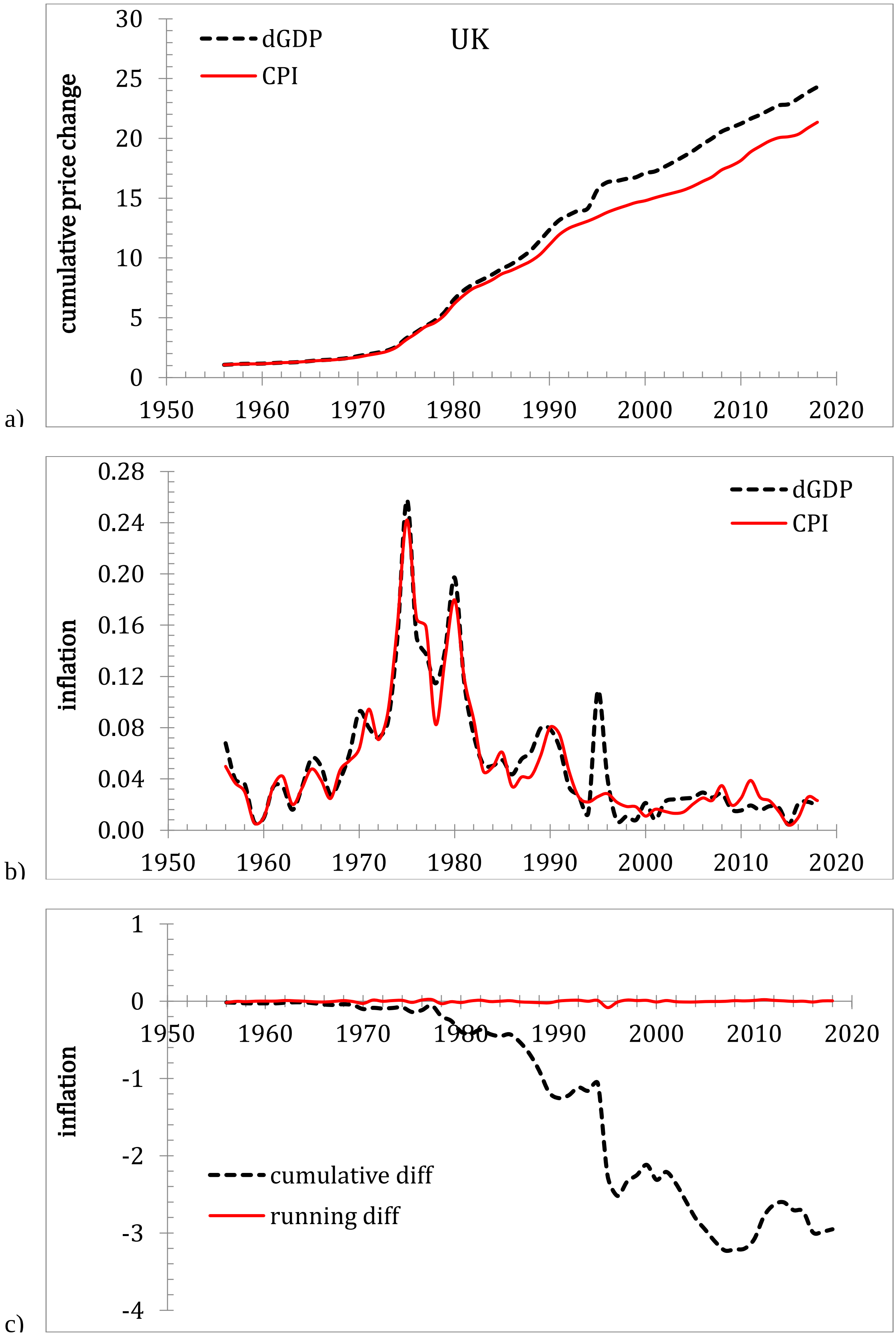

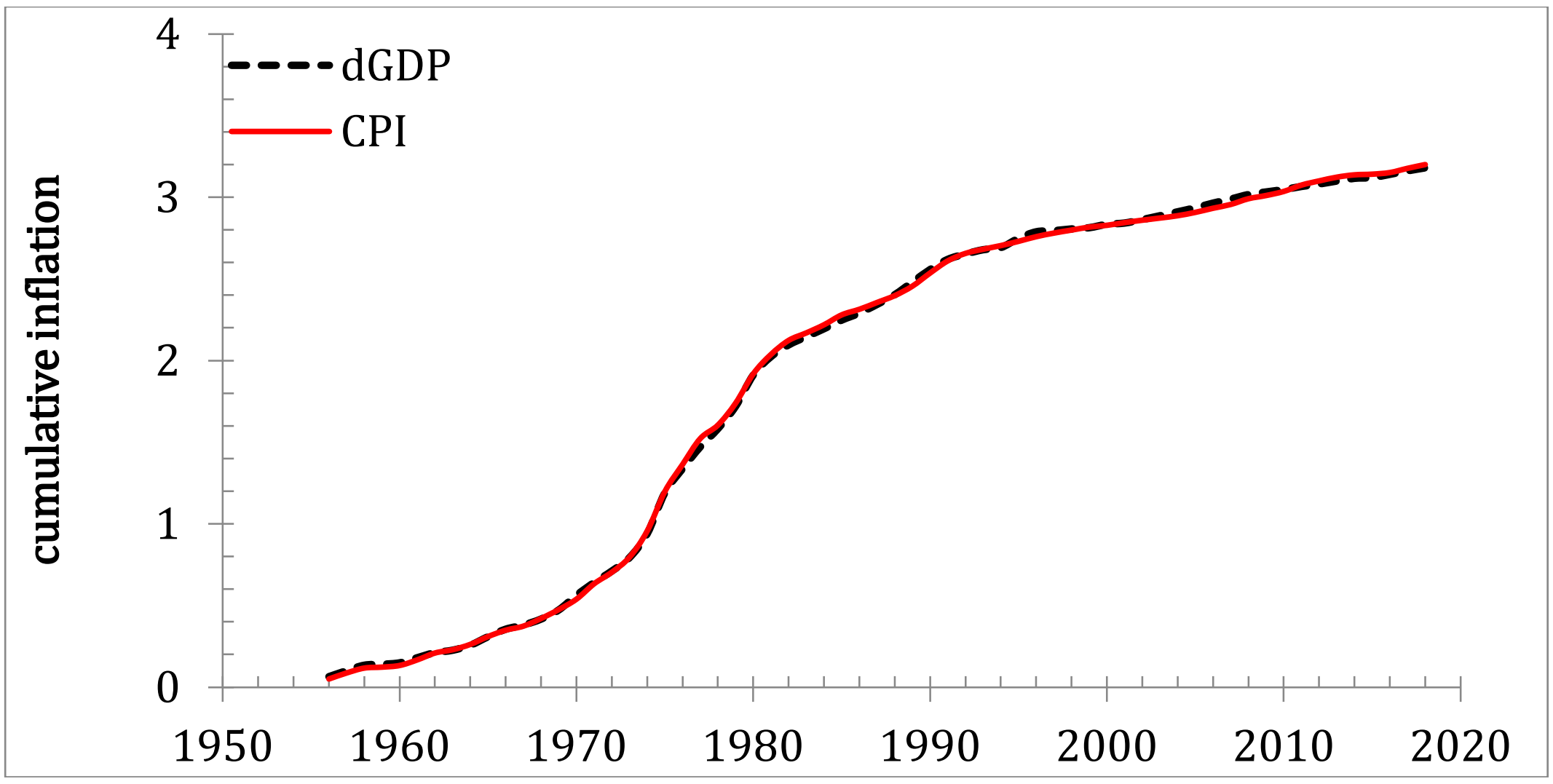

d)

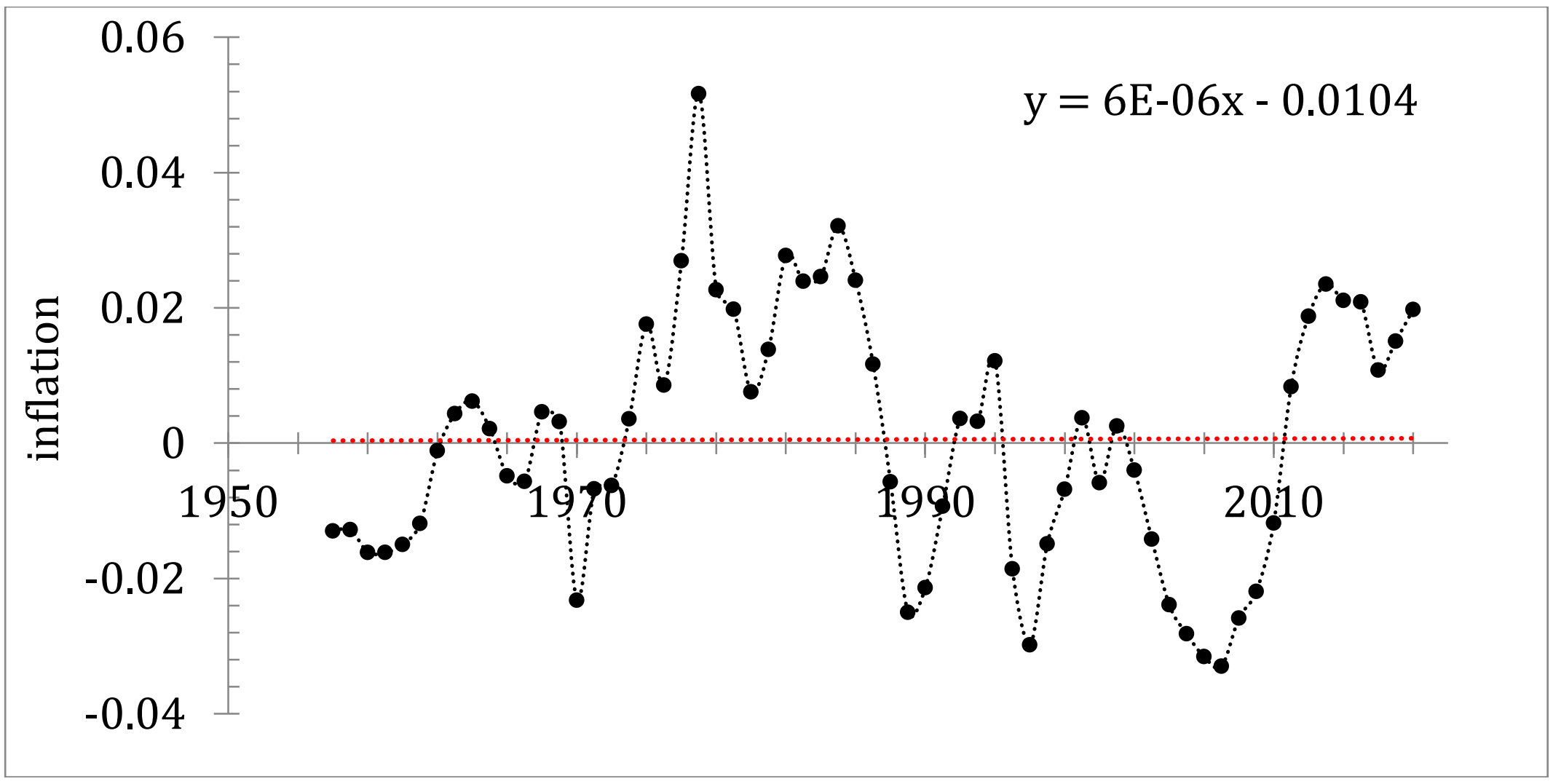

e)

Figure 12. a) The evolution of the cumulative inflation (the sum of annual inflation rates) as defined by the *CPI* and *dGDP* between 1955 and 2018. Both variables are normalized to their respective values in 1955. b) Inflation rates for the *CPI* and *dGDP*. c) The difference between the CPI and dGDP curves in panels a) and b). d) The fit between the *CPI* and the *dGDP* corrected according to relationship (6). e) The residual error of model (6).

The break years obtained from the *dGDP* are used as start points in the search of the best fit in our version of Okun's law for the UK. We are using standard LSQR procedure to estimate a new set of coefficients for the period between 1961 and 2018. The data before 1961 are not used because the rate of inflation is very close to 1%, and is close to the model accuracy. Preliminary analysis gives the following model:

$$du_p = -0.63dlnG + 1.75, \quad 1987>t\geq1963$$
$$du_p = -0.42dlnG + 0.64, \quad 2010\geq t\geq1988$$
$$du_p = -0.39dlnG - 0.13, \quad t\geq2011 \qquad (7)$$

Figure 13 illustrates the model predictive power. In the upper panel, the measured rate of unemployment in the UK between 1962 and 2018 is compared with the rate predicted by model (7) with the real GDP per capita published in the Maddison Project Database. The rate of unemployment is borrowed from the OECD. In the middle panel, the model residual errors are presented with the standard deviation of 0.91% and the mean unemployment rate of 5.99%. Lower panel depicts the linear regression of the measured and predcited time series with $R^2$=0.91. Hence, the new set of coeffcients provides a good match between the measured and predicted values, i.e. the model linking the change in unemployment rate and the change in real GDP per capita.

The UK is able to retain the integrity of data in longer time periods. The overall sensitivity of unepmloyment to real GDP per capita seems to decrease with time. As in the USA, the most recent period has a negative constant term and the rate of unemployment in the UK has to decrease by 0.13% per year even in the absence of real economic growth. Such a dependence was not observed in our models before 2010. It is not clear what is the economic mechnism behind the uneployment rate increase or decrease without economic growth. One of the oppoertunities is that this constant term is an artificial parameter created by inherent defficieny in the definition of unemployment and/or economic growth. As we discussed in Section 2.1, the difference in the GDP per capita estimates might be a linear function of time and the constant term expresses the fact that the MPD underestimates the GDP per capita and the US economy actually grows in real term as described by the TED.

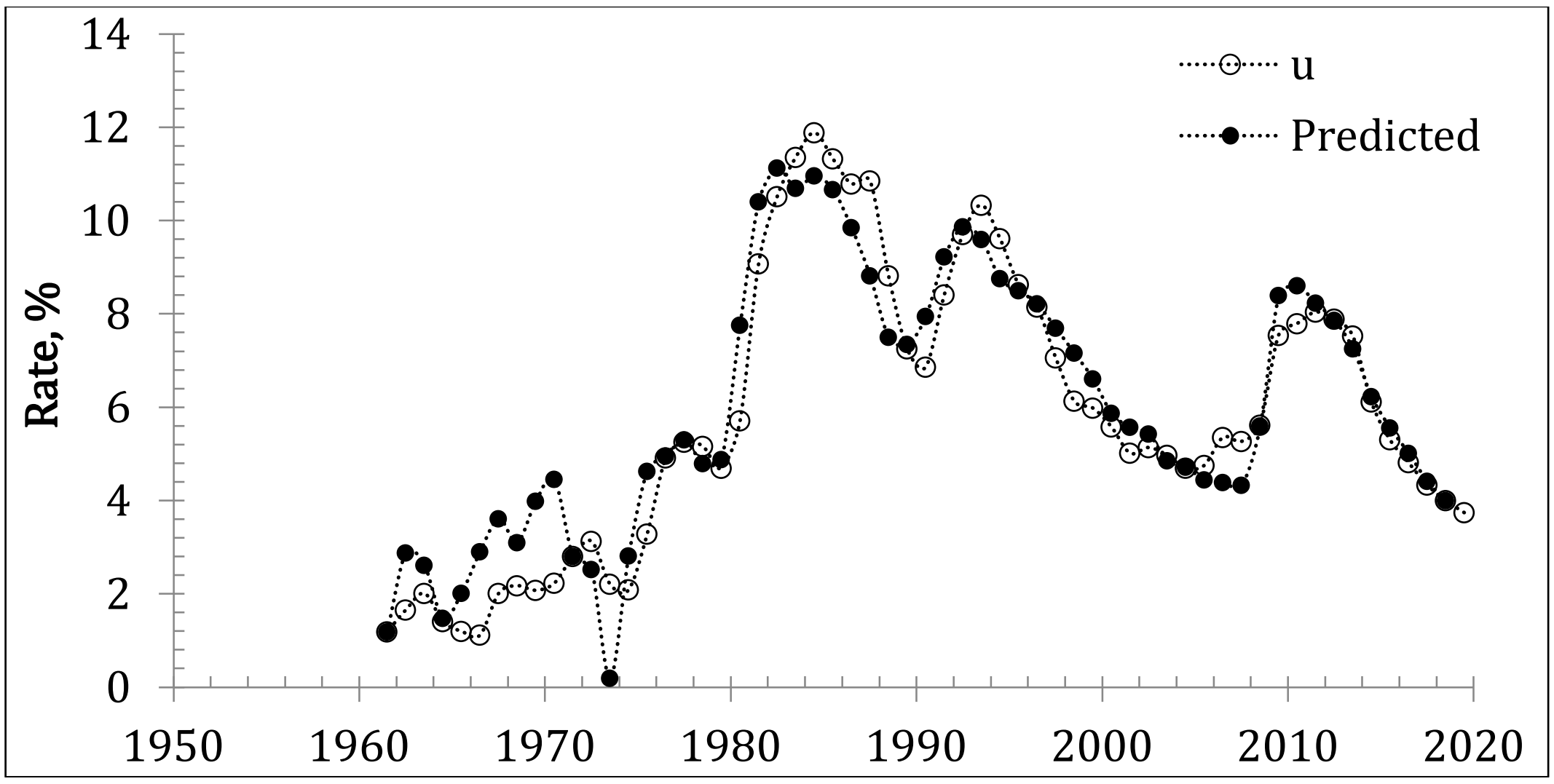

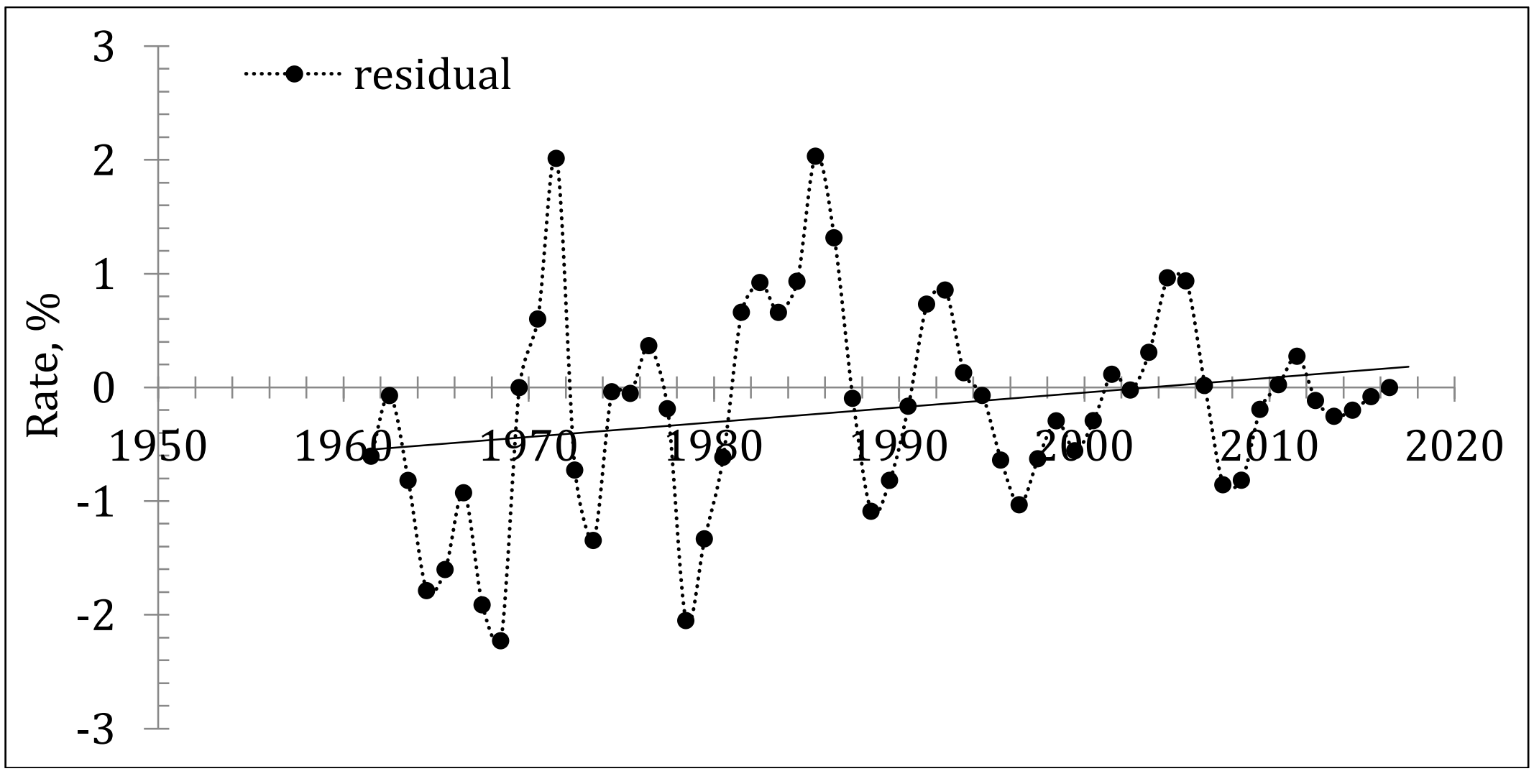

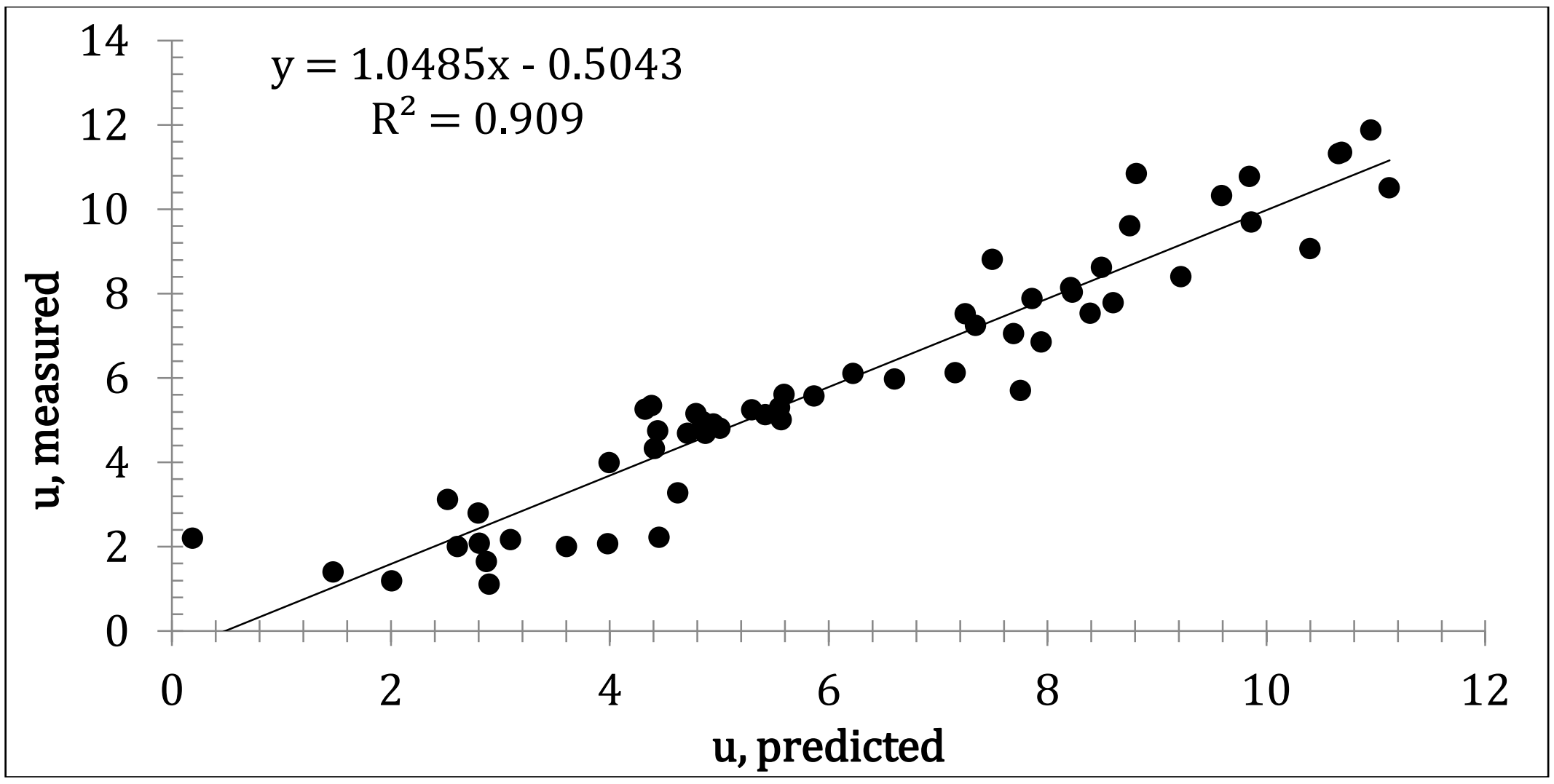

Figure 13. Upper panel: The measured rate of unemployment in the UK between 1962 and 2018 and the rate predicted by model (7) with the real GDP per capita published by the MPD. Middle panel: The model residual, stdev=0.91% Lower panel: Linear regression of the measured and predcited time series. $R^2$ = 0.91.

## 2.3. *France*

In the upper panel of Figure 14, we present the evolution of the cumulative inflation (the sum of annual inflation estimates) in France. There are two curves as defined by the CPI and dGDP between 1955 and 2018. Both variables are normalized to their respective values in 1955. Since 1985, the dGDP curve is above the CPI one. In the middle panel, the inflation rates are shown for both variables. In the lower panel, we present the difference between the CPI and dGDP curves in the upper and middle panels. One can see that the difference between the cumulative curves has several quasi-linear segments. The change in the slope between these segments in most likely related to the multiple revision to the dGDP definition. The years of breaks in the dGDP time series are not easy to estimate from the lower panel of Figure 14 and we allow the LSQR method to find these years when minimizing the RMS residuals. For France, the piece-wise linear relations between the *CPI* and *dGDP* is not presented because the break years estimation is the best illustrated by the difference of the cumulative curves in the lower panel.

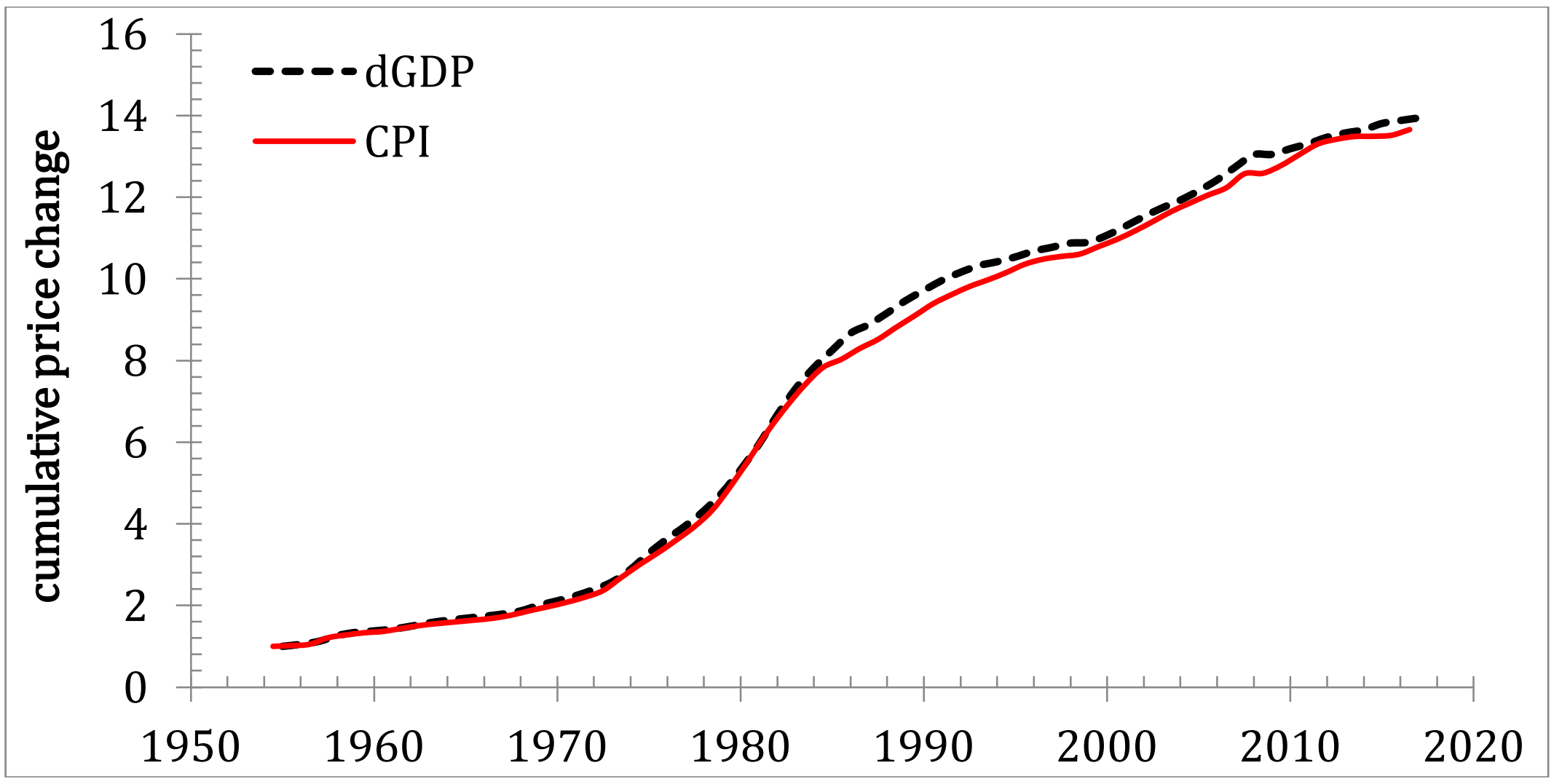

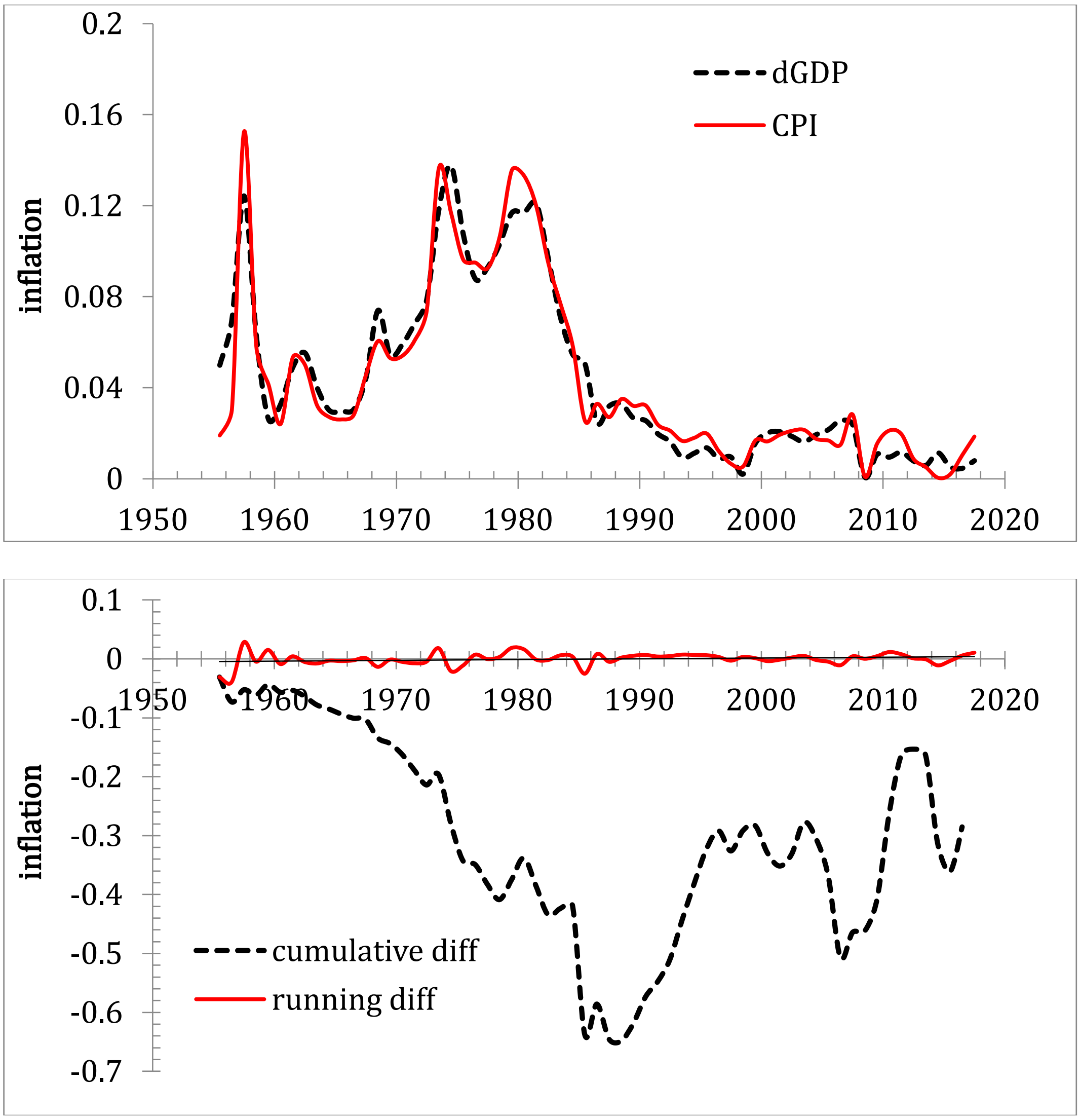

Figure 14. Upper panel: The evolution of the cumulative inflation (the sum of annual inflation estimates) as defined by the *CPI* and *dGDP* between 1955 and 2018. Both variables are normalized to their respective values in 1961. Middle panel: The dGDP and CPI inflation estimates. Lower panel: The difference between the CPI and dGDP curves in the upper and middle panels.

As before, we minimize the model residuals and determine the break years together with the regression coefficients. For France, the best fit model between 1962 and 2018 is as follows:

$$du_p = -0.134dlnG + 0.750,\ 1962>t\geq1984$$
$$du_p = -0.255dlnG + 0.620,\ \ 1985\geq t\geq1999$$
$$du_p = -0.520dlnG + 0.355,\qquad t\geq2000 \qquad (8)$$

Two break years are determined automatically: 1985 and 2000. The 2010 break crucial for the USA and UK models is absent in the France model. Figure 15 presents the measured and predicted rate of unemployment (upper panel), the model residual error (middle panel), and the regression of the measured and predicted time series (lower panel). The average rate of unemployment between 1960 and 2018 is 5.55%, and a standard deviation of the residual error is 0.55%. The overall fit ($R^2$=0.98) is excellent by all means, with the break years close to those expected from Figure 14. One of possible reasons is that France has a good set of methods and procedures to measure/estimate economic parameters.

This perfect result does not resolve the data incompatibility problem, however, and statistical analysis needs extra efforts to distinguish between the actual economic structural breaks and ignorance of basic procedures. The importance of data quality is best illustrated by an example in Figures 16, where two GDP per capita estimates from the OECD, MPD, and TED are compared. One can see that these three agencies provide quite different estimates with the highest estimates reported by the OECD. The MPD and TED estimates are equal before 1979. The OECD/MPD ratio reaches 1.082 in 1991 and then falls to 1.064 in 2014. Therefore, the use of the OECD estimates would change the statistical model.

Statistically, the French model is best in terms of coefficient of determination and it is also parsimonious with only two breaks between 1960 and 2018. The difference between the observed and predicted values is most likely related to the measurement errors, as Figure 16 suggests.

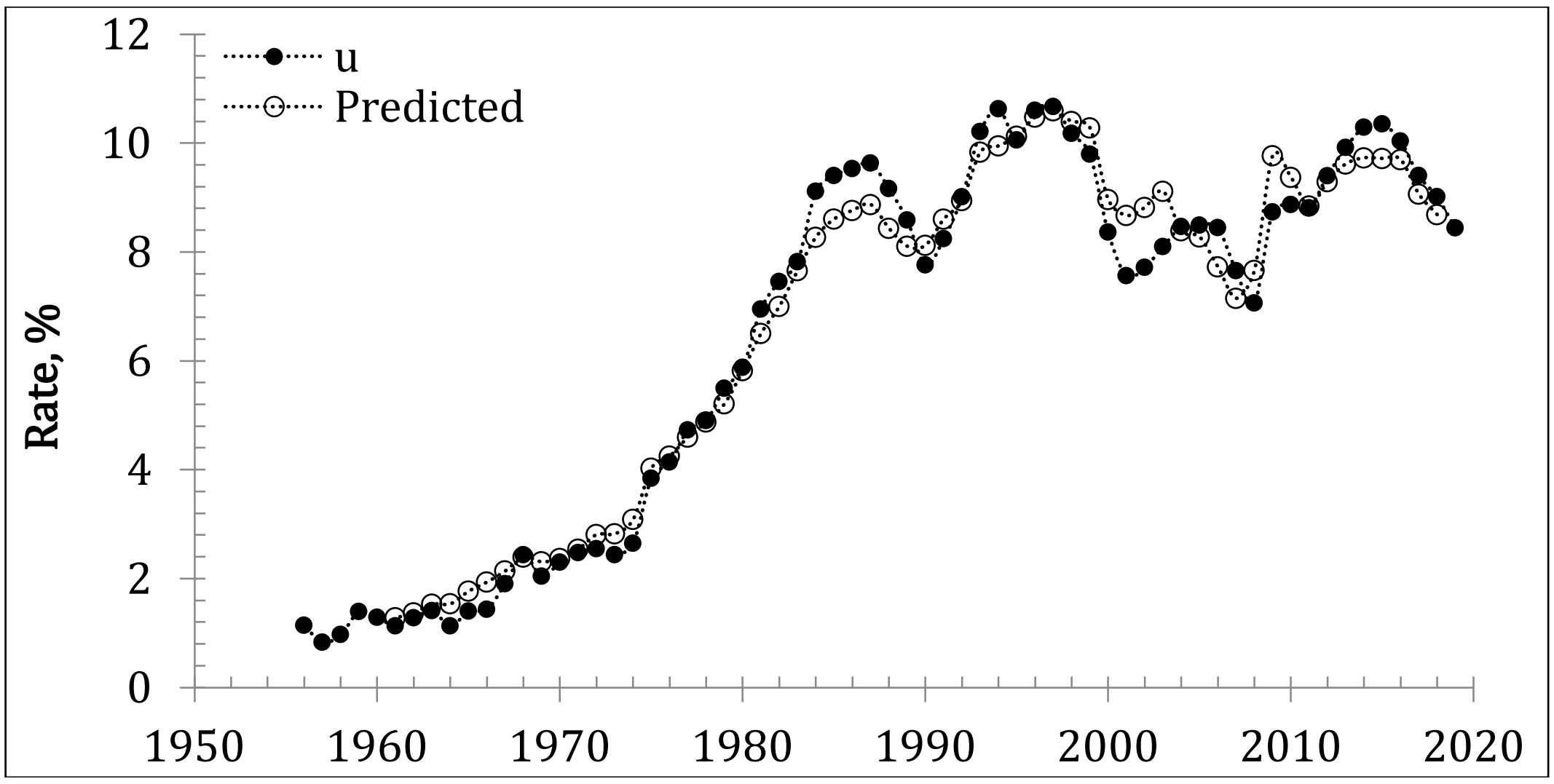

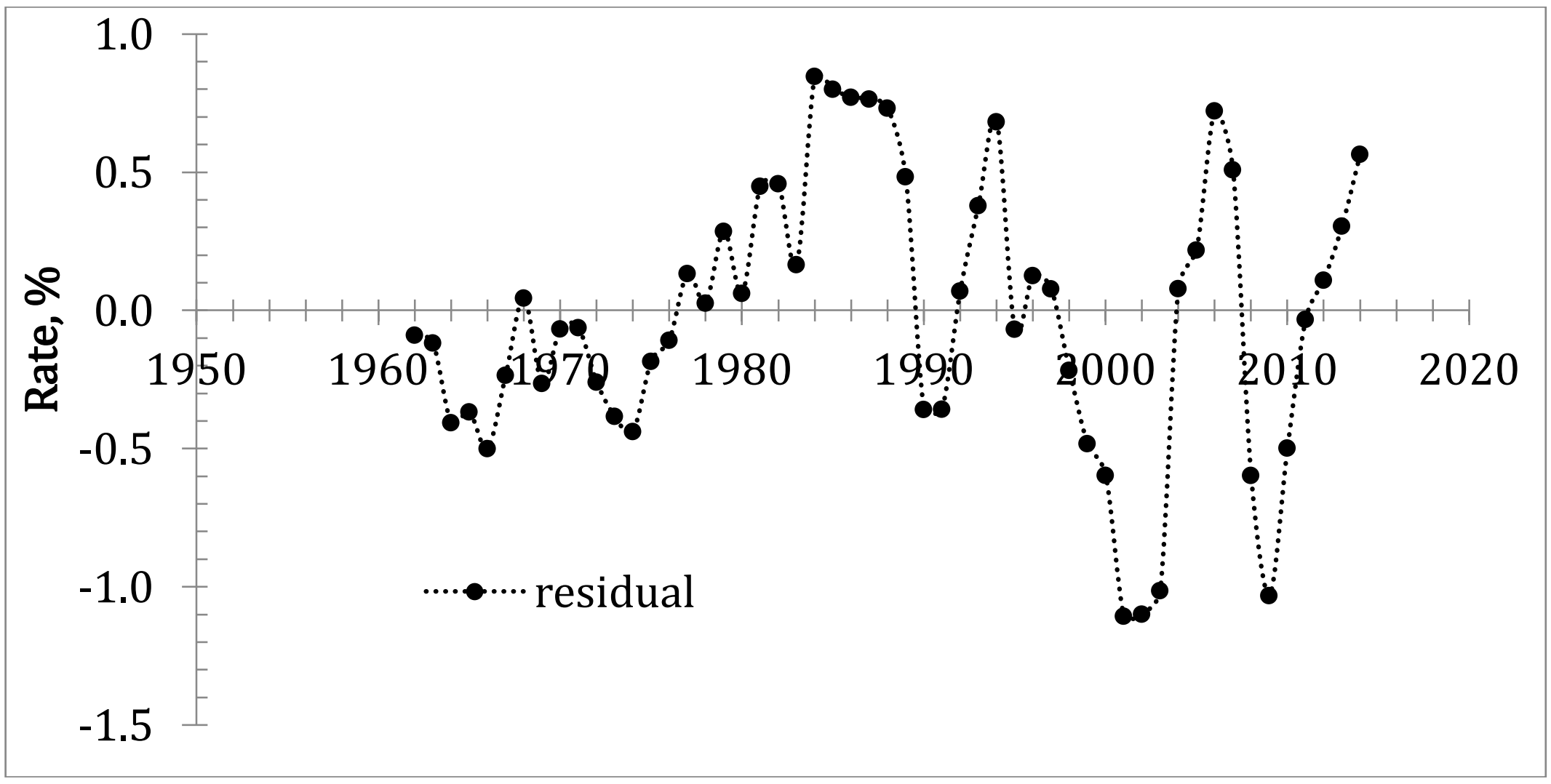

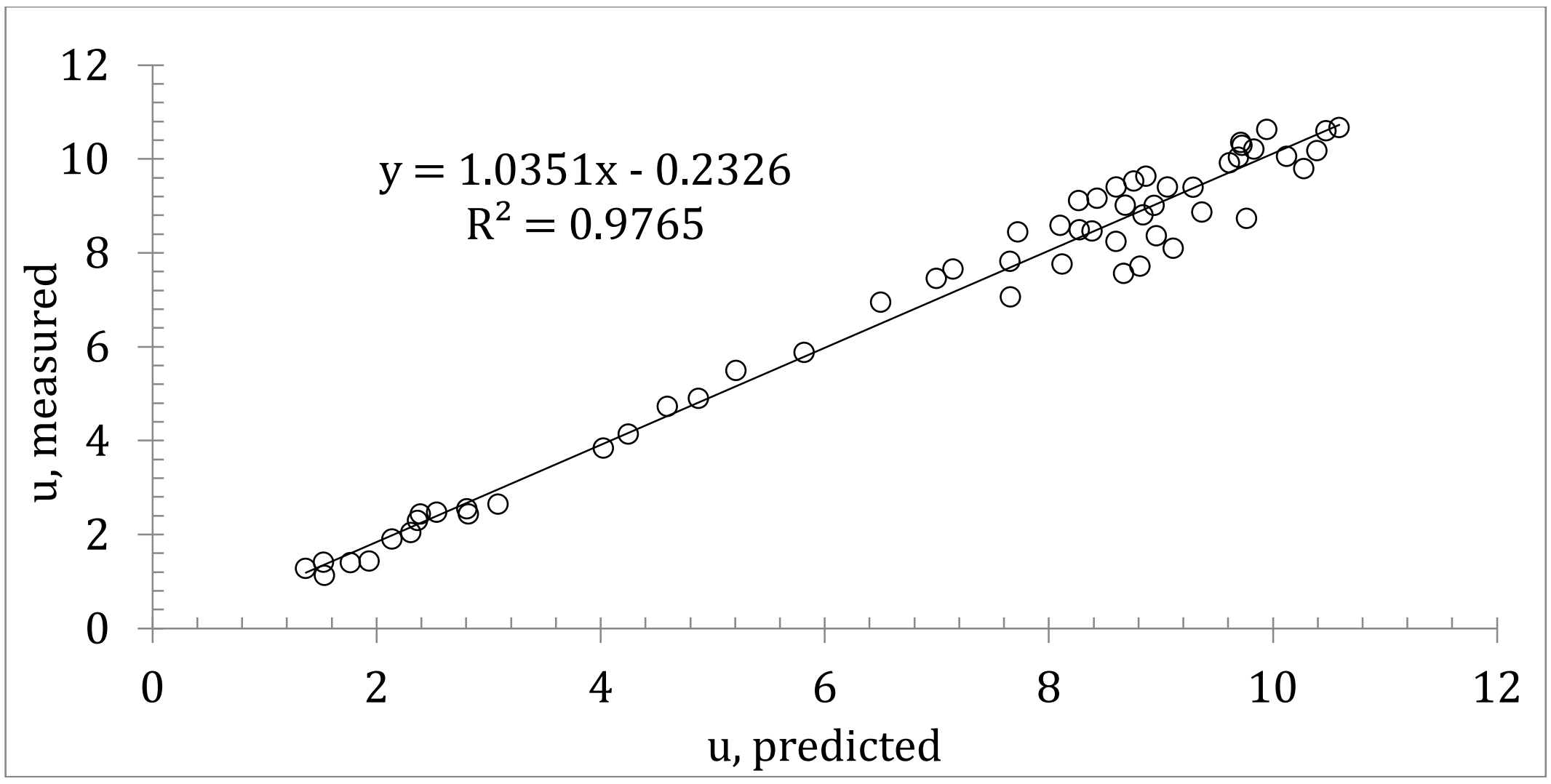

Figure 15. Upper panel: The measured rate of unemployment in France between 1960 and 2018, and the rate predicted by model (8) with the unemployment rate published by the OECD. Middle panel: The model residual: stdev=0.50%. Lower panel: Linear regression of the measured and predicted time series. $R^2$=0.98.

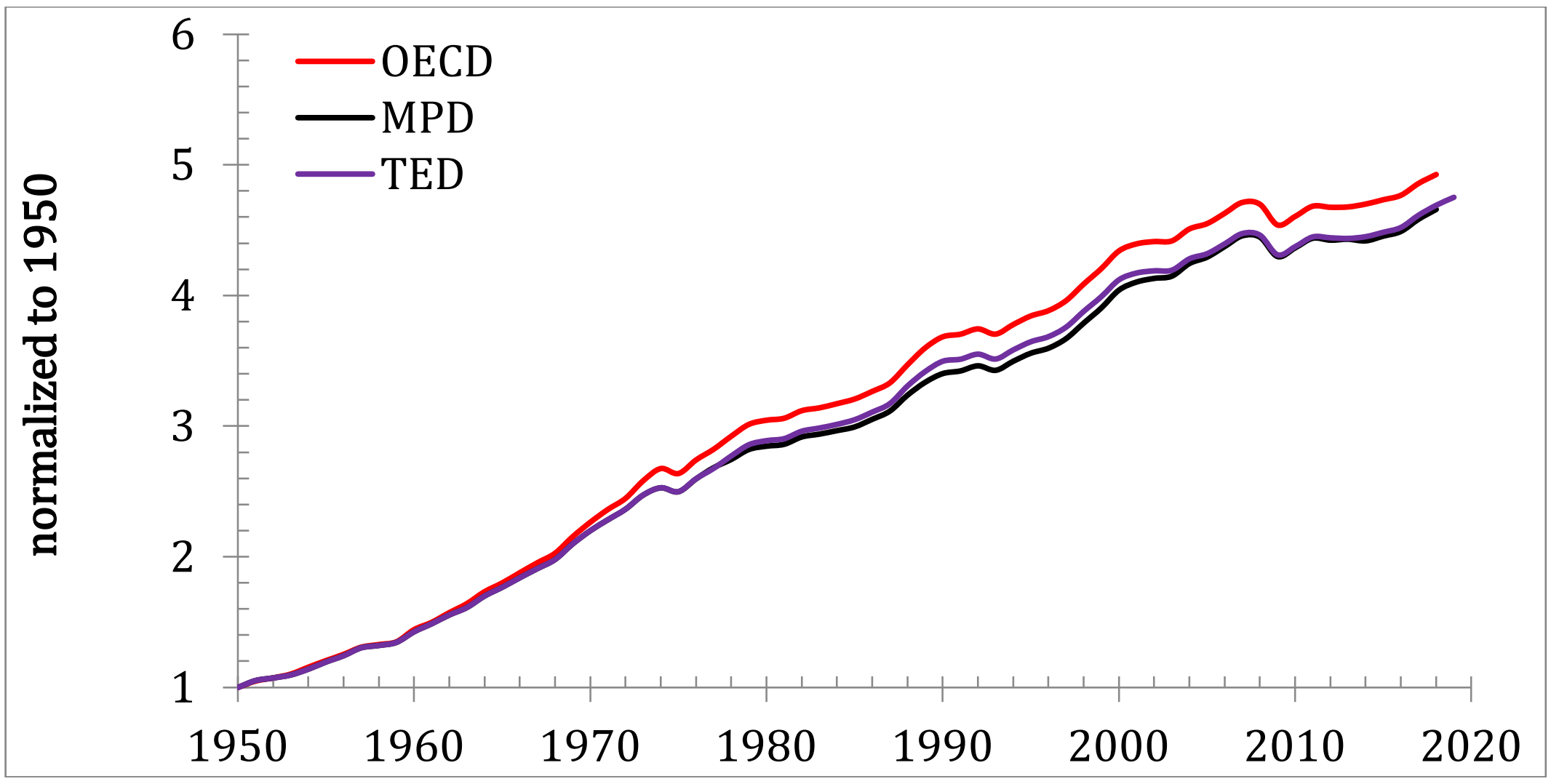

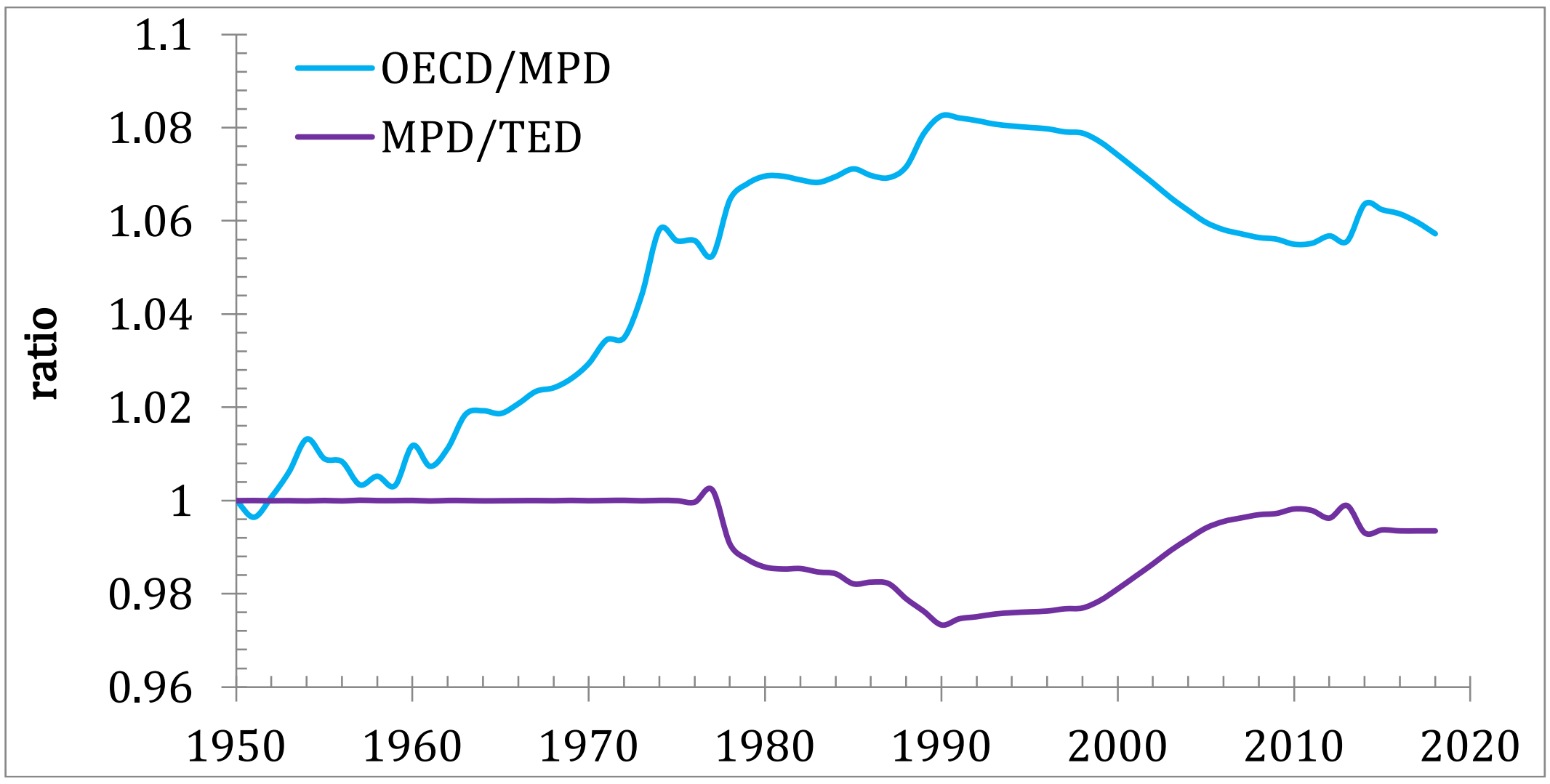

Figure 16. Upper panel: Three times series (MPD, OECD, and TED) normalized to their respective levels in 1950. The OECD provide the highest estimates and is quite different from the other three. Lower pnale: Pair-wise ratios. The MPD and TED are equal before 1979. The OECD/MPD ratio reaches 1.082 in 1991 and then falls to 1.064 in 2014.

## 2.4. *Germany*

By design, Germany is a synthetic case. The 1991 reunification not only changed the economy and population size, it also changed all economic statistics. One may suggest that the economic data for the years before 1991 are obtained by weighted averaging. The result of such a procedure does not reveal any major discontinuities. In the upper panel of Figure 17, we present the evolution of the cumulative inflation (the sum of annual inflation estimates). There are two curves as defined by the *CPI* and *dGDP* between 1970 and 2018, as provides by the OECD database. Both variables are normalized to their respective values in 1970. Since 1996, the dGDP curve is above the CPI. In the middle panel, the inflation rates are shown for both variables. In the lower panel, we present the difference between the CPI and dGDP curves. The difference between the cumulative curves has several quasi-linear segments. The change in the slope between these segments is most likely related to the multiple revisions to the dGDP definition. The years of breaks in the dGDP time series are not easy to estimate from the lower panel of Figure 17 and we allow the LSQR method to find these years when minimizing the RMS residuals.

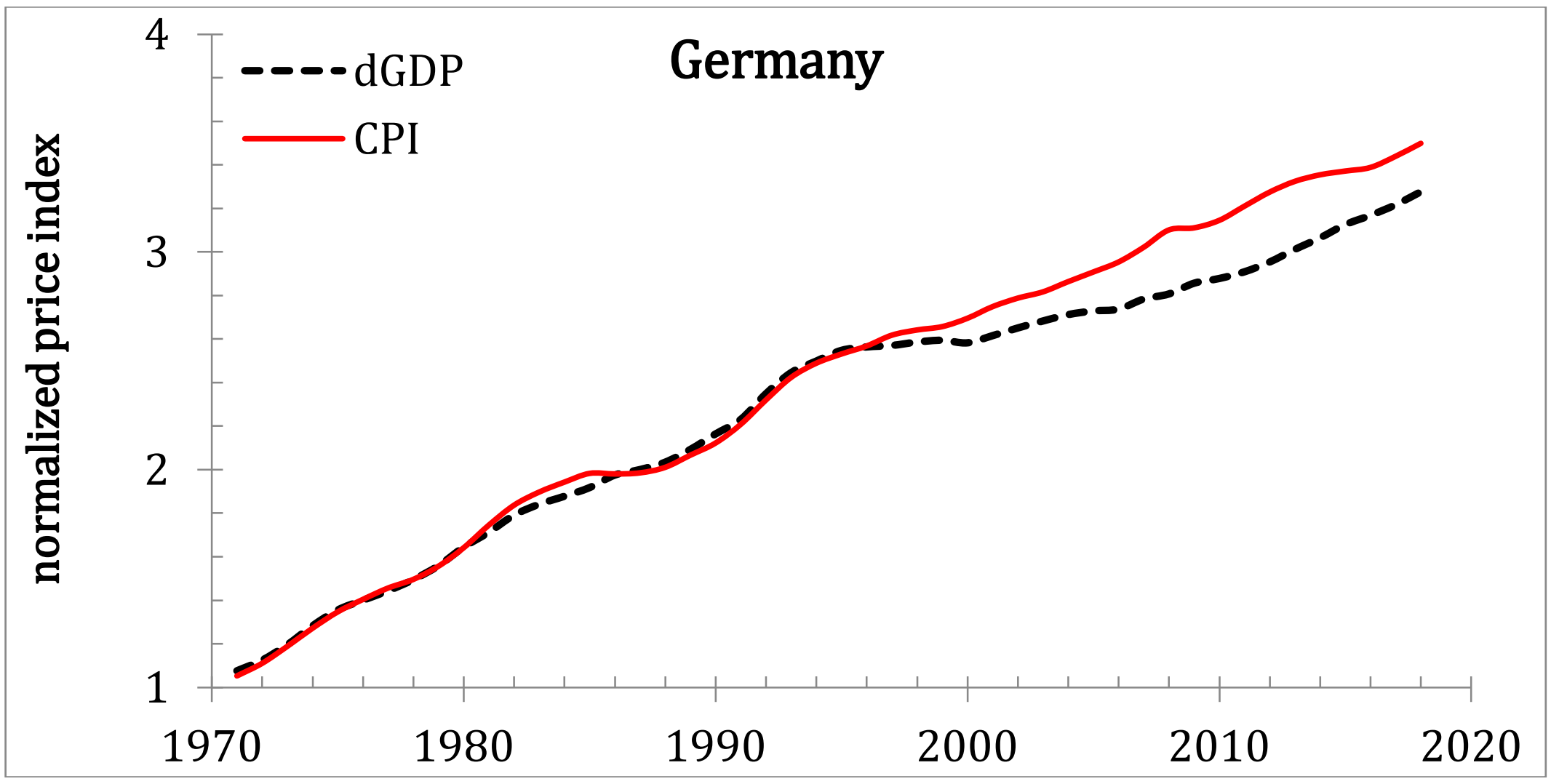

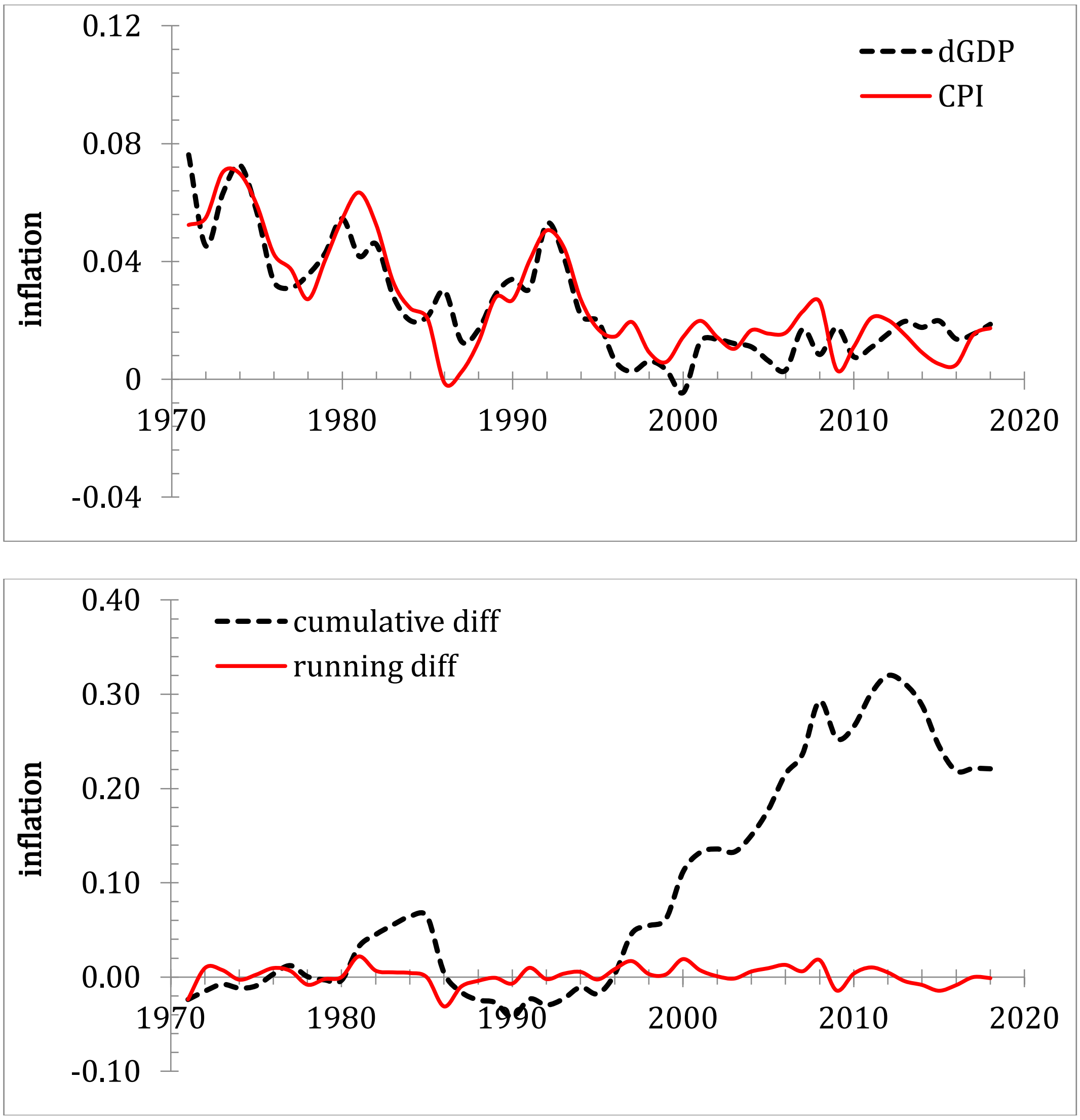

Figure 17. Upper panel: The evolution of the cumulative inflation (the sum of annual inflation estimates) as defined by the *CPI* and *dGDP* between 1970 and 2018. Both variables are normalized to their respective values in 1970. Middle panel: The *dGDP* and *CPI* inflation estimates. Lower panel: The difference between the *CPI* and *dGDP* curves in the upper and middle panels.

As for other countries, we minimize the model residuals, i.e. determine the break years together with the regression coefficients. For Germany, the best fit model between 1971 and 2018 is as follows:

$$
\begin{aligned}
du_p &= -0.420 dlnG + 1.50, && 1970 > t \geq 1984 \\
du_p &= -0.555 dlnG + 0.700, && 1985 \geq t \geq 1992 \\
du_p &= -0.450 dlnG + 1.300, && 1993 \geq t \geq 2006 \\
du_p &= -0.450 dlnG + 0.400, && t \geq 2007
\end{aligned} \quad (9)
$$

The break years are determined automatically: 1984, 1993, 2007. The break in 1993 is likely associated with the 1991 reunification process but does not match this year exactly. The LSQR procedure has to include the spike in 1991 in one of the two adjacent segments, and decided to retain the spike in the earlier segment. The cause of this break is likely different from a simple revision to the dGDP definition.

Figure 18 presents the measured and predicted rate of unemployment (upper panel), the model residual error (middle panel), and the regression of the measured and predicted time series. The overall fit ($R^2$=0.88) is good with the break years close to those expected from Figure 17. One of the largest model errors in the residual time series was observed in 1990. This is most likely related to the reunification and merging of two time series belonging to two different economies. When this spike is excluded, the standard deviation fall from 0.99% to 0.76%, and $R^2$ increases from 0.88 to 0.92. Therefore, the modified Okun's law linking the change in the unemployment rate and the change in the real GDP per capita is validated by new data for Germany for the period between 2010 and 2019.

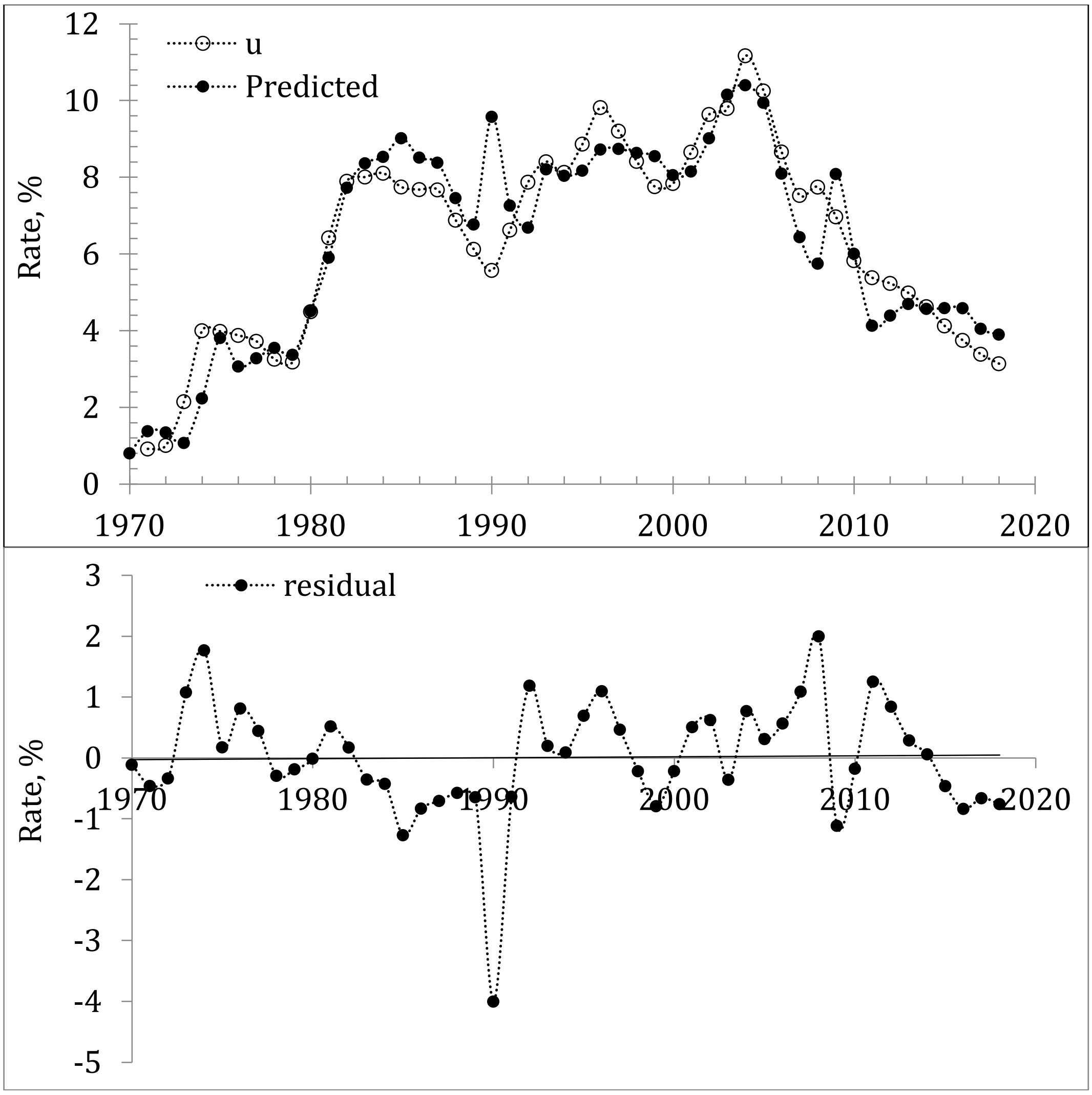

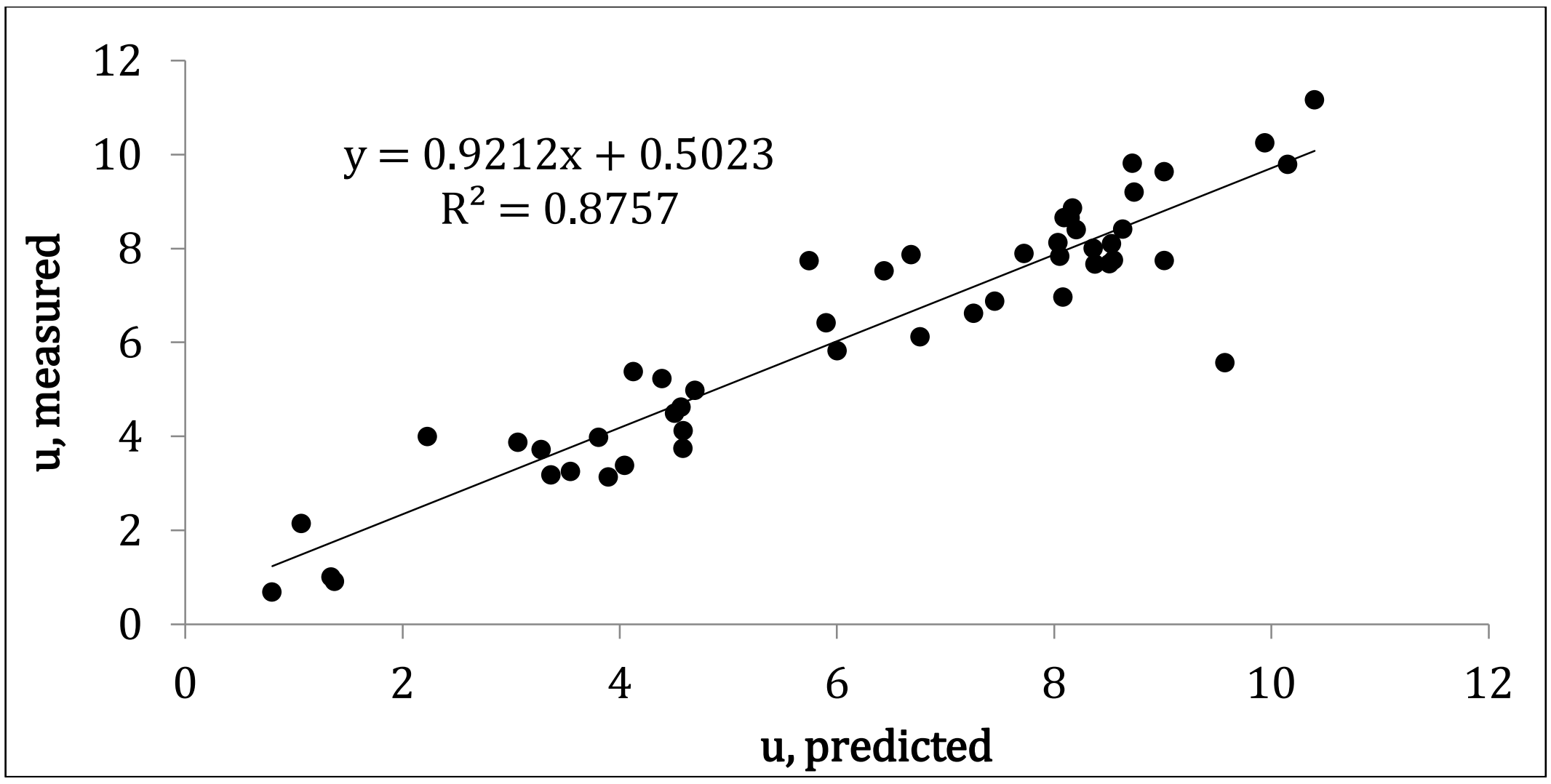

Figure 18. Upper panel: The measured rate of unemployment in Germany between 1970 and 2018, and the rate predicted by model (9) with the real GDP per capita and the unemployment rate published by the MPD. Middle panel: The model residual: stdev=0.99%. When the 1991 reading is excluded, stdev=0.76%. Lower panel: Linear regression of the measured and predicted time series. $R^2$=0.88.

2.5. *Canada*

In this Section, we apply the same approach to Canada and start with the CPI and GDP deflator difference, which is used to reveal definitional breaks in the dGDP estimates. Obviously, such breaks in the dGDP create breaks in the real GDP per capita estimates, and thus, in the statistical estimates associated with our model. One needs to find such breaks and allow the model to compensate corresponding disturbances. In the upper panel of Figure 19, we present the evolution of the cumulative inflation (the sum of annual inflation estimates) as defined by the CPI and dGDP between 1962 (we use the OECD data for the unemployment rate since 1961) and 2018. Both variables are normalized to their respective values in 1961. From the very beginning, the dGDP curve is above the CPI one. In the middle panel, the inflation rates are shown for both variables. In the lower panel, we present the fit between the CPI and the dGDP cumulative inflation curves after correction of the latter: from 1962 (coefficient 0.8), from 1977 (0.8*1.4=1.12), and from 2003(0.8*1.4*0.77=0.86). Hence, the break years are 1977 and 2003.

In our model for Canada, we are looking for breaks in the linear relationship near the years revealed by the revisions to the dGDP definition and obtain the following intervals and coefficients:

$$du_p = -0.270dlnG + 1.130, \quad 1977>t\geq1970$$
$$du_p = -0.281dlnG + 0.303, \quad 2000\geq t\geq1978$$
$$du_p = -0.280dlnG + 0.505, \quad 2009\geq t\geq2001$$
$$du_p = -0.350dlnG + 0.180, \quad t\geq2010 \qquad (10)$$

The break years are slightly different from those estimated from the inflation curves in Figure 19. This is likely due to much higher sensitivity of the predicted unemployment rate to the coefficients in (1). The cumulative inflation curves in the upper panel of Figure 19 are both synchronously corrected in many revisions through their whole length. The estimates of unemployment rate are obtained in the Current Population Surveys and represent independent estimates. The unemployment values are also corrected in the revisions to unemployment definition and when new population controls estimated after the decennial censuses. The original estimates cannot be not changed, but rather synchronously corrected.

The rate of unemployment is an independent economic variable obtained in independent measurements. The predicted rate of unemployment depends on the integral value of the real GDP per capita. This makes the predicted value to be very sensitive to the GDPpc evolution. In other words, the current prediction, $u_p$, depends on the initial value, $u(t_0)$ and the whole path of the GDPpc between $t_0$ and the current time. This is 49 years for Canada and 68 for the USA. The new readings of unemployment rate and GDPpc (2010 to 2019) validate the model, which links the change in the rate of unemployment and the relative growth rate of the real GDP per capita in Canada. It is worth noting that France, Germany and Canada have positive constant terms in the most recent period. Inflation will be growing without economic growth estimated by the MPD.

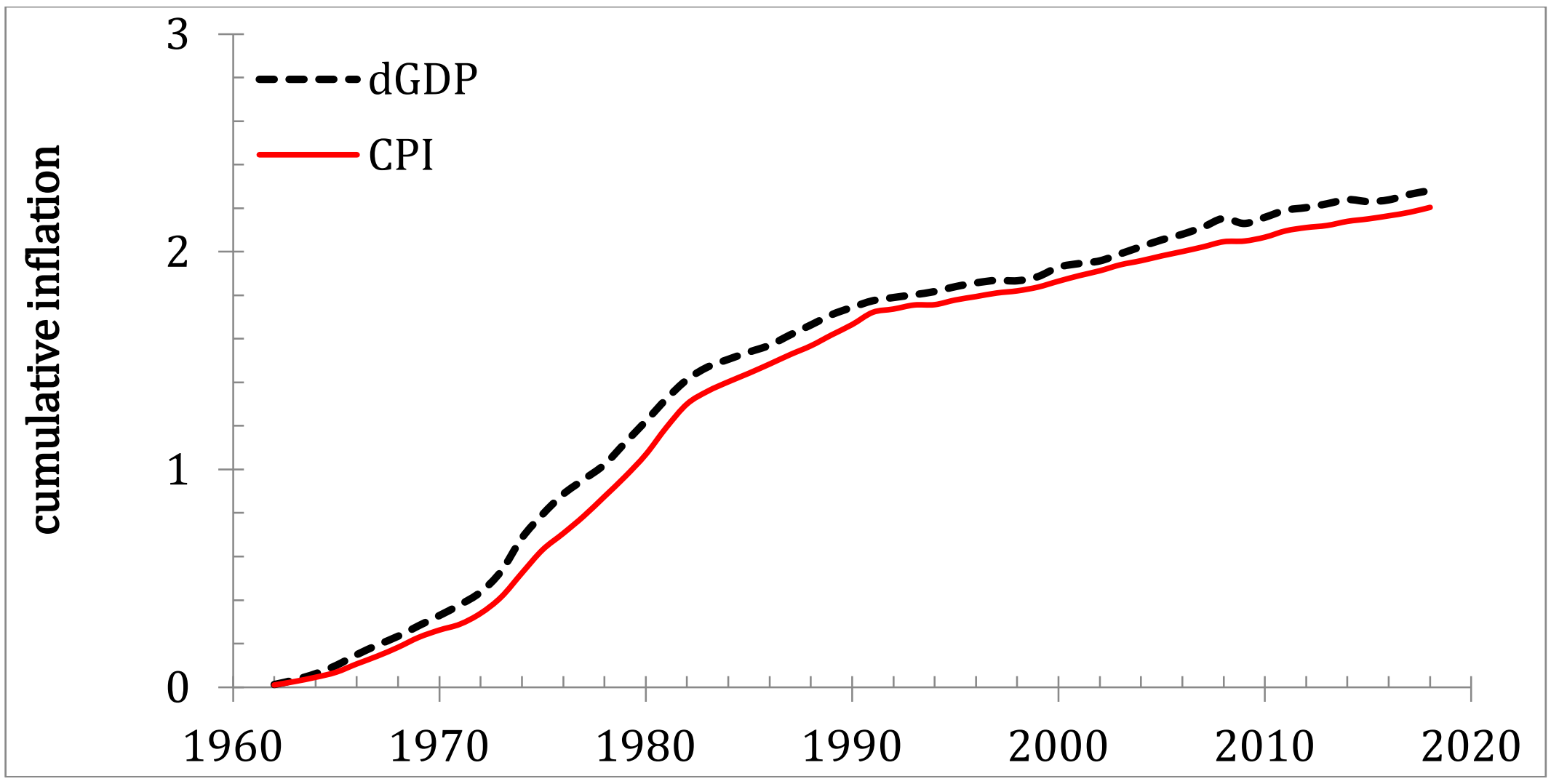

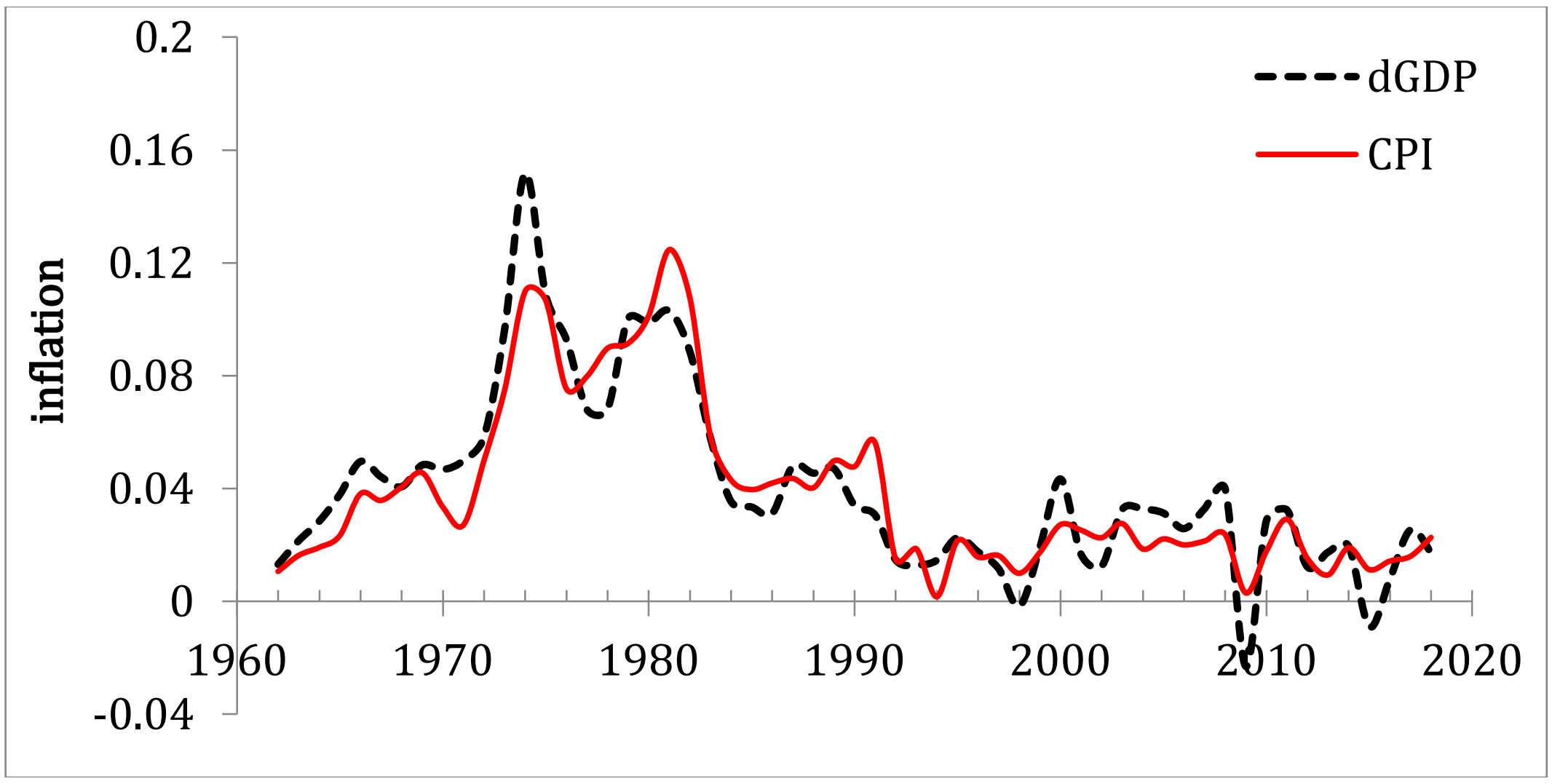

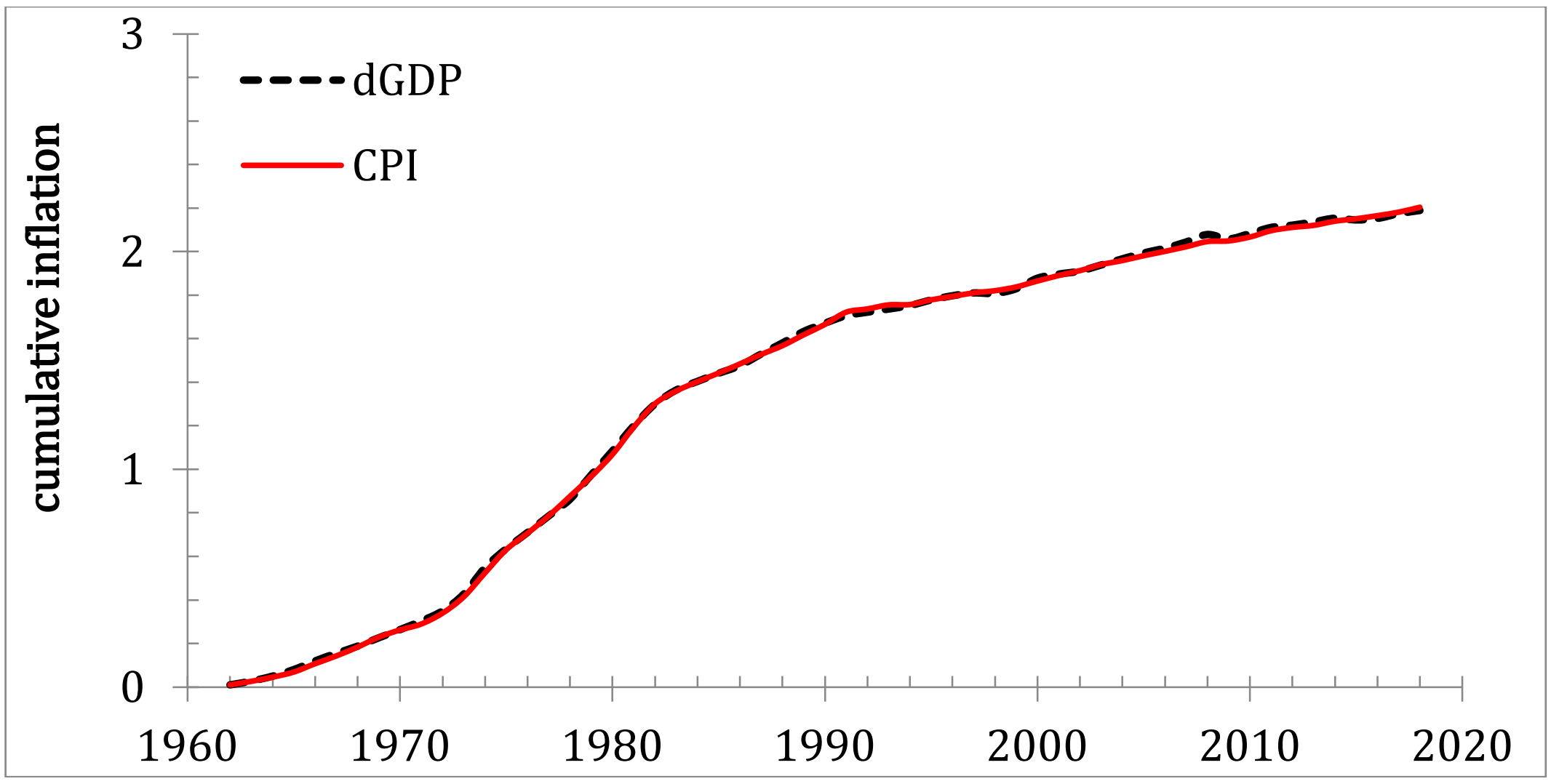

Figure 19. Upper panel: The evolution of the cumulative inflation (the sum of annual inflation estimates) as defined by the CPI and dGDP between 1961 and 2018. Both variables are normalized to their respective values in 1961. Middle panel: The dGDP and CPI inflation estimates. Lower panel: The fit between the *CPI* and the *dGDP* cumulative inflation curves after correction of the latter in 1962, 1977, and 2003.

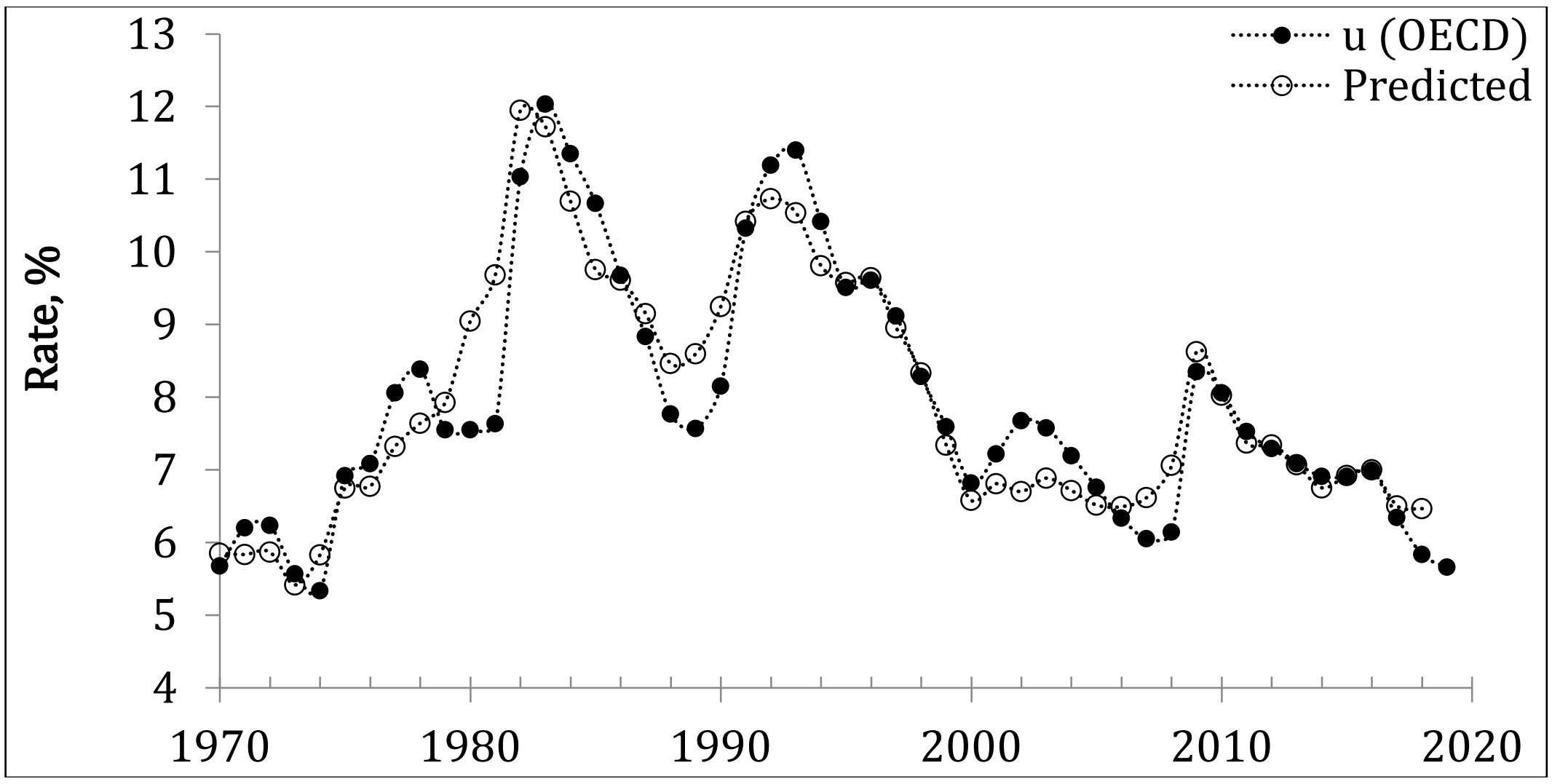

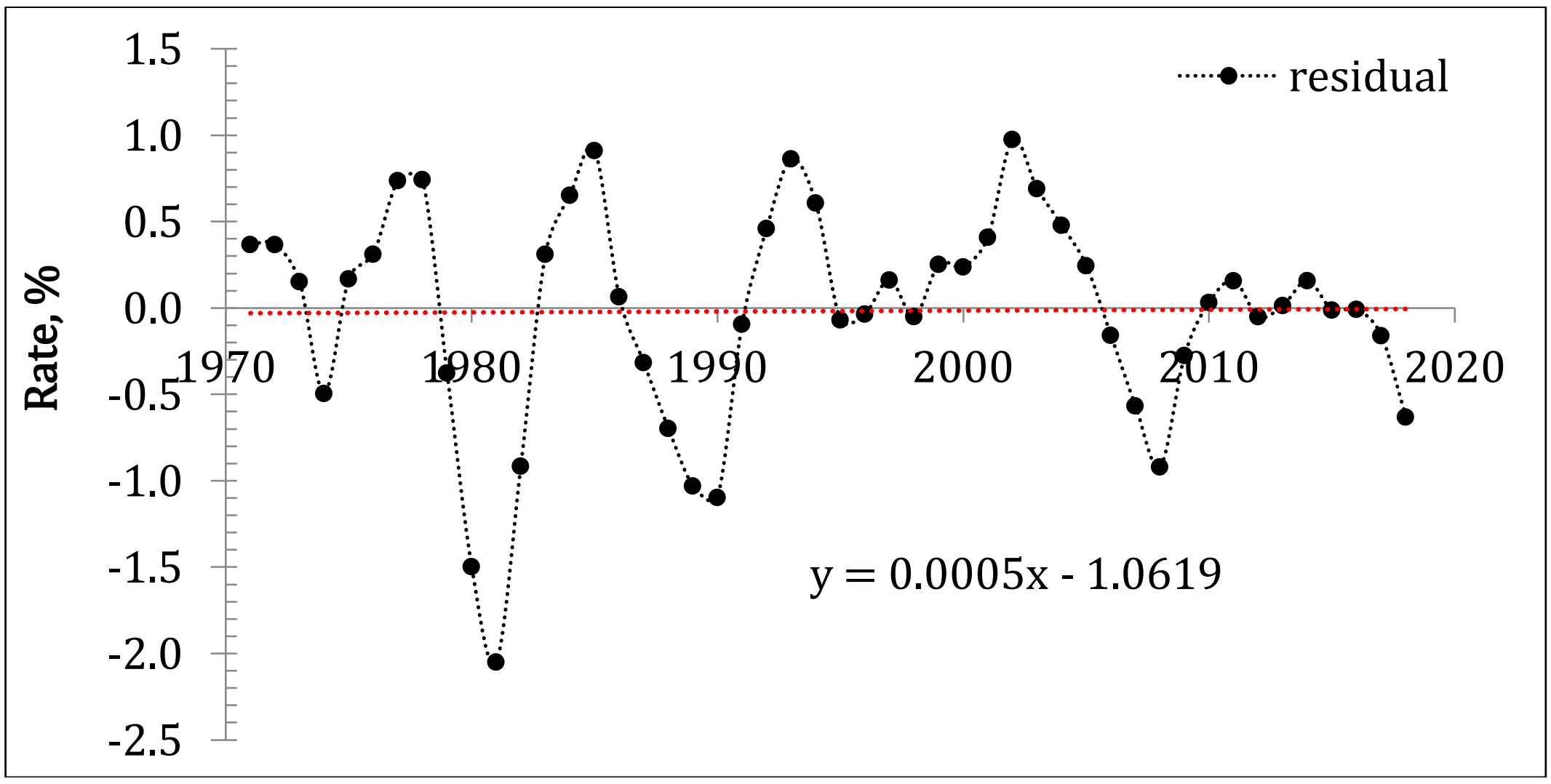

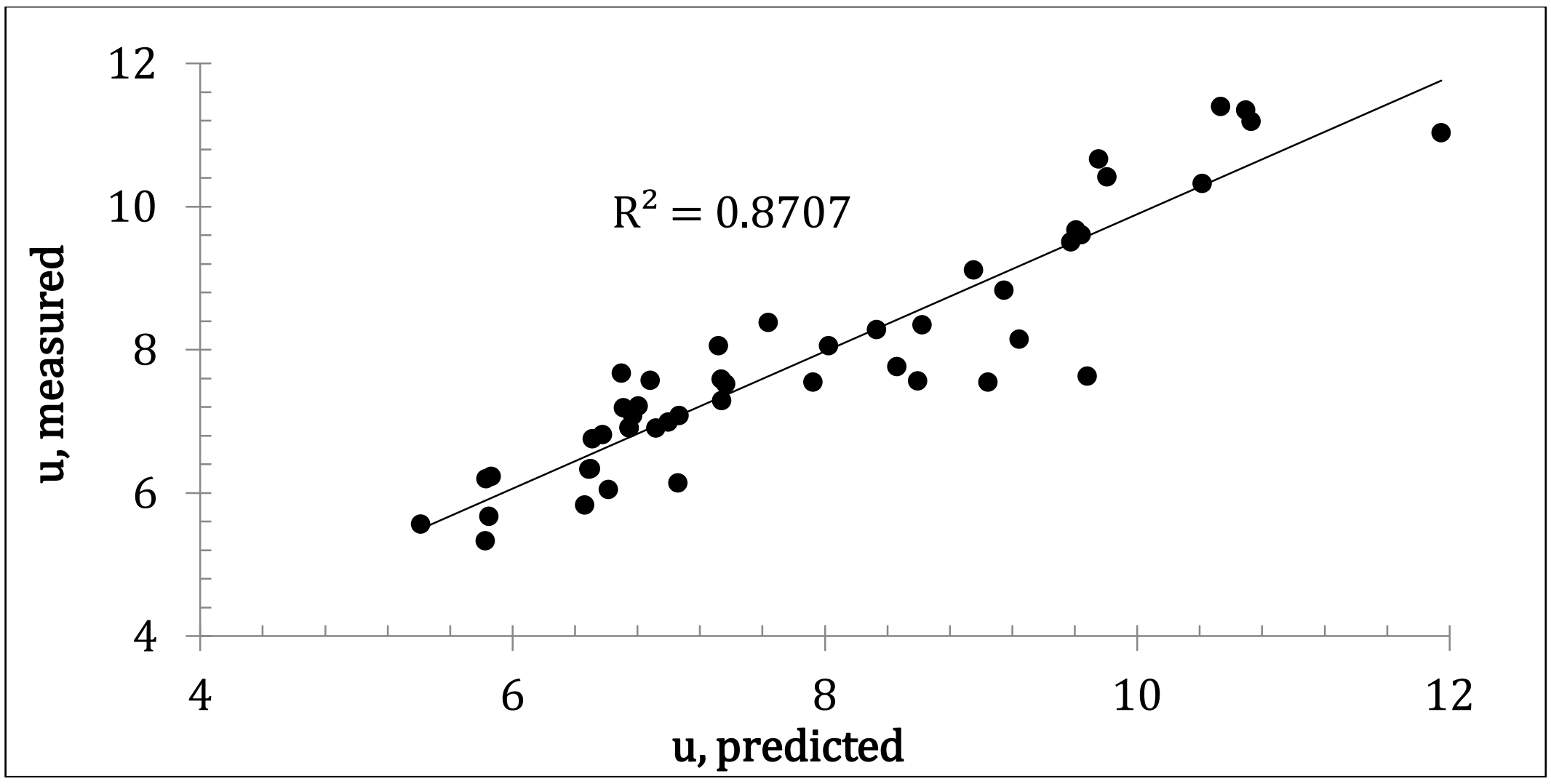

Figure 20. Upper panel: The measured rate of unemployment in Canada between 1970 and 2019, and the rate predicted by model (10) with the real GDP per capita published by the MPD and the unemployment rate reported by the OECD. Middle panel: The model residual: stdev=0.62%. Lower panel: Linear regression of the measured and predicted time series. $R^2$= 0.87.

2.6. *Australia*

According to the established procedure, for Australia we first present the breaks in the GDP deflator, *dGDP*. The difference between the *CPI* and *dGDP* clearly reveals the definitional breaks in the dGDP estimates. In the upper panel of Figure 21, we present the evolution of the cumulative inflation (the sum of annual inflation estimates) as defined by the *CPI* and *dGDP* between 1962 (we use the OECD data for the unemployment rate since 1961) and 2018. Both variables are normalized to their respective values in 1961. In the middle panel, inflation rate is presented for both indices, and the lower panel displays the differences of the curves in the upper and middle panel. The difference between cumulative price change estimates has complex structure with many pivot points. In such a complex structure, the estimated break years might be no so reliable due to larger uncertainty in the modelled parameters.

For Australia, we obtain the following intervals and coefficients:

$$
\begin{aligned}
du_p &= -0.76dlnG + 1.50, && 1993>t\geq 1977 \\
du_p &= -0.35dlnG + 0.75, && 2006\geq t\geq 1993 \\
du_p &= -0.76dlnG + 1.25, && 2013\geq t\geq 2007 \\
du_p &= -0.36dlnG + 0.25, && t\geq 2014 \qquad (11)
\end{aligned}
$$

The break years are present in the inflation difference curves in Figure 21, but not all pivot points are reflected in model (11). This is likely due to much higher sensitivity of the predicted unemployment rate to the coefficients in (11). There are 3 pivot points (breaks) in the linear dependence: 1993, 2006, and 2013. Nevertheless, the overall fit is good ($R^2$=0.87) as the lower panel in Figure 22 demonstrates. The revised model for Australia is successful and validated the original model of unemployment change.

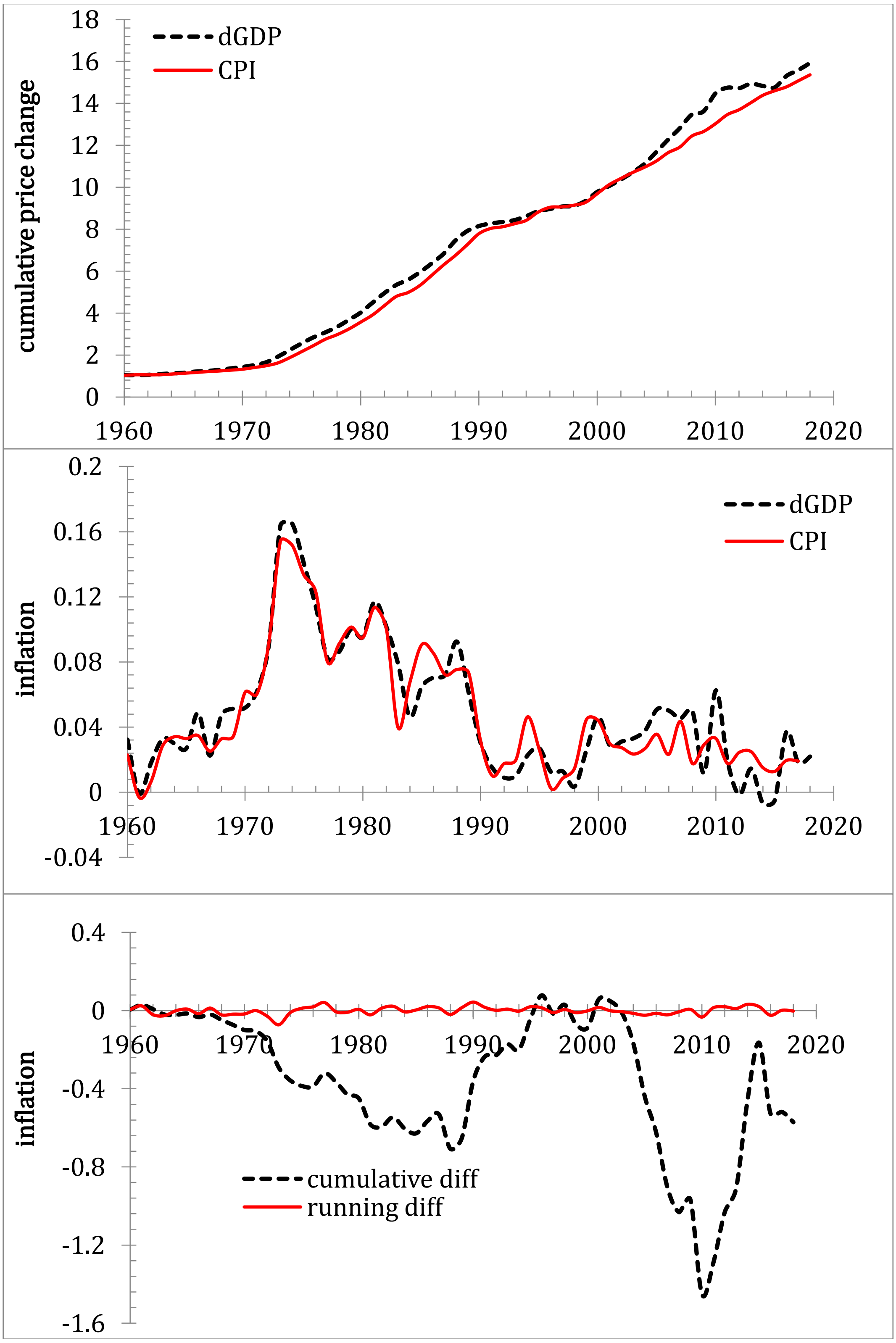

Figure 21. Upper panel: The evolution of the cumulative inflation (the sum of annual inflation estimates) as defined by the *CPI* and *dGDP* between 1961 and 2018. Both variables are normalized to their respective values in 1961. Middle panel: The *dGDP* and *CPI* inflation estimates. Lower panel: The difference between the CPI and the dGDP curves in the upper and middle panels.

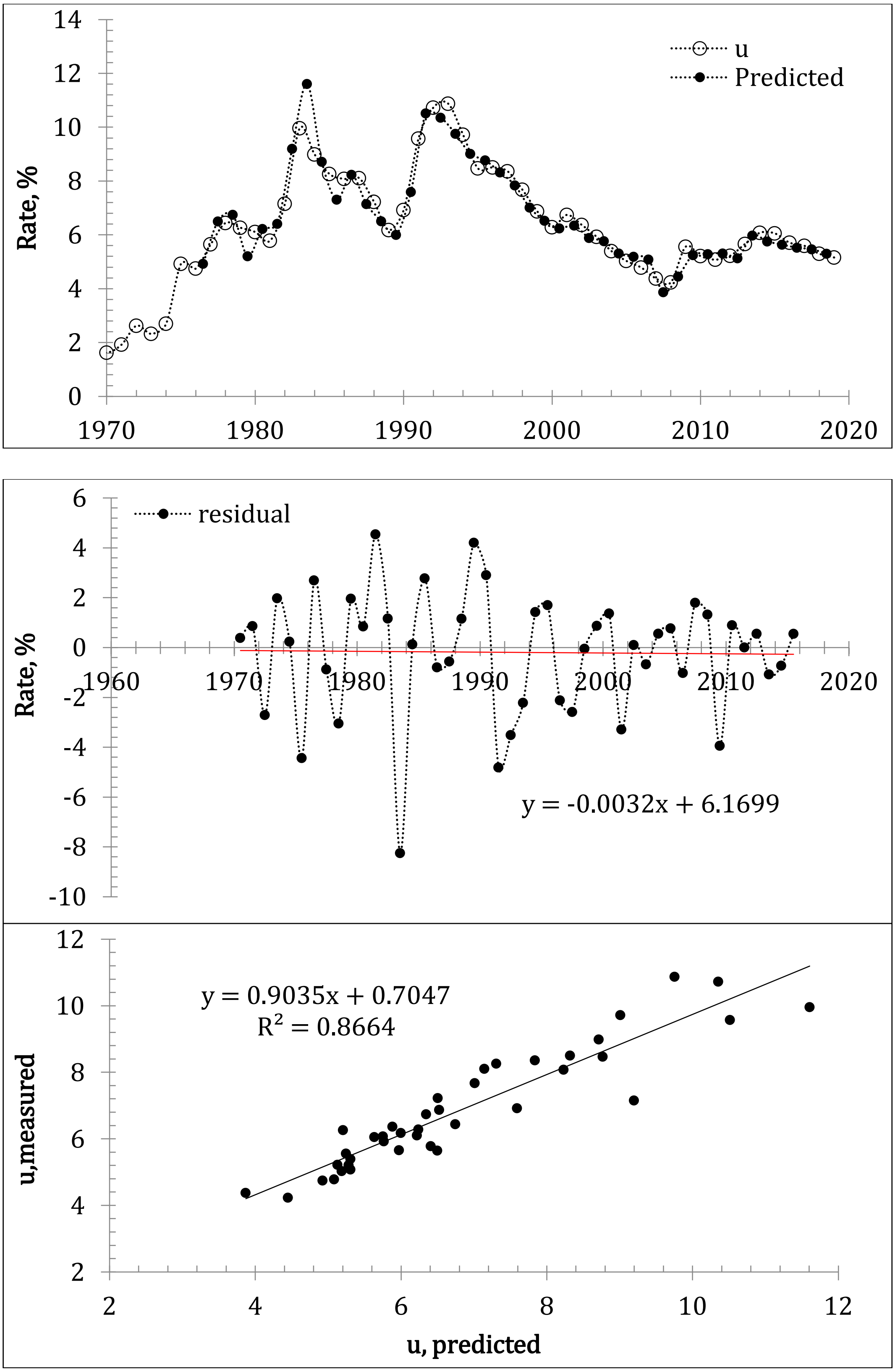

Figure 22. Upper panel: The measured rate of unemployment in Australia between 1977 and 2018, and the rate predicted by model (1) with the real GDP per capita published by the MPD and the unemployment rate reported by the OECD. Middle panel: The model residual: stdev=2.5%. Lower panel: Linear regression of the measured and predicted time series. $R^2$= 0.87.

2.7. *Spain*

The rate of unemployment in Spain deserves a special study. It was above 10% since 1980 and reached its peak of 27.8% in 2013. Larger variations in amplitude are important for an accurate statistical estimation of the model parameters and we revise the model for Spain with new data between 2010 and 2018 with the peak value to model.  The CPI and GDP deflator difference is used to reveal potential definitional breaks in the dGDP estimates. One has to find potential breaks and allow the model to compensate corresponding disturbances and to provide unbiased statistical estimates of the defining parameters. For the Spain model, no dummy variables are used and the best fit is achieved only with the structural breaks, i.e. the change in the coefficients of linear regression.

In the upper panel of Figure 23, we present the evolution of the cumulative inflation (the sum of annual inflation estimates) as defined by the *CPI* and *dGDP* between 1970 and 2018 (the OECD data). Both variables are normalized to their respective values in 1970. The *dGDP* curve is close to the *CPI* before 1995 and then a significant deviation is observed. There is a low amplitude deviation between 1980 and 1990. After 1996, the deviation increases in amplitude and the *dGDP* is first above the *CPI* curve and then dives below the *CPI* line in 2012. In the middle panel, the rates of price inflation are shown for both indices. In the lower panel, we present the difference between the *CPI* and the *dGDP* cumulative inflation curves in the upper and middle panels. One can suggest the presence of breaks in 1979, 1985, 1995, 2007, and 2014. This is for the model to decide, however, when the breaks result in the bets LSQR fit.

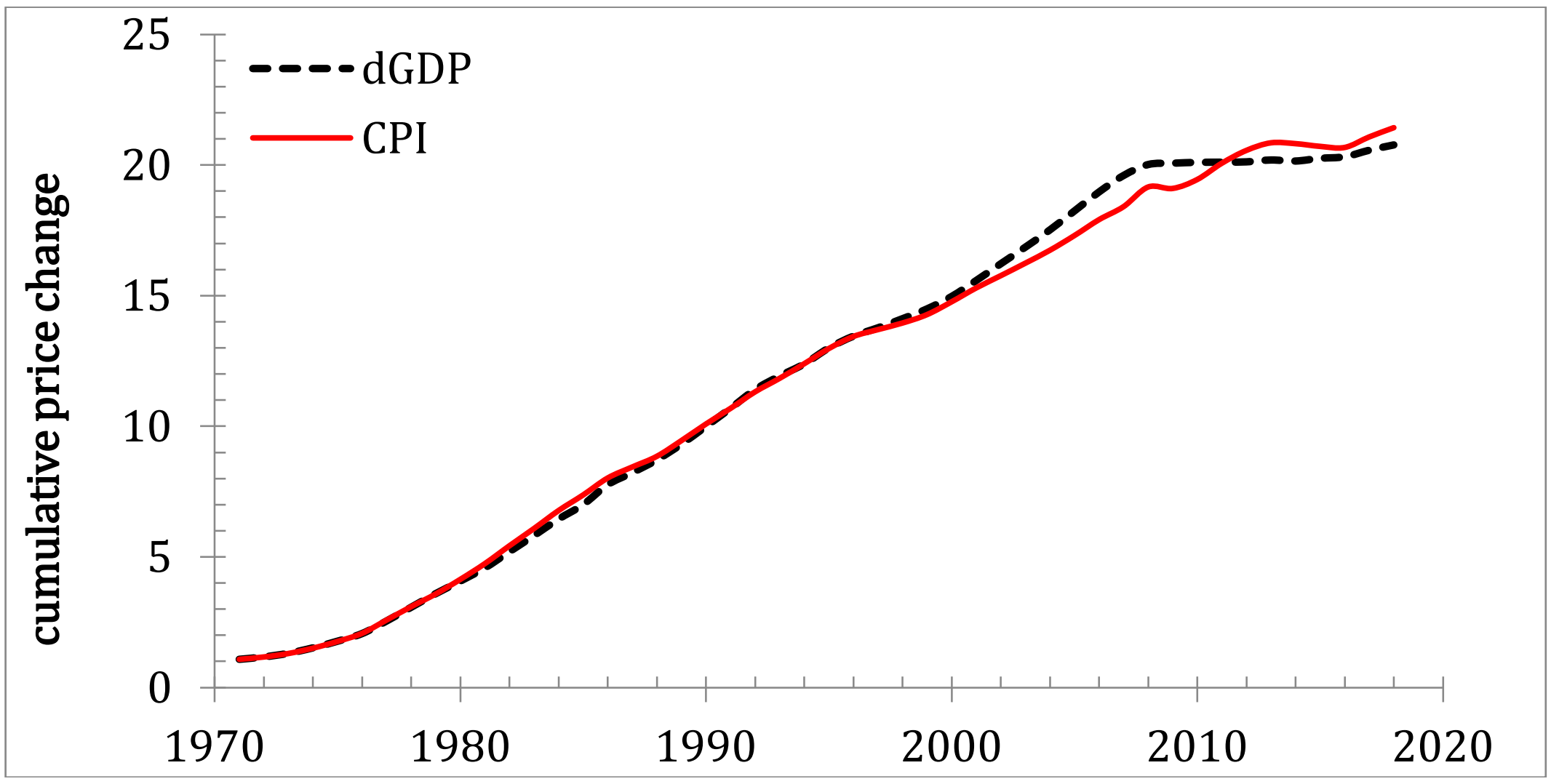

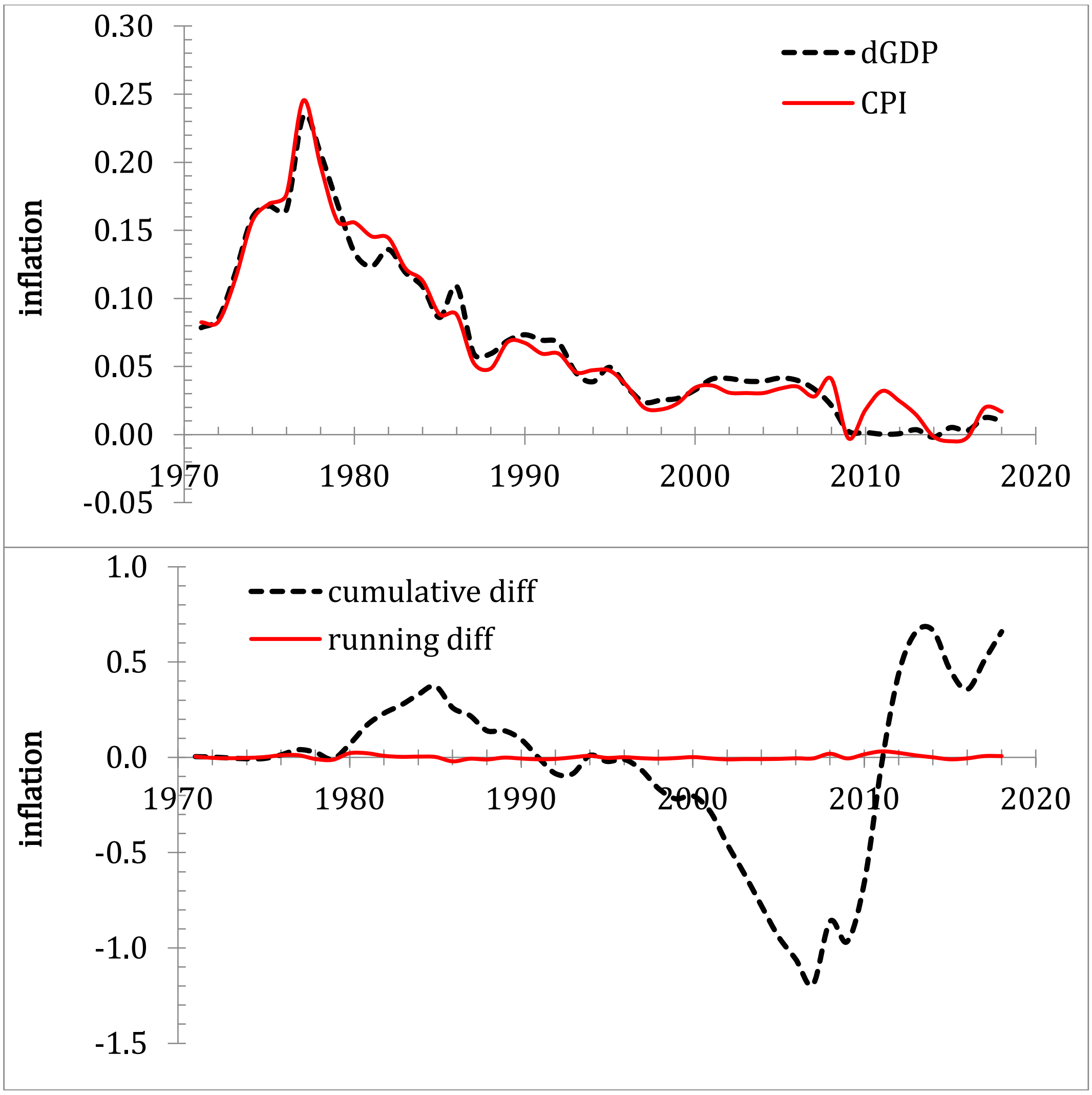

Figure 23. Upper panel: The evolution of the cumulative inflation (the sum of annual inflation estimates) as defined by the *CPI* and *dGDP* between 1970 and 2018. Both variables are normalized to their respective values in 1970. Middle panel: The *dGDP* and *CPI* inflation estimates. Lower panel: The difference between the curves in the upper and middle panels. One can observe the breaks in the difference between the cumulative curves. We suggest potential breaks in 1979, 1985, 1995, 2007, and 2014.

The modified model for Spain is:

$$du_p = -0.40dlnG + 2.11, \quad 1995>t\geq1970$$
$$du_p = -0.95dlnG + 2.03, \quad 1996\geq t\geq2013$$
$$du_p = -0.50dlnG - 2.10, \quad t\geq2014 \qquad (12)$$

The break years in Figure 24 are close to those estimated from the inflation curves in Figure 23, but not all potential breaks are used. The overall fit shown in the upper panel in excellent, as confirmed by the residual errors in the middle panel and the regression ($R^2$=0.96) of the predicted and measured employment between 1973 and 2018. We retain in mind that the estimates of unemployment rate are obtained in the population surveys, i.e. they are independent measurements. At the same time, the unemployment values are also corrected in the revisions to unemployment definition and the change in the population controls after censuses.

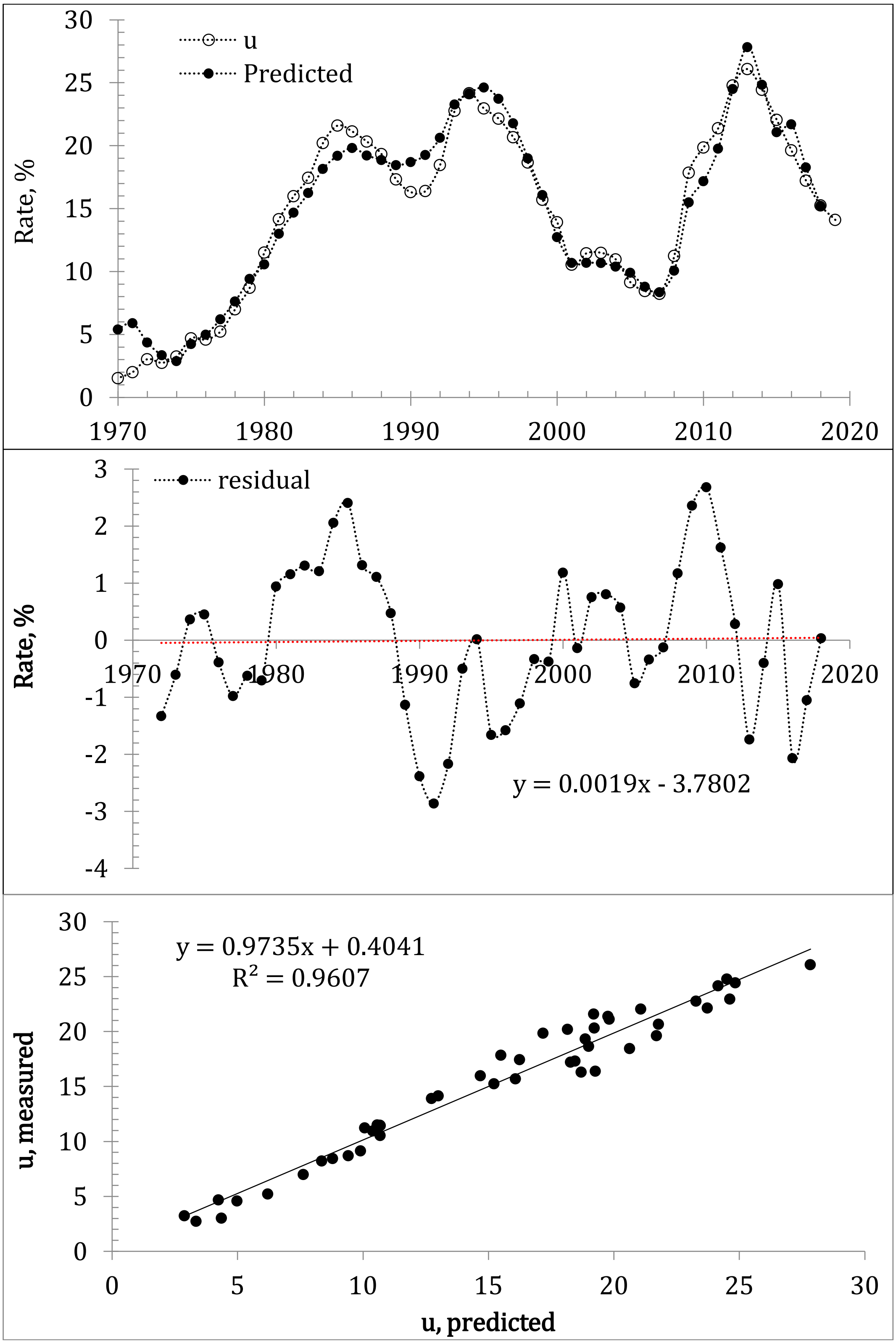

Figure 24. Upper panel: The measured rate of unemployment in Spain between 1970 and 2018, and the rate predicted by model (12) with the real GDP per capita published by the MPD and the unemployment rate reported by the OECD. Middle panel: The model residual: stdev=1.3%. Lower panel: Linear regression of the measured and predicted time series. $R^2$= 0.96.

Considering the fall in real GDP growth caused by the COVI-19 pandemic one could expect that the rate of unemployment in Spain may increase according to equation (12) and probably will stay at the elevated level. Coefficient -0.5 in (12) predicts that 1% decrease in real GDP per capita is converted in a 0.5% increase in the rate of unemployment. For Spain with its extremely high historical unemployment rate this is a problem.

2.8. *Austria*

In the previous Sections, we revisited and validated our version of Okun's law for the USA, UK, France, Germany, Canada, Australia, and Spain with new GDP and unemployment data for the years between 2010 and 2019. In all seven countries, the revised model accurately described the new data, i.e. the original model is validated. Austria is a much smaller economy dependent on the neighbouring countries, and the model validation for Austria is of practical and theoretical importance.

In order to reach the best fit between the measured and predicted unemployment rates, we introduce structural breaks related to the change in real GDP definition. To illustrate the breaks in the real GDP data (i.e. nominal GDP corrected for the price change) we compare price inflation estimates as defined by the *GDP* deflator and *CPI*. The latter is considered as a reference. It is also important that the goods and services in the CPI are parts of the GDP deflator, *dGDP*. In the past, the *CPI* and *dGDP* were almost equivalent. In panel a) in Figure 25, we present the evolution of the cumulative inflation as defined by the *CPI* and *dGDP* between 1970 and 2018 (the OECD data). Both variables are normalized to their respective values in 1970. The *dGDP* curve is close to the *CPI* one before 1982 and then some low-amplitude deviations are observed. After 1995, the deviation increases in amplitude and the *dGDP* is below the *CPI*. In panel b), the inflation rates are shown for both variables. Panel c) presents the difference between the *CPI* and the *dGDP* cumulative inflation curves in panels a) and b) and suggest the presence of breaks in 1982, 1996, and 2007. Panel d) in Figure 25 presents the original *CPI* curve and the corrected *dGDP* curve similar to that in Figure 19. The fit is good. As a result, our modified Okun's law model is allowed to have breaks near 1982, 1996 and 2007.

For Austria, we obtained the following intervals and coefficients:

$$du_p = -0.25dlnG + 0.60, \quad 1982>t\geq1970$$
$$du_p = -0.36dlnG + 0.97, \quad 2006\geq t\geq1982$$
$$du_p = -0.40dlnG + 0.34, \quad t\geq2007 \qquad (13)$$

The break years in Figure 26 are the same as estimated from the inflation curves in Figure 25. The overall fit shown in the upper panel in excellent, as confirmed by the residual errors in the middle panel and the regression ($R^2$=0.92) of the predicted and measured employment between 1970 and 2018. The unemployment values are also corrected in the revisions to unemployment definition (e.g., Austria did not include in unemployment those who had no job before, e.g. graduates). Considering the fall in real GDP growth caused by the COVI-19 pandemic one could expect that the rate of unemployment in Austria may increase according to equation (13) and probably will stay at an elevated level. Coefficient -0.4 in (13) predicts that 1% of drop in real GDP per capita growth is converted in a 0.4% increase in the rate of unemployment.

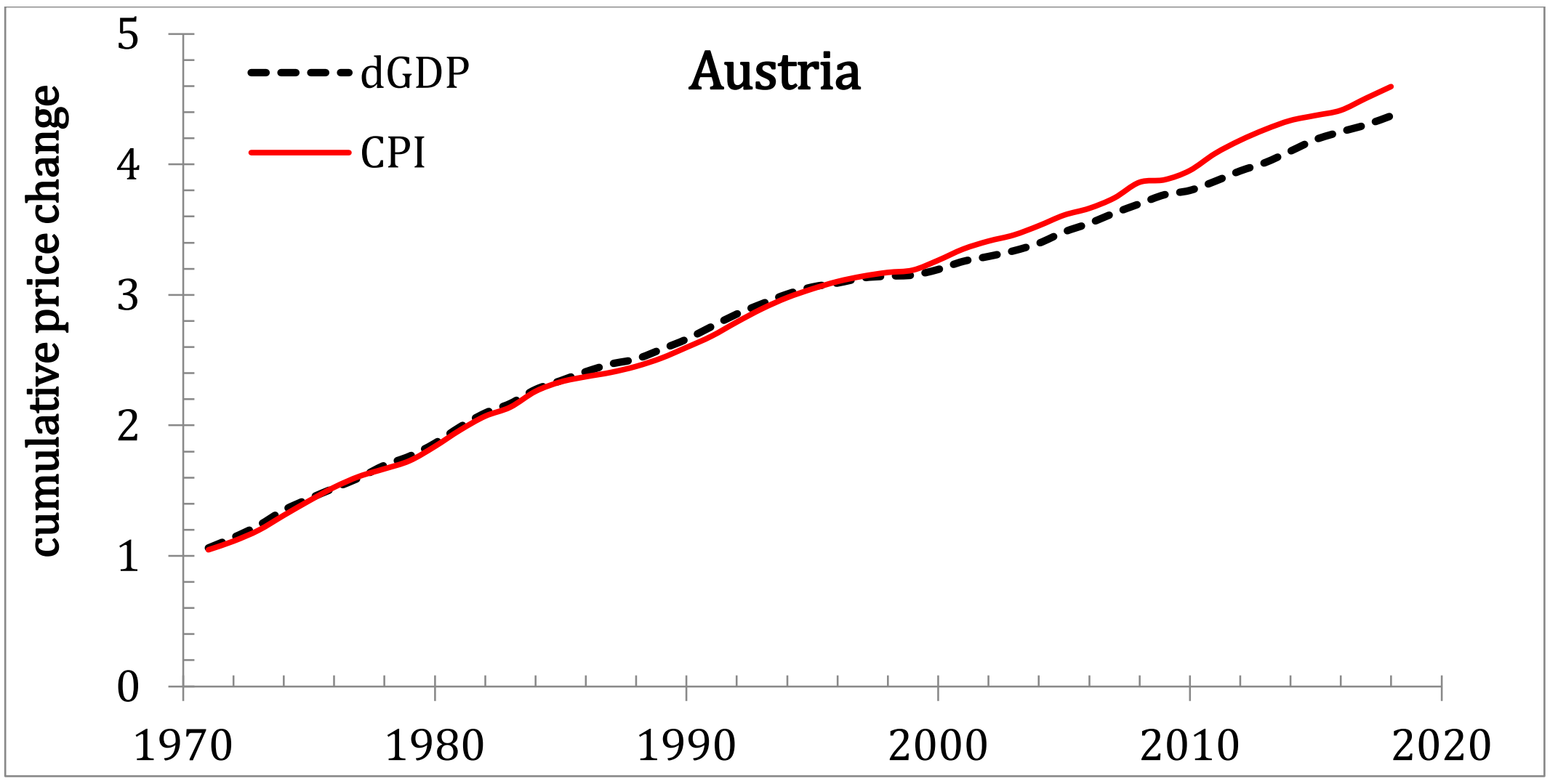
a)

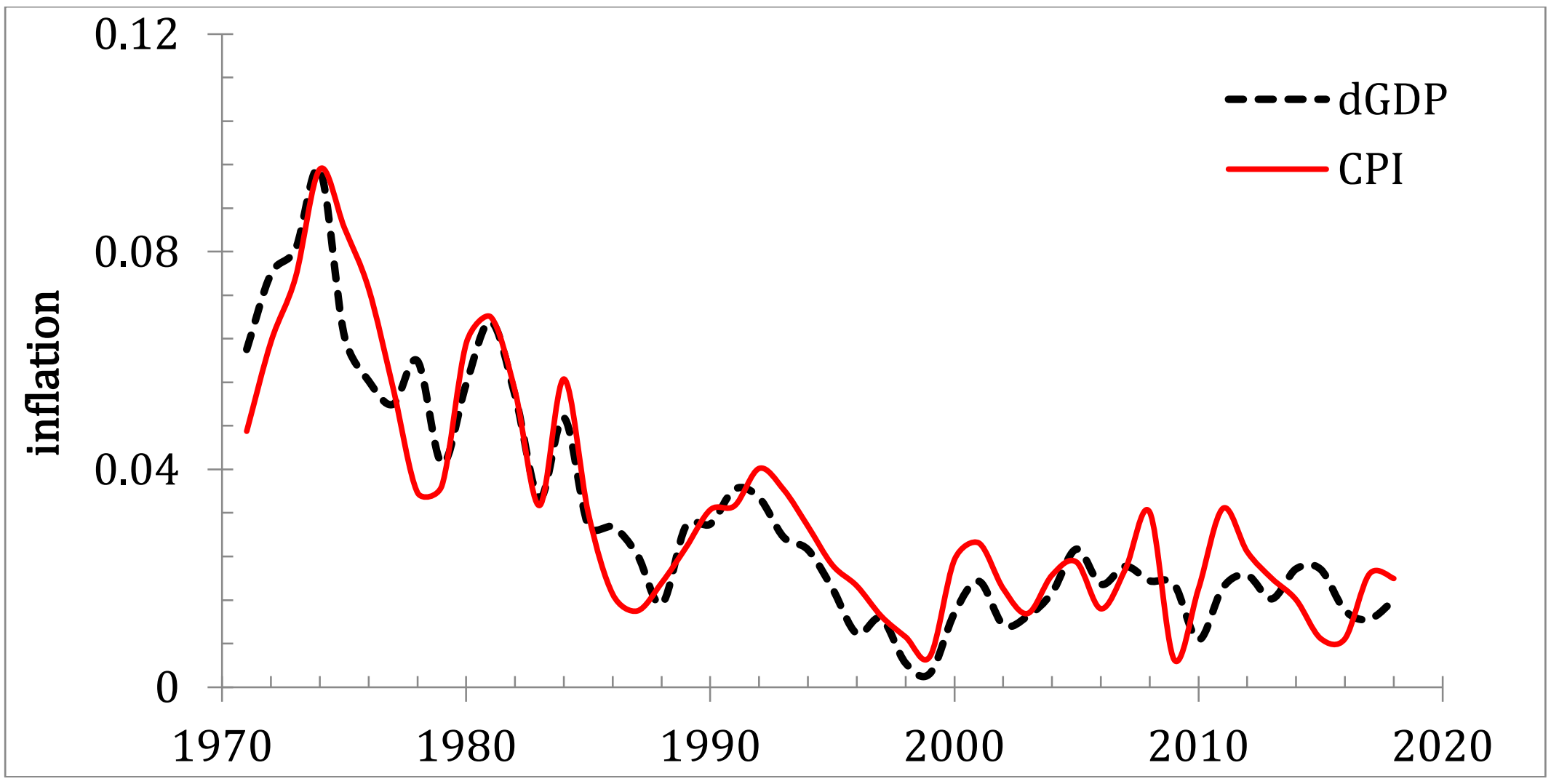
b)

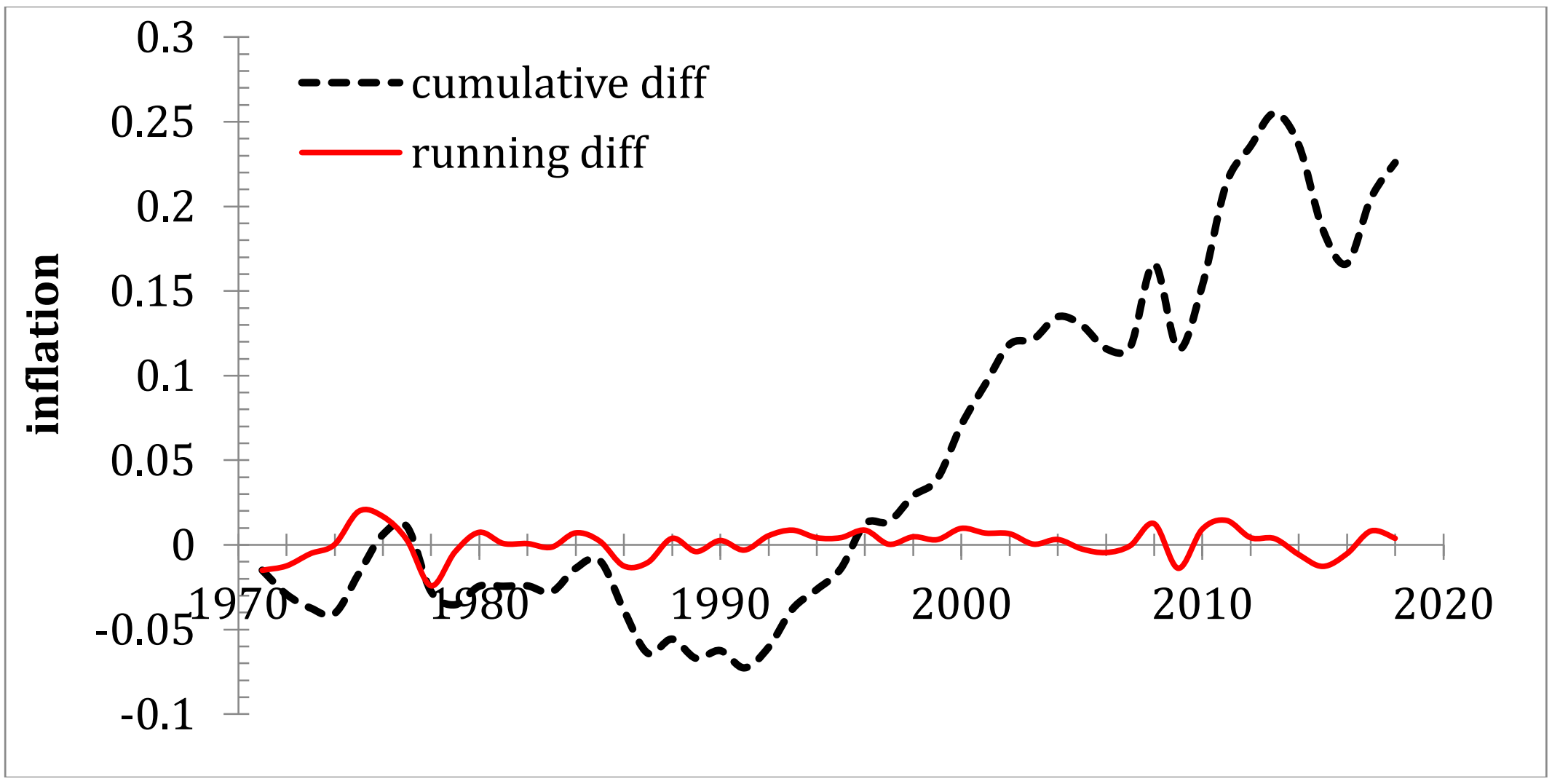
c)

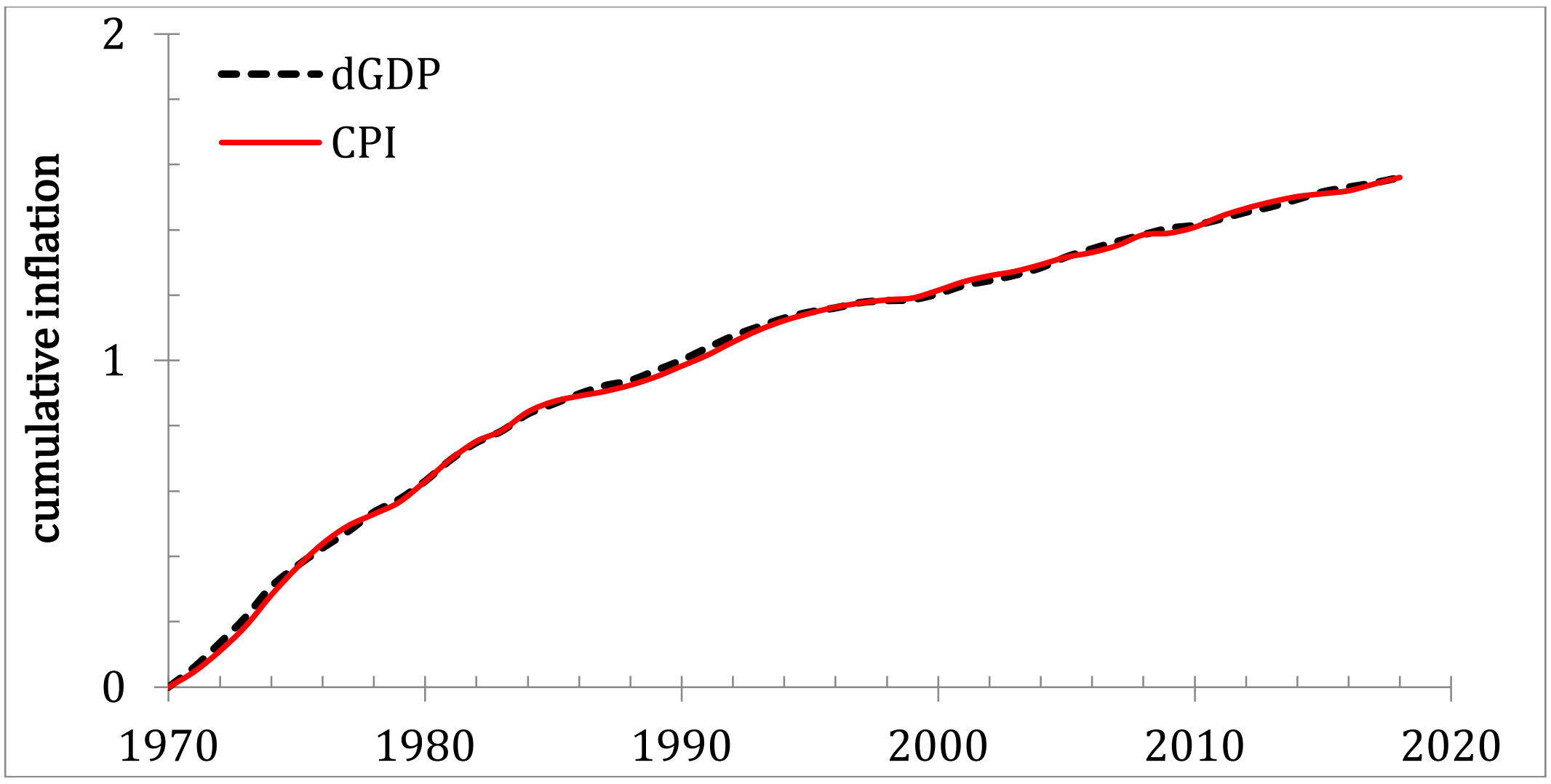

d)

Figure 25. a) The evolution of the cumulative inflation (the sum of annual inflation estimates) as defined by the *CPI* and *dGDP* between 1970 and 2018. Both variables are normalized to their respective values in 1961. b) The *dGDP* and *CPI* inflation estimates. d) The difference between the curves in panels a) and b). One can observe the breaks in the difference between cumulative curves. We propose the breaks in 1982, 1996 and, 2007. d) The fit between the CPI and the dGDP cumulative inflation curves after correction of the latter in 1982, 1996 and, 2007.

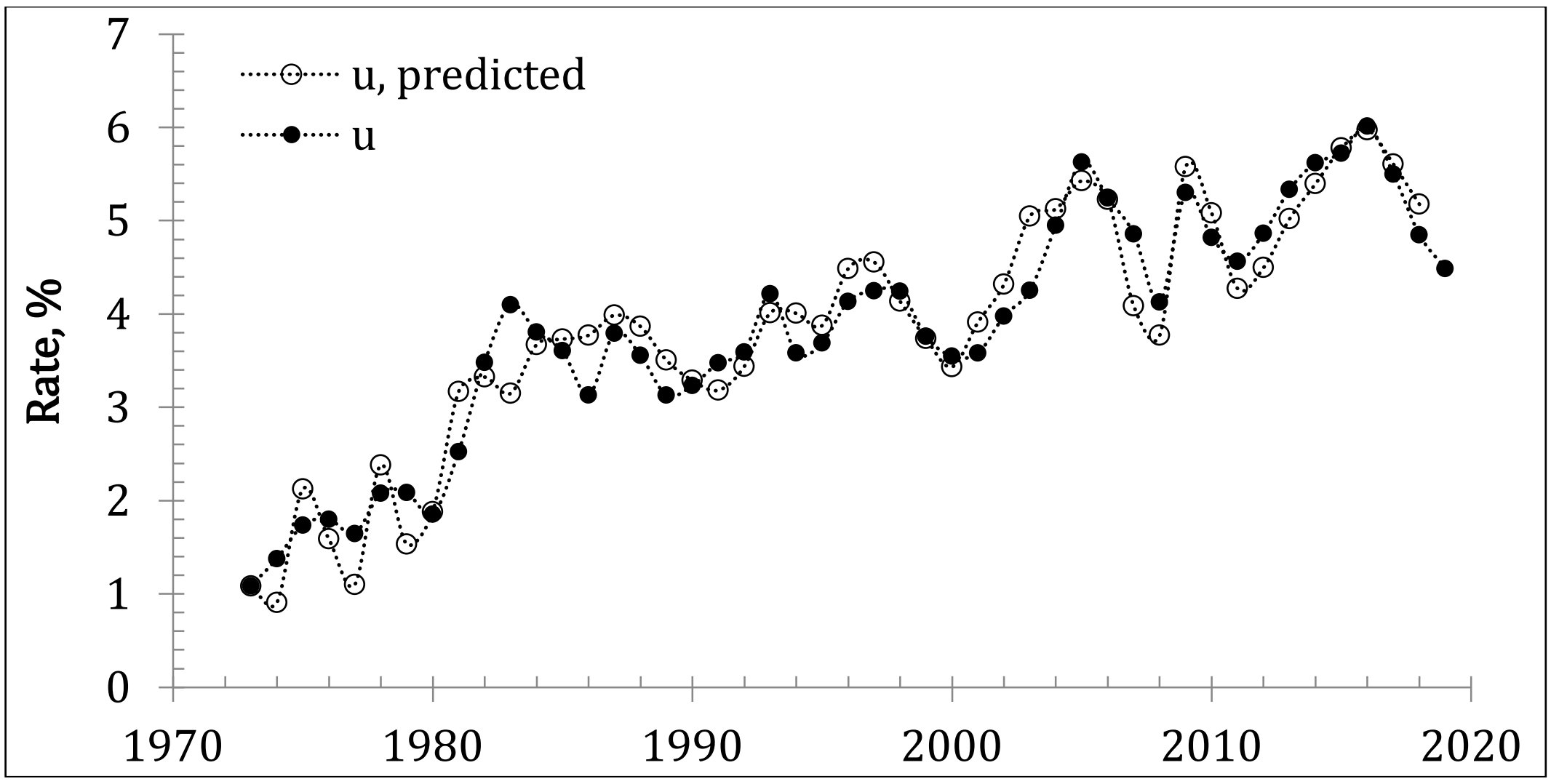

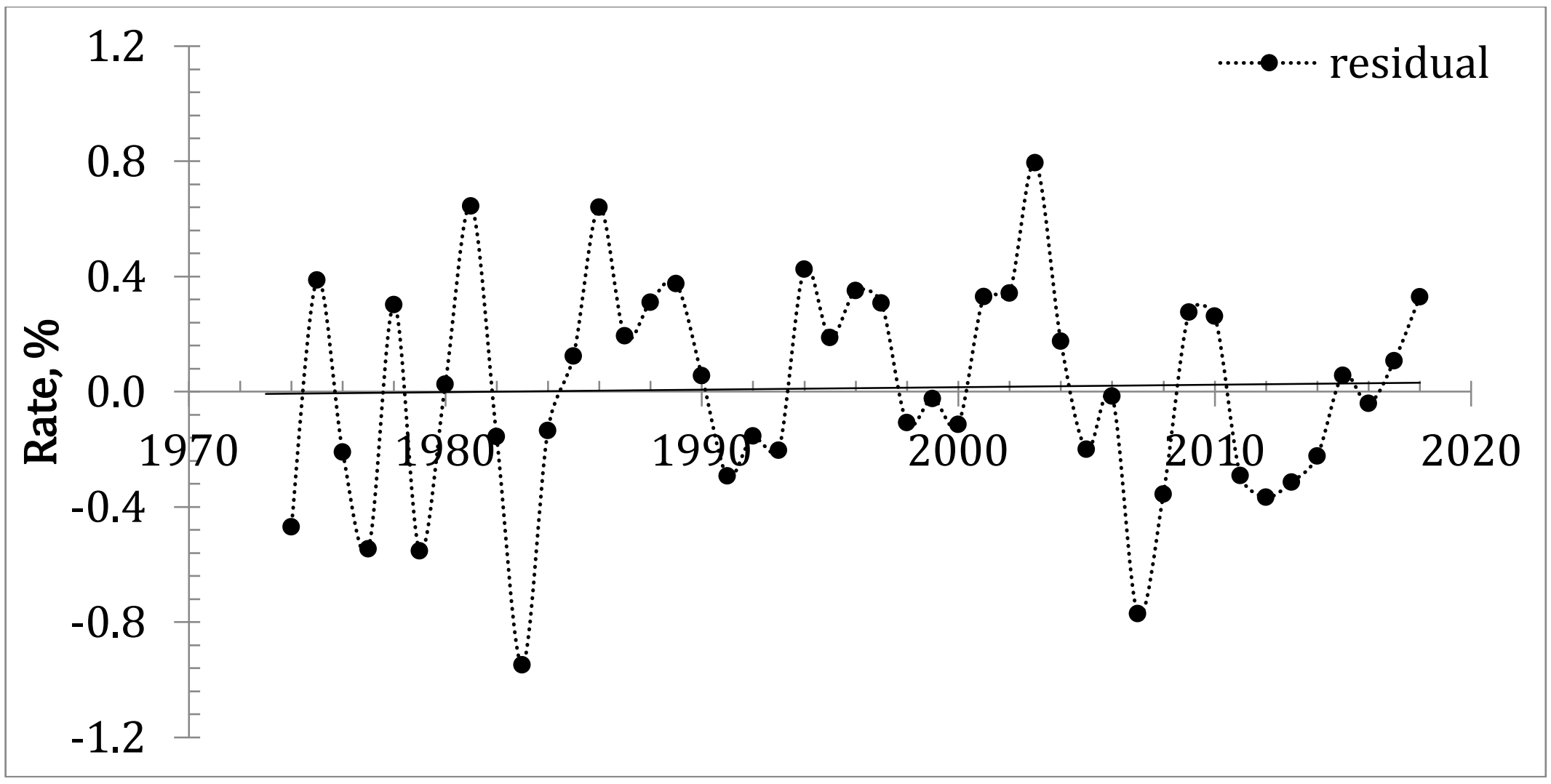

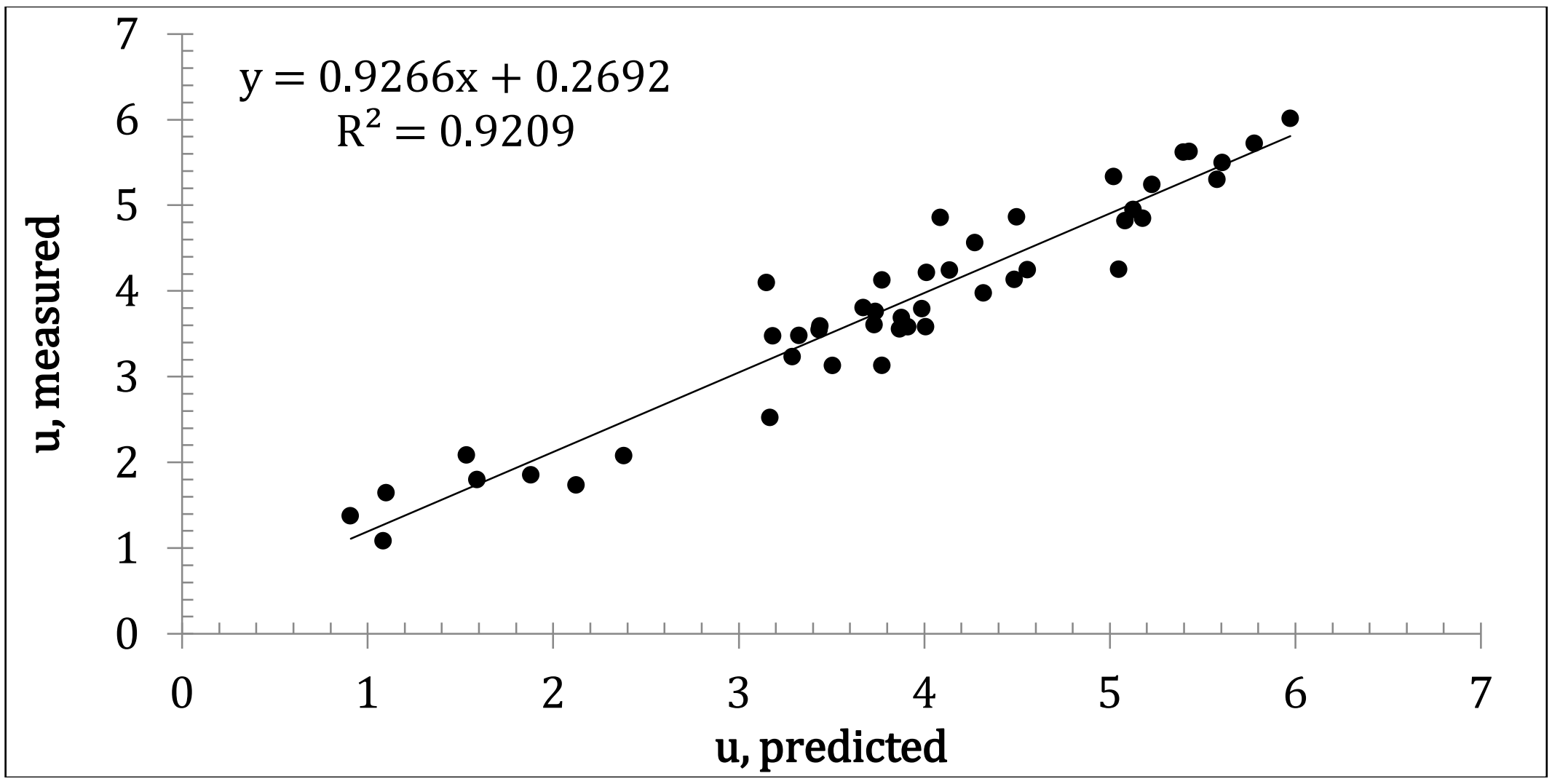

Figure 26. Upper panel: The measured rate of unemployment in Austria between 1970 and 2018, and the rate predicted by model (1) with the real GDP per capita published by the MPD and the unemployment rate reported by the OECD. Middle panel: The model residual: stdev=0.37%. Lower panel: Linear regression of the measured and predicted time series. $R^2$ = 0.92.

3. **Problems with the real GDP per capita estimates**

In this paper, we revisit our original model (modified version of Okun's law) linking the rate of unemployment and the change rate of real GDP per capita in developed countries. The modification used in our approach is just conversion of the link between the rate of unemployment and real GDP per capita (as a measure of output gap) into a differential form. Then the integral change in the unemployment rate is predicted by the GDP per capita growth. For this reason, the real GDP per capita data are needed as the major term of the differential equation. We have already reported on the definitional revisions (problems) to the GDP deflator which makes our model piecewise to match these revisions. However, we have also found and reported another problem – the real GDP per capita estimates provided by various sources (BEA, OECD, Total Economy Database, and Maddison Project Database) are quite different. There was no reason to classify these differences in conspiratorial sense and we used them without prejudice as fully interchangeable.  Economic measurements and data analysis is a serious professional occupation, and some features revealed during the practical use of the GDPpc estimates from various sources give us an impression of not random bias which makes us to think in non-economic words.

We have presented several examples of biased estimates. The reference years in economic time series are related to real GDP change to later dates with the major/comprehensive revisions. One cannot directly compare the real GDP per capita estimates from two sources when the reference years are different. Therefore, we normalize all time series to the same year, usually to the start year of the shortest time series. Obviously, the relative change in the real GDP per capita has to be the same, when all time series are normalized to the same year and we can consider any difference as related to definitions used in the corresponding estimation procedure. The economics is a developing science in both theoretical and experimental (measurement) parts and we understand the necessity of different approaches as an important methodological aspect of the overall progress. However, the differences between the normalized time series reveal high bias in the estimation of real GDP growth, specifically, in the countries related to the preparation of these estimates.

The upper panel in Figure 9 in Section 2.1 displays the evolution of real GDP per capita estimates for the USA borrowed from four sources: the Bureau of Economic Analysis (BEA - USA), Organization of Economic Cooperation and Development (OECD, Headquarters – Paris), Maddison Project Database (MPD – Netherlands), and Total Economy Database (TED – USA, China, …). In the past, the MPD was also in contact with the Conference Board publishing the TED. Currently, the MPD and TED are two different databases and underlying methodologies. One can see that the TED gives the highest growth in the GDP per capita since 1970. The MPD provides the lowest estimates. In the lower panel, several pair-wise ratios are presented in order to illustrate the relative differences in the four time series. It is worth noting that the BEA and OECD provide the same estimates except the most recent period, which is subject to further revisions, however.

Figure 27 is similar to Figure 9 and illustrates the case of Germany. The best result to Germany is given by the MPD – the total growth in the real GDP per capita since 1970 is 2.67. The OECD is less generous and gives the factor of 2.41. The TED gives the worst estimate – 2.09. The US based source with tight connections to China does not present Germany as a country with a healthy economic growth.

The OECD headquarters is in Paris, France. Figure 16 in Section 2.3 displays the real GDP per capita estimates and their ratios. It proves that assumption that the OECD is in favor of France in terms of the rate of economic growth since 1950. The OECD estimate is a factor of 4.93 between 1950 and 2018, which is much higher than 4.66 from the MPD and 4.69 from the TED. The OECD curve is above the other two sources from the very beginning.

There is potential bias in the real GDP per capita estimates for the UK. Figure 28 shows that the largest growth is estimated by the Office of national statistics (ONS). The ONS is a national source and its bias is not unexpected. The OECD gives almost the same estimates as the ONS. The TED is in favor of modest economic growth in the UK, and the MPD is the least generous.

Summarizing the observations in Figures 9, 16, 27, and 28 one can conclude that the data origin defines the method of real GDP estimation most appropriate for the related country to have the largest real economic growth. Such an approach definitely introduces serious bias in the estimates of various economic variables used for quantitative economic analysis. The latter becomes vulnerable to non-economic forces and likely suffers larger problems with statistical estimates in the mainstream economic models.

The cases of Japan (Figure 29) and Switzerland (Figure 30) must be a joke. For Japan, the TED provides the estimates showing the real GDP per capita growth by a factor of 3.25 since 1970, with the OECD estimate of only 2.58. In Switzerland, the total increase reported by the MPD is 2.56, and the other two sources give approximately 1.6. It is not clear how the TED or MPD estimates are so different if all economic agencies use similar sets of original data and methodologies. Such differences should not acceptable for the professional economic community. One cannot assess statistical performance of the real economic growth models using fully incompatible estimates of basic economic variables.

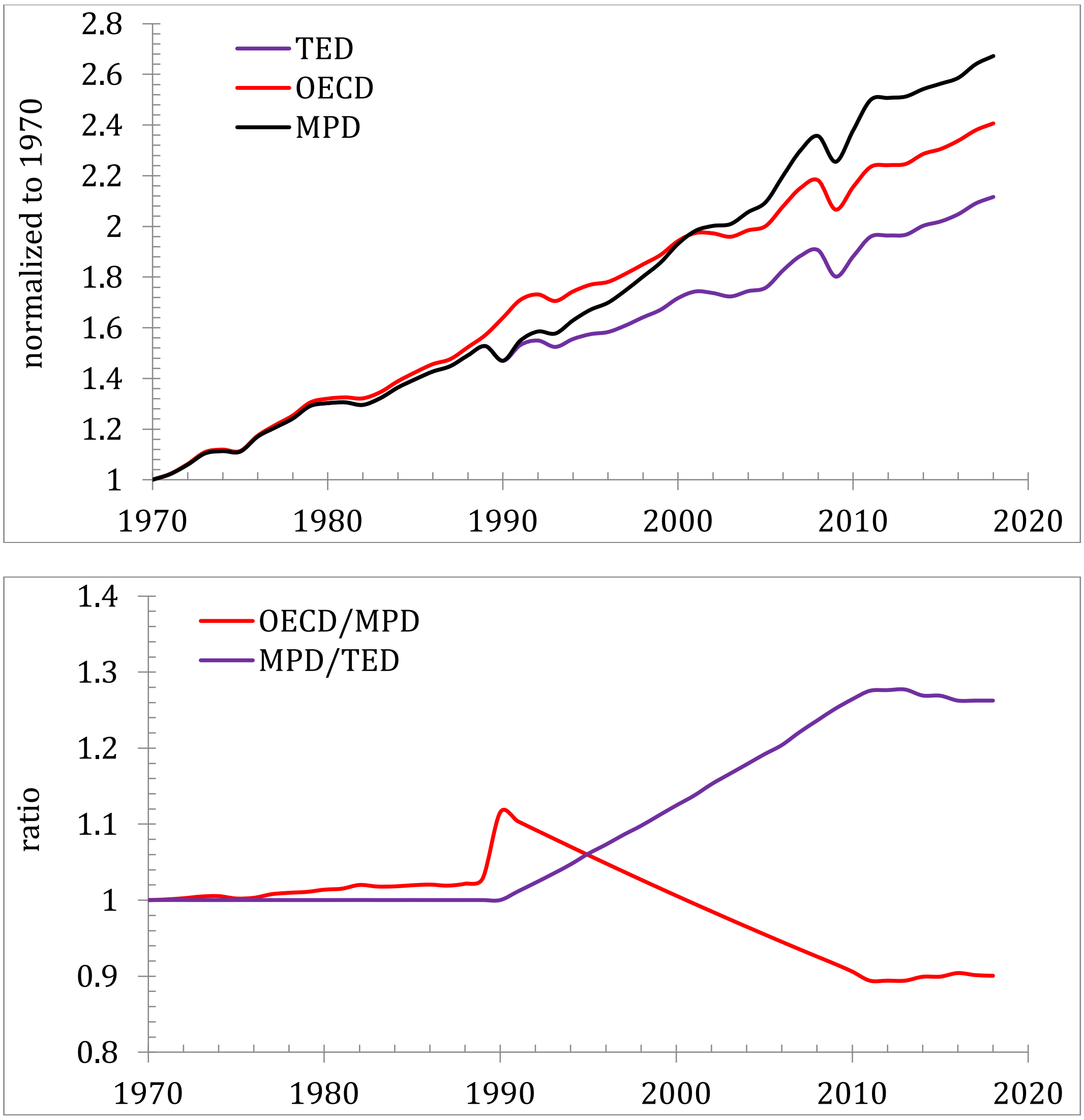

Figure 27. Same as in Figure 9 for Germany

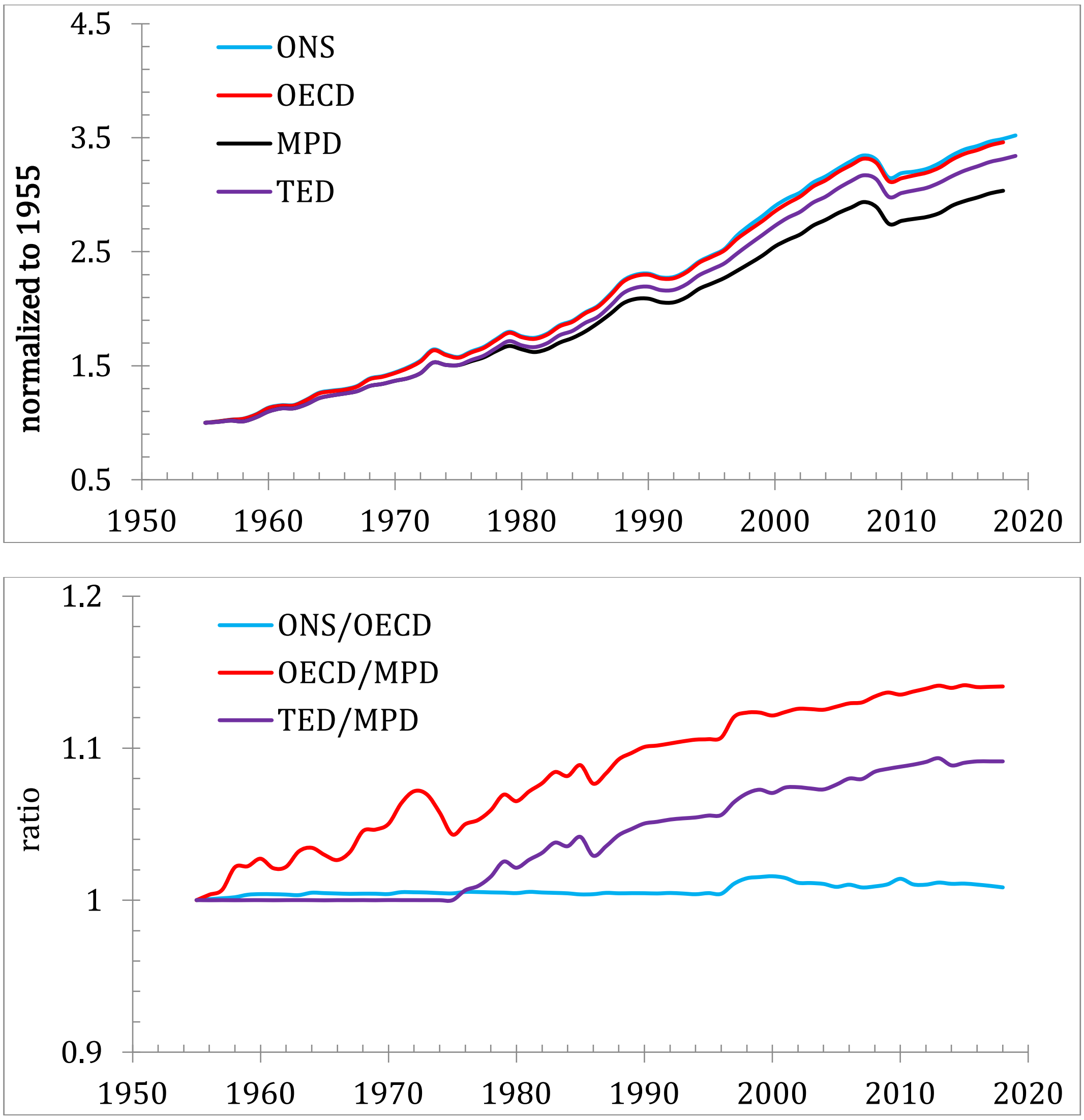

Figure 28. Same as in Figure 9 for the UK

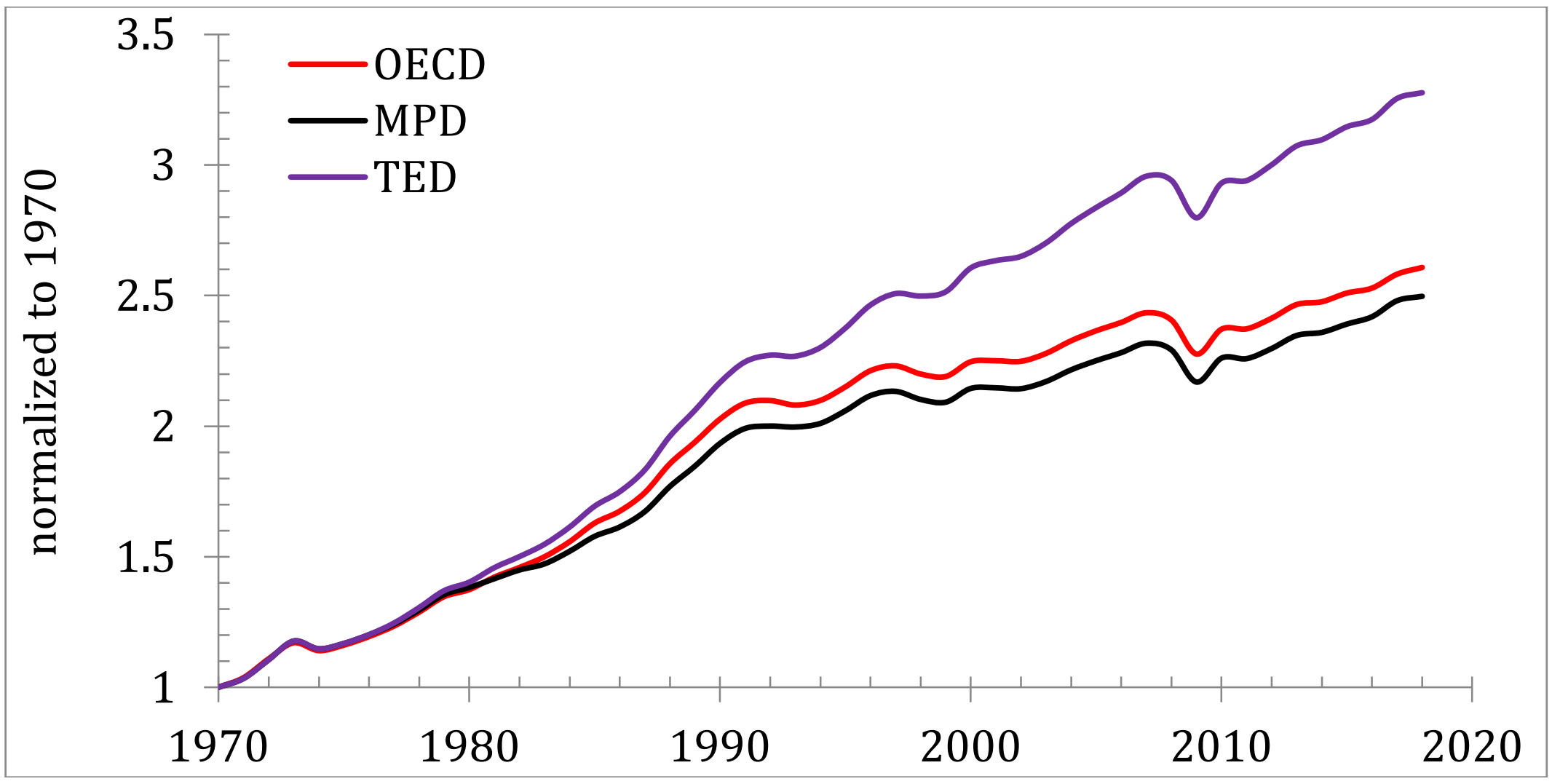

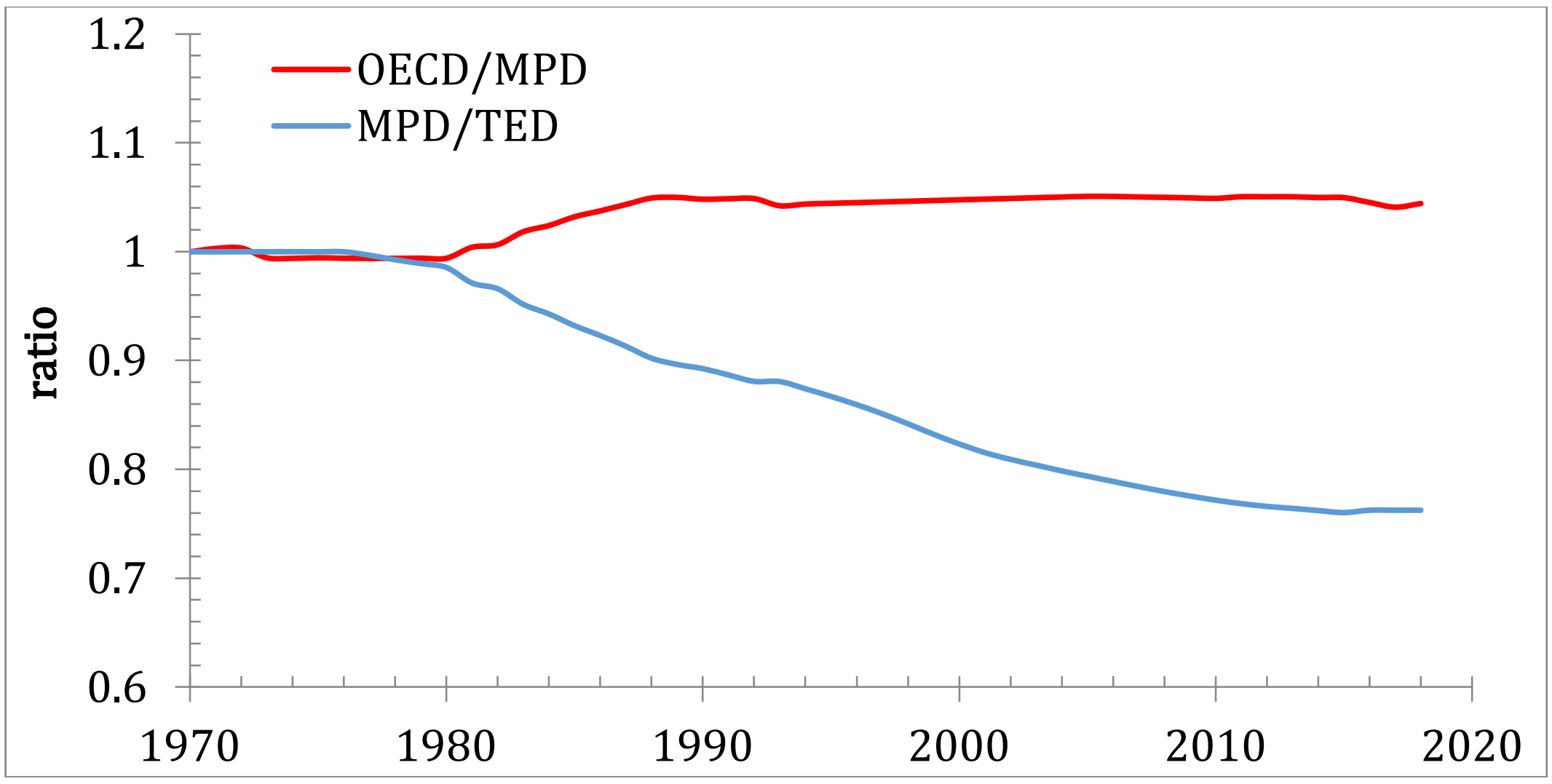

Figure 29. Same as in Figure 27 for Japan

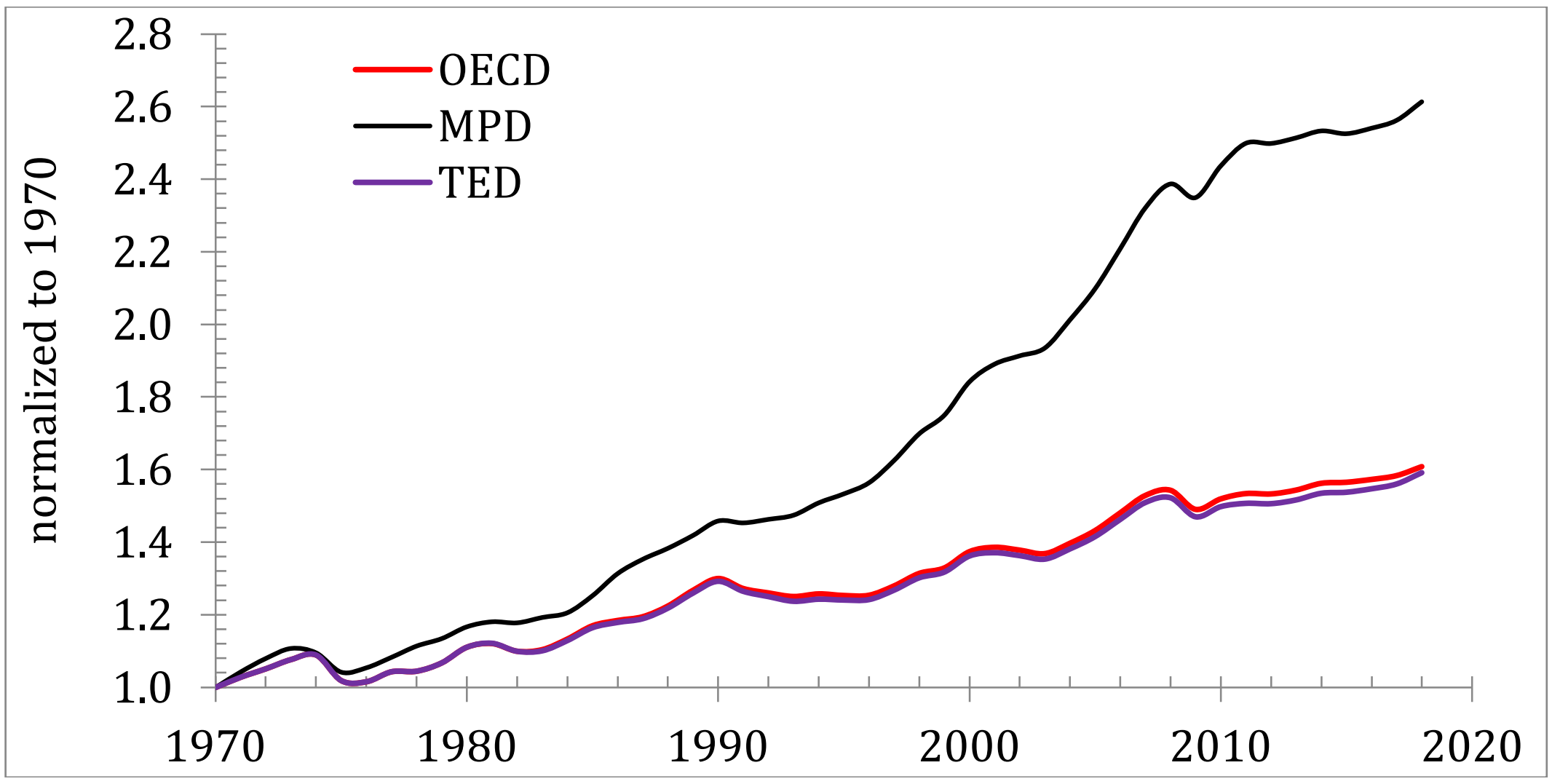

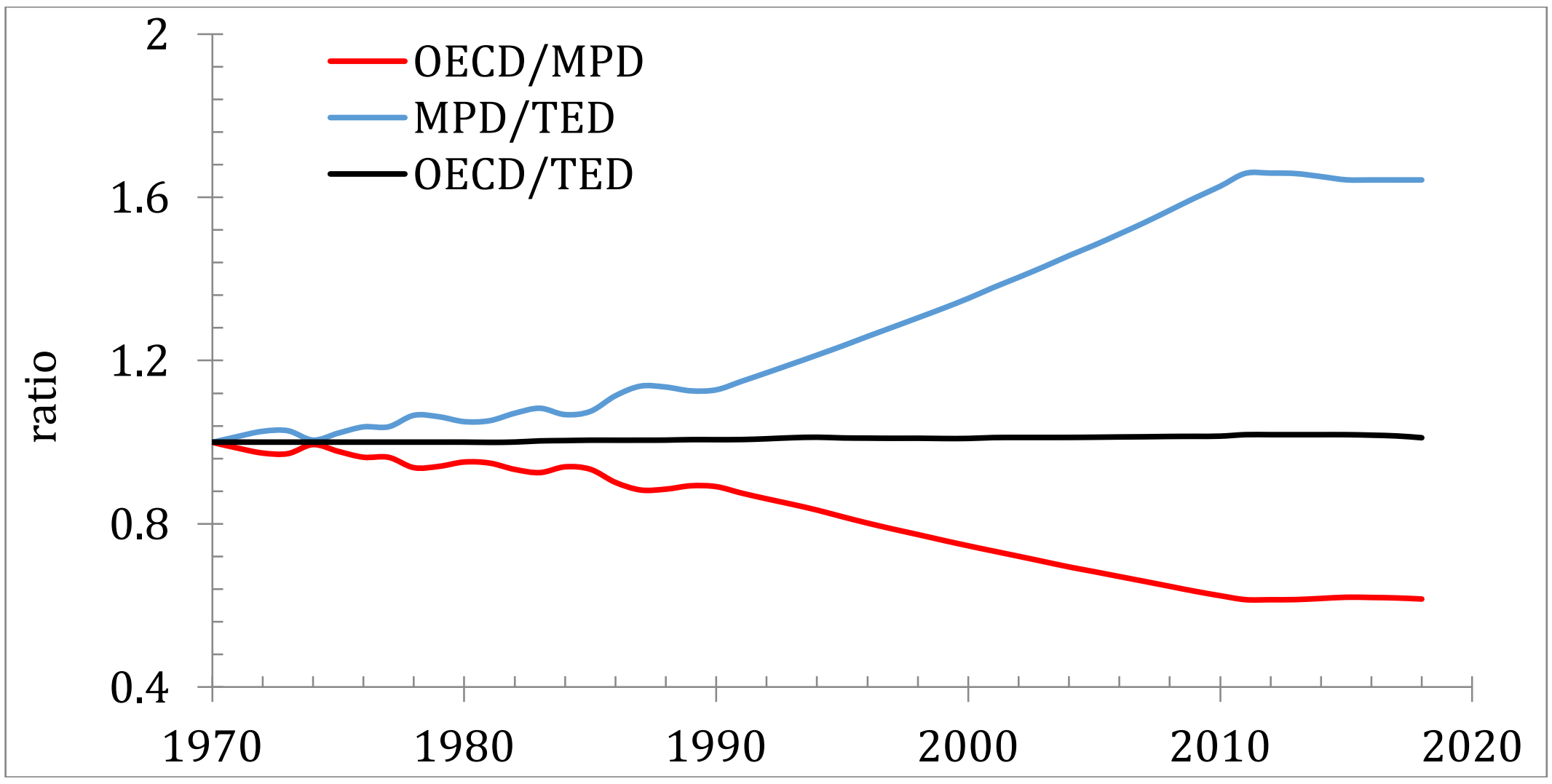

Figure 30. Switzerland. Upper panel: Three time series are normalized to their respective levels in 1970. Lower panel: pair-wise ratios of the normalized curves.

## Conclusion

We have re-estimated eight models of the link between the rate of unemployment and real GDP per capita. Retaining the ideas of Okun's law, for quantitative analysis we used real GDP per capita instead of the overall GDP. With ten new years of data added after the publication of the previous study, the modified Okun's law demonstrates an extraordinary predictive power and validates the models for eight countries: the USA, the UK, France, Germany, Canada, Australia, Spain, and Austria. One can accurately describe the dynamics of unemployment since the 1960s (at least from 1970).

Despite the accurate predictions with the data borrowed from the OECD and MPD as well as from national sources we found striking incompatibility of the GDP data from three sources: the OECD, TED, and MPD. Such differences should not be acceptable for the professional economic community. One cannot assess statistical performance of the real economic growth models using fully incompatible estimates of basic economic variables.